\documentclass[a4paper,11pt]{article}
\pdfoutput=1 % if your are submitting a pdflatex (i.e. if you have
\usepackage{jheppub} % for details on the use of the package, please
\usepackage[utf8]{inputenc}

\usepackage{amsmath,amssymb}
\usepackage{mathtools}
\usepackage[dvipsnames]{xcolor}
\usepackage{physics}
\usepackage{bm}
\usepackage{tikz}
\usepackage{siunitx}
\usepackage{comment}

\newif\ifModC
\ModCtrue

\newcommand{\Ord}[1]{O( #1 )}
\newcommand{\VecZero}{\mathbf{0}}
\renewcommand{\va}[1]{\bm{#1}}

\usetikzlibrary{decorations.markings}
\usetikzlibrary{calc}

\tikzset{every picture/.style={line width=0.3mm}}

\definecolor{pred}{RGB}{238,28,37}
\definecolor{pblue}{RGB}{48,49,146}
\definecolor{pgreen}{RGB}{00,163,80}

\def\centerarc[#1](#2)(#3:#4:#5)% Syntax: [draw options] (center) (initial angle:final angle:radius)
    { \draw[#1] ($(#2)+({#5*cos(#3)},{#5*sin(#3)})$) arc (#3:#4:#5); }

\newcommand{\LegFontSize}{\scriptsize}

\begin{document}

\title{Critical dynamics in a real-time formulation of the functional renormalization group}

%\newcommand{\JLU}{Institut f\"ur Theoretische Physik, Justus-Liebig-Universit\"at, 35392 Giessen, Germany}   
%\newcommand{\HFHF}{Helmholtz Research Academy Hesse for FAIR (HFHF), Campus Giessen, 35392 Giessen, Germany}

%\author{Johannes V. Roth}
%\affiliation{\JLU}

%\author{Lorenz von Smekal}
%\affiliation{\JLU}
%\affiliation{\HFHF}

\author[a]{Johannes V. Roth,}
\author[a,b]{Lorenz von Smekal}

\affiliation[a]{Institut f\"ur Theoretische Physik,\\
  Justus-Liebig-Universit\"at, %Heinrich-Buff-Ring 16, 
  35392 Giessen, Germany}
\affiliation[b]{Helmholtz Research Academy Hesse for FAIR (HFHF),\\Campus Giessen, 35392 Giessen, Germany}

\emailAdd{johannes.v.roth@physik.uni-giessen.de}
\emailAdd{lorenz.smekal@physik.uni-giessen.de}

\abstract{We present first calculations of critical spectral functions of the relaxational Models A, B, and C in the Halperin-Hohenberg classification using a real-time formulation of the functional renormalization group (FRG).
We revisit the prediction by Son and Stephanov that the linear coupling of a conserved density to the non-conserved order parameter of Model~A gives rise to critical Model-B dynamics.
We formulate both 1-loop and 2-loop self-consistent expansion schemes in the 1PI vertex functions as truncations of the effective average action suitable for real-time applications, and analyze in detail how the different critical dynamics are properly incorporated in the framework of the FRG on the closed-time path.
We present results for the corresponding critical spectral functions, extract the dynamic critical exponents for Models A, B, and C, in two and three spatial dimensions, respectively, and compare the resulting values with recent results from the literature.
%Moreover, we discuss the limitations arising on a regulator when one imposes the causality structure of the Keldysh action to hold at all FRG scales during the flow.
%As a main result, we find that a causal regulator is necessarily not UV finite, i.e.~that one can not cut off contributions from large frequencies in the regulated loop integrals without violating causality.
}


%The following command hides all subsubsections in the table of contents
\setcounter{tocdepth}{2}

\maketitle

%\begin{keyword}
%dynamical critical behavior \sep functional renormalization group \sep closed time path \sep universality
%\end{keyword}

%\linenumbers

%\tableofcontents

\section{Introduction}

Real-time quantities such as spectral functions can serve as useful tools to examine the near-equilibrium dynamics of a system as they encode the possible excitations in a medium at finite temperature and density. In the context of probing the phase diagram of QCD with relativistic heavy-ion collisions the electromagnetic spectral function, for example, describes thermal photon and dilepton rates and hence provides direct access to these experimentally measurable penetrating probes~\cite{Weldon:1990iw,HADES:2019auv,Tripolt:2022hhw}. In order to identify signatures of the critical endpoint (CEP) in QCD, one searches for universal critical behavior in event-by-event fluctuations. In addition to its static universality, in the class of the three dimensional Ising model \cite{Halasz:1998qr,Berges:1998rc}, also the dynamic critical behavior of QCD matter near the CEP is of particular interest for a more direct connection with the phenomenology of heavy-ion collisions. Signatures of critical dynamics and non-equilibrium phase transitions are expected to be observable in the cumulants of critical fluctuations near the QCD critical point \cite{Mukherjee:2016kyu,Mukherjee:2017kxv}. 

The relevant dynamic universality classes were classified a long time ago by Hohenberg and Halperin \cite{RevModPhys.49.435}. With the diffusive dynamics of the conserved baryon density coupled to the two diffusive transverse shear modes of the energy-momentum tensor, according to Son and Stephanov \cite{Son:2004iv} the dynamic universality class of the QCD CEP belongs to that of a liquid-gas transition in a pure fluid, called Model H after Hohenberg and Halperin. Without the coupling to the energy-momentum tensor, the dynamics reduces to that of  the coupled dissipative and diffusive fluctuations of chiral order parameter and baryon density according to Models A and B in this classification. Trading the conserved baryon density for a conserved energy density one obtains Model C, on the other hand, which had originally also been suggested for the QCD CEP \cite{Berdnikov:1999ph}. Incentive enough for us to start studying critical spectral functions with Model A, B and C dynamics. 
Since the spectral functions reflect the underlying scale-invariant physics, they also show the corresponding universal infrared behavior as encoded, e.g.~in the dynamic critical exponent $z$~\cite{Berges:2009jz,Schlichting:2019tbr} and new dynamic scaling functions \cite{Schweitzer:2020noq,Schweitzer:2021iqk}.

It is therefore highly desirable to further develop suitable real-time methods for non-perturbative calculations of spectral functions \cite{Roth:2021nrd}.
Available methods range from classical-statistical lattice simulations~\cite{Berges:2009jz,Schlichting:2019tbr,Schweitzer:2020noq,Schweitzer:2021iqk}, over 2PI effective action and Dyson-Schwinger equation studies \cite{Roder:2005vt,Mueller:2010ah,Fischer:2020xnb,Horak:2020eng} to the functional renormalization group~\cite{Floerchinger:2011sc,Kamikado:2013sia,Tripolt:2013jra,Tripolt:2014wra,Wambach:2014vta,Strodthoff:2016pxx,Pawlowski:2017gxj,Huelsmann:2020xcy,Tripolt:2020irx,Tripolt:2021jtp,Jung:2021ipc,Braun:2022mgx,Horak:2022aza}.
%Nicht SFs: Berges:2000ew,Pawlowski:2005xe,

In the context of the functional renormalization group (FRG) there is a distinction between calculations of spectral functions from analytically continued Euclidean (aFRG) flows \cite{Floerchinger:2011sc,Kamikado:2013sia,Tripolt:2013jra,Tripolt:2014wra,Wambach:2014vta,Strodthoff:2016pxx,Pawlowski:2017gxj,Tripolt:2020irx,Tripolt:2021jtp,Jung:2021ipc}, and formulations of the FRG  on the Schwinger-Keldysh closed-time path (CTP) directly in Minkowski spacetime, where the latter are of course more flexible not only to describe the various forms of critical dynamics of interest here, but also for systems fully out of equilibrium~\cite{Canet_2007,Gasenzer:2007za,Gasenzer:2010rq,Berges:2012ty,PhysRevLett.110.195301,Mesterhazy:2013naa,Mesterhazy:2015uja,Duclut:2016jct,Corell:2019jxh}.

Spectral functions at a quantum critical point were calculated in Refs.~\cite{Rose:2015bma,PhysRevB.89.180501} using the analytic continuation from the Euclidean FRG. Following the setup for real-time FRG studies of spectral functions as applied to the anharmonic oscillator in quantum mechanics in Ref.~\cite{Huelsmann:2020xcy}, $O(4)$-model  spectral functions in $3+1 $ dimensions were calculated in Ref.~\cite{Tan:2021zid}.
For a recent overview on FRG applications in  QCD, see~\cite{Fu:2022gou}.

In this paper we start from a single-component real scalar $\varphi^4$-theory with no conservation laws for the dissipative Model A dynamics in $d$ spatial dimensions governed by Langevin equations of motion. We formulate two distinct self-consistent expansions of the effective average action in the 1PI vertex functions around two different expansion points. We calculate the  critical spectral functions and extract the corresponding dynamic critical exponents $z$ in $d=2$ and  $ 3$ spatial dimensions, in order to explicitly verify the quantitative validity of our approach.
As we will see, our  results are generally in good agreement with existing results from the literature. We then  implement dynamical descriptions for systems containing conserved densities which couple to the non-conserved order parameter fluctuations on the level of the Landau-Ginzburg-Wilson  free energy. As a first step towards a description of the full Model H dynamics relevant for the QCD CEP, according Son and Stephanov \cite{Son:2004iv}, this coupling represents the mixing of the chiral order parameter field with the fluctuations of the conserved baryon density where the coupling between the two is linear which yields the diffusive dynamics of Model B.
The resulting infrared (IR) dynamics changes yet again, on the other hand, when the conserved field represents an energy density and is hence coupled to the square of the order parameter field. 
The model then becomes the prototypical example for Model C dynamics in the classification by Halperin and Hohenberg \cite{RevModPhys.49.435} as in Hamiltonian systems without any other conservation laws.

%and a different dynamic critical exponent is obtained \cite{Mesterhazy:2013naa}.

This paper is organized as follows:
In Section~\ref{sct:frg} we briefly summarize the FRG on the CTP and establish our notations. We then discuss regulators and truncation schemes suitable for real-time applications of the FRG.
In Section~\ref{sct:critDynamicsNoConservedQuant} we summarize the basics of Model~A from the Halperin-Hohenberg classification, and introduce two different truncations of the effective average action in 1PI vertex functions which both rely on simultaneous expansions in loops and spatial gradients in Sections \ref{sct:ExpAroundScaleDepMinforA} and \ref{sct:ExpAroundIRMinforA}.
In Section~\ref{sct:ModB} we formulate Model~B after Son and Stephanov \cite{Son:2004iv} on the CTP, and straightforwardly extend our truncation scheme from Section~\ref{sct:ExpAroundScaleDepMinforA} to include the linear coupling to a conserved density via Gaussian integration.
In Section~\ref{sct:ModC} we change the coupling from linear to quadratic in the order parameter field to arrive at critical Model-C dynamics with a conserved energy density \cite{RevModPhys.49.435}. Based on 
the scheme introduced in Section~\ref{sct:ExpAroundIRMinforA} for Model A, we
formulate a simple but suitable truncation which already suffices, however, to see the critical dynamics in the spectral functions.
In Section~\ref{sct:results} we discuss our results on the critical spectral functions, and the resulting dynamic critical exponents for the three Models.
We conclude with a brief summary and our outlook for future work in Section~\ref{sct:conclusion}.
Further details on causality and Kramers-Kronig relations, the FRG flow equations and the numerical implementation are provided in several appendices.

%All theories we are inspecting are defined classically.
%Hence the UV cutoff $\Lambda$ of the FRG is physically identified with a scale where a stochastic hydrodynamic description in terms of mesoscopic effective degrees of freedom as an effective field theory (EFT) for the underlying microscopic dynamics of QCD becomes applicable.
%For classical dynamics to be a good approximation we necessarily need the (mesoscopic) fields to obey classical statistics, such that the distribution function is essentially given by the Rayleigh-Jeans distribution.
%Therefore, when we speak about the `classical limit' in this work, it is always understood as that the relevant scales in the system and, importantly, the UV cutoff $\Lambda$ are small compared to the temperature, $ \Lambda \ll T $.

%\section{Causal regulators}

%We propose that the most general regulator that respects the causal structure of the Keldysh action has the spectral representation
%\begin{align}
%    \label{causalRegGeneral}
%    R^{R/A}_k(\omega, \va{p}) &= - \int_0^\infty \frac{d\omega'}{2\pi} \frac{2 \omega' \, J_k(\omega',\va{p})}{ (\omega \pm i \varepsilon)^2 - \omega'^2 }  - \Delta M_k(\va{p}) \, ,
%\end{align}
%with spectral density
%$J_k(\omega,\va{p})$ which is an anti-symmetric function of the frequency $J_k(-\omega,\va{p}) = -J_k(\omega,\va{p})$ and positive-definite for $\omega >0$, and with the constant mass-like shift $ \Delta M_k(\va{p}) > 0$.

\section{Functional renormalization group on the closed-time path}
\label{sct:frg}

Wetterich's FRG flow equation \cite{Wetterich:1992yh,Berges:2000ew,Pawlowski:2005xe} for the so-called effective \emph{average} action implements Wilson's idea \cite{Wilson:1973jj} of successively integrating out fluctuations.
It assumes that the effective average action $\Gamma_k$ is known at some momentum scale~$k$, where fluctuations of all modes with momenta $|\va{p}| \lesssim k$ are effectively suppressed.
For $k \to \Lambda$ the effective average action  reduces to an initial bare action $\Gamma_\Lambda \cong S$ at some ultraviolet (UV) cutoff scale $\Lambda$, while in the limit $k \to 0$ one formally obtains the full effective action of the theory, $\lim_{k\to 0} \Gamma_k = \Gamma$.
In its formulation on the closed-time path it is given by \cite{Berges:2012ty,Pawlowski:2015mia,Huelsmann:2020xcy}
\begin{align}
    \partial_k \Gamma_k[\phi^c, \phi^q] = \frac{i}{2} \mathrm{Tr}\left\{ \partial_k R_k \circ \left( R_k + \Gamma^{(2)}_k[ \phi^c, \phi^q ] \right)^{-1} \right\} \, , \label{wetterichEq}
\end{align}
where the classical and quantum field components are defined by a Keldysh 
rotation 
\begin{align}
    \phi^{c,q}(x) = \frac{1}{\sqrt{2}} \left( \phi^+(x) \pm \phi^-(x) \right) \, , \label{keldyshRot}
\end{align}
of the fields $\phi^{\pm}(x)$ on the forward $(+)$ and backward $(-)$ branches of the closed-time path.

The full field-dependent propagator is in compact matrix notation given by
\begin{align}
    G_k[\phi^c,\phi^q] = -\left(R_k + \Gamma^{(2)}_k[ \phi^c, \phi^q ] \right)^{-1} \, , \label{eq:fullPropDef}
\end{align}
in terms of matrix valued regulator $R_k$ and two-point function $\Gamma^{(2)}_k$.
As usual, the field-dependent propagators on the closed-time path are given by the following connected 2-point correlation functions (in presence of sources $j^{c,q}$),
\begin{align}
    G_k(x,x') &\equiv  \begin{pmatrix}
        i \langle \phi^c(x) \phi^c(x') \rangle_j &  i \langle \phi^c(x) \phi^q(x') \rangle_j \\
        i \langle \phi^q(x) \phi^c(x') \rangle_j & 
        i \langle \phi^q(x) \phi^q(x')\rangle_j
    \end{pmatrix} 
   =  \begin{pmatrix}
        G_k^K( x,x' ) & G_k^R( x,x' ) \\
        G_k^A( x,x' ) &  G_k^{\widetilde{K}}( x,x' )
    \end{pmatrix}
    \,,
\end{align}
where the causality structure entails that the lower right `anomalous' $G_k^{\widetilde{K}}=i \langle\phi^q \phi^q\rangle_j$ correlation function must vanish for $j^q=0$, in order 
to ensure that the effective average action stays zero for a vanishing  expectation value 
%$\phi^q(x) = 0$~\cite{kamenev_2011}.
$\phi^q = 0$ of the  response-field~\cite{kamenev_2011}.
At this point, it is convenient to follow Refs.~\cite{Huelsmann:2020xcy,Roth:2021nrd} and introduce the shorthand notations
\begin{subequations}
\begin{align}
    B_k^R(x,x') &= \int_{yy'} G_k^R(x,y)\, \partial_k R_k^R(y,y')\, G_k^R(y',x') \, , \label{BRDef} \\
    B_k^A(x,x') &= \int_{yy'} G_k^A(x,y)\, \partial_k R_k^A(y,y')\, G_k^A(y',x') \, , \label{BADef} \\
    B_k^K(x,x') &= \int_{yy'} \Big[ G_k^R(x,y)\, \partial_k R_k^K(y,y')\, G_k^A(y',x') + G_k^R(x,y)\, \partial_k R_k^R(y,y')\, G_k^K(y',x') \nonumber \\
    &\hspace{2.0cm} + G_k^K(x,y)\, \partial_k R_k^A(y,y')\, G_k^A(y',x') \Big] \, , \label{BKDef}  
\end{align}
\end{subequations}
for propagators $B_k^R(x,x')$, $B_k^A(x,x')$, and $B_k^K(x,x')$ which have the same legs as their counterparts $ G_k^R(x,x')$, $G_k^A(x,x')$, and $G_k^K(x,x')$, but with all possible regulator derivatives inserted in between.
We also use the shorthand notations
\begin{align*}
     \int_x \equiv \int d^D x\, , \;\;\mbox{or} \hspace{0.5cm} \int_q \equiv \int \frac{d^D q}{(2\pi)^D} 
\end{align*}
for integrals in coordinate or momentum space in $D\equiv d+1$ dimensional  spacetime.

Correlation functions in thermal equilibrium satisfy the Kubo-Martin-Schwinger (KMS) condition~\cite{doi:10.1143/JPSJ.12.570,PhysRev.115.1342}, which implies a fluctuation-dissipation relation (FDR) between the Fourier-transformed propagators in a translationally  invariant system,
\begin{align}
    G_{k}^K(\omega,\va{p}) &= \coth\left(\frac{\omega}{2T}\right) \left( G_{k}^R(\omega,\va{p}) - G_{k}^A(\omega,\va{p}) \right) \,, \label{FDRForG}
\end{align}
and correspondingly between the different components of the 2-point functions,
\begin{align}
    \Gamma_{k}^{qq}(\omega,\va{p}) &= \coth\left(\frac{\omega}{2T}\right) \left( \Gamma_{k}^{qc}(\omega,\va{p}) - \Gamma_{k}^{cq}(\omega,\va{p}) \right) \,, \label{FDRForGam2}
\end{align}
and in general also for higher order $n$-point functions \cite{Wang:1998wg}, provided that the regulator also shares the symmetry of thermal equilibrium \cite{Sieberer:2015hba}.
We can insert the FDR's \eqref{FDRForG} and \eqref{FDRForGam2} to find a corresponding FDR between the $B$'s,
\begin{align}
    B_k^K(\omega,\va{p}) = \coth\left(\frac{\omega}{2T}\right) \left( B_k^R(\omega,\va{p}) - B_k^A(\omega,\va{p}) \right) \, , \label{FDRForB}
\end{align}
analogous to Eq.~\eqref{FDRForG}.
In classical-statistical systems, the bosonic distribution function in all these relations is of course replaced by its Rayleigh-Jeans limit, $\coth(\omega/2T) \to 2T/\omega$.

For the numerical results in this work, we have chosen a frequency-independent version of the optimized Litim regulator~\cite{Litim:2001up},
\begin{align}
    R_k^{R/A}(\omega,\va{p}) &= -Z_k^\perp (k^2-\va{p}^2) \theta(k^2-\va{p}^2) \,, \hspace{0.5cm} R_k^K(\omega,\va{p}) = 0\,, \label{eq:litimReg}
\end{align}
as it is often done in finite-temperature field theory, here with the spatial wave function renormalization factor $Z_k^\perp$ included to ensure that no artificial scale is introduced \cite{Berges:2000ew}. 
This choice of a frequency independent regulator is the arguably simplest way to maintain the causal structure of the Keldysh action, and it turns out to be sufficient for the practical applications considered below. If a frequency dependent regulator is required for some reason, however, the question of how to maintain causality and thus the K\"all\'en-Lehmann spectral representation of the propagators during the flow becomes much more subtle \cite{Floerchinger:2011sc,Pawlowski:2015mia,Duclut:2016jct,Pawlowski:2017gxj,Braun:2022mgx}. The field-theory generalization of our physics motivated and intuitive solution to this problem in quantum-mechanical systems \cite{Roth:2021nrd} in the context of the Keldysh closed-time path formalism is presented in the next subsection.

\subsection{Causal regulators}
\label{sct:causalReg}

Assume adding some regulator part to the Keldysh action in the functional integral on the CTP that is bilinear in the fields $\Phi = (\phi^c,\phi^q)^T $ and spacetime translation invariant,
\begin{equation}
  \Delta S_k[\Phi] = \frac{1}{2} \int_{xy}  \Phi(x)^T R_k(x-y) \Phi(y) \, . \label{regSk}
\end{equation}
In order to maintain the causal structure of the Keldysh action, the regulator matrix $R_k$ is required to be of the form of a self-energy, i.e.~(after Fourier transformation)
\begin{align}
    R_k(\omega,\boldsymbol{p}) &= \begin{pmatrix}
        0&R_k^{R}(\omega,\boldsymbol{p}) \\
        R_k^{A}(\omega,\boldsymbol{p})&R_k^{K}(\omega,\boldsymbol{p})
    \end{pmatrix} \,.  \label{regRA}
\end{align}
As any self-energy it can then always be represented, via Hubbard-Stratonovich linearization on the CTP, by a linear coupling of the fields $\Phi $ to a Gaussian ensemble of bosonic degrees of freedom. The modelling of such a causal self-energy regulator can therefore generally be shifted into the modelling of the FRG scale $k$ dependent spectral distribution $J_k(\omega, \boldsymbol p) $ of this ensemble to represent the regulator $R_k(\omega,\boldsymbol{p})$ via dispersion relations, as we will discuss explicitly below. Although one might thus intuitively think of an artificial scale-dependent heat bath to provide the damping of low frequency and momentum modes in a causal manner, there is more flexibility to choose such an artificial spectral distribution as it 
does not necessarily have to represent an ensemble of physical degrees of freedom, with a positive and normalizable spectral distribution. 

In order to demonstrate this explicitly, 
we now assume that the regulator (\ref{regRA})  on the closed-time path depends on frequency $\omega$ \emph{without} violating causality, so that we can discuss the subsequent constraints that arise on its structure.\footnote{Although we use the Keldysh formalism on the CTP here, with causality, any conclusion on the structure of the regulator \eqref{regRA} can be mapped one-to-one to a corresponding Euclidean regulator $R_k^E(\omega_E,\boldsymbol{p})$ with real Euclidean frequency $\omega_E$ by analytic continuation, via $R_k^E(\omega_E,\boldsymbol{p}) = -R_k^R(i|\omega_E|,\boldsymbol{p}) = -R_k^A(-i|\omega_E|,\boldsymbol{p})$.}
As a starting point for our discussion and as a general guiding principle, note the structural resemblance of  the definition of the full propagator in the real-time FRG (here for a vanishing response-field expectation value $\phi^q(x) = 0$),
\begin{align}
    %- G_k^{-1} &= \Gamma_k^{(2)} + R_k \\
    -\begin{pmatrix}
        G_k^K & G_k^R \\
        G_k^A & 0
    \end{pmatrix}^{-1} &=
    \begin{pmatrix}
        0 & \Gamma_k^{cq} \\
        \Gamma_k^{qc} & \Gamma_k^{qq}
    \end{pmatrix} +
    \begin{pmatrix}
        0 & R_k^A \\
        R_k^R & R_k^K
    \end{pmatrix} \,, \label{eq:dysonEqWithReg}
\end{align}
with a Dyson equation.
A quite non-trivial feature of the Keldysh technique is that the self-energy matrix, as defined by such a Dyson equation, generally inherits the causality structure of the Keldysh action \cite{kamenev_2011}.
It is thus a sufficient (if not even necessary) condition that the regulator also inherits this causal structure for causality to be conserved during the FRG flow.
We can in fact turn this argument around, and interpret the statement that the regulator has the causal structure of a self-energy matrix, defined by a Dyson equation, as the technical definition of `complying with causality.'
Imposing the causal structure on the regulator tightly restricts its frequency dependence, as we shall see in the following.

First, this causal structure requires that retarded and advanced parts are connected by complex conjugation, i.e.~${R_k^R}^*(\omega,\boldsymbol{p}) = R_k^A(\omega,\boldsymbol{p})$.
%$R_k^R(-\omega,\boldsymbol{p}) = R_k^A(\omega,\boldsymbol{p})$.
Moreover,
in order to have real regulators for real scalar fields in the time domain, so that $\Delta S_k[\Phi]$ in (\ref{regSk}) is real, we must also  have $R_k^R(-\omega,\boldsymbol{p}) = {R_k^R}^*(\omega,\boldsymbol{p}) $. This implies that the real/imaginary parts are even/odd in $\omega $.

Because such a regulator thus has all the necessary analyticity properties of a retarded/advanced self-energy, we can therefore also write down a spectral representation. In order to include the class of frequency-independent regulators, we decompose the regulator in a frequency-dependent and a frequency-independent part. This can be done in various ways: If the regulator has a finite, non-vanishing and unique limit for $\omega$ towards complex infinity, this limit will necessarily be real and we can subtract it to define a standard spectral representation for the remainder which then vanishes for $\omega\to \pm\infty $. While this assumption might seem reasonable for regulators, self-energies in interacting theories typically do have singularities at infinity. Therefore, we alternatively assume here that the imaginary part of $ R_k^{R/A}(\omega,\boldsymbol{p})$ vanishes for $\omega \to 0$, implying that the regulator does not introduce artificial massless excitations which could otherwise introduce infrared divergences. We call this the assumption of \emph{infrared finiteness} which was one of the main original motivations for the Euclidean FRG \cite{Berges:2000ew}.

\paragraph{IR finiteness:}
If the regulator is analytic at $\omega=0$, we can define a real and momentum dependent {\em mass shift} via 
\begin{align}
    \Delta M_k^2(\boldsymbol{p}) \equiv  - R_k^{R/A}(0,\boldsymbol{p}) \, . \label{massShiftIR}
\end{align}
This frequency-independent part is trivially causal and can be chosen as convenient, with the usual properties that any regulator should have \cite{Gies:2006wv}. In particular,
it must necessarily be positive for all FRG scales~$k$ in order to properly regulate IR modes without introducing artificial acausal regulator singularities \cite{Roth:2021nrd}, i.e.\
$   \Delta M_k^2(\va{p}) >  0$.
Because it does not vanish at infinity in the complex $\omega$-plane, we then need to write down a subtracted spectral representation for $ R_k^{R/A}(\omega,\boldsymbol{p})$,  based on the analytic properties of $(R_k^{R/A}(\omega,\boldsymbol{p}) -  R_k^{R/A}(0,\boldsymbol{p}))/\omega $ and assuming that $R_k^{R/A}(\omega,\boldsymbol{p})$  is analytic at $\omega = 0$, see  Appendix~\ref{sct:KramersKronigRels} where the corresponding Kramers-Kronig relations are given as well. It amounts to writing
\begin{equation}
  R^{R/A}(\omega, \va{p}) \equiv - \Delta M_k^2(\boldsymbol{p})  + \Sigma_k^{R/A}(\omega,\boldsymbol{p}) \, , \label{regulatorSPDef}
  \end{equation}
where the frequency-dependent part $\Sigma_k^{R/A}(\omega,\boldsymbol{p})$, which we will refer to as the `spectral part' of the regulator in the following, is given by
\begin{equation}  
  \Sigma_k^{R/A}(\omega,\boldsymbol{p})= 
  - \int_{0}^\infty \frac{d\omega'}{2\pi} \, \frac{2 \omega^2 J_k(\omega', \va{p})}{\omega'((\omega \pm i\varepsilon)^2 - \omega'^2)} \, . \label{sigmaRADef}
\end{equation}
It is expressed in terms of an FRG-scale $k$ dependent spectral density $J_k(\omega,\va{p})$ which is in turn given by the imaginary part of the regulator itself,
\begin{align}
    J_k(\omega,\boldsymbol{p}) = \pm 2\Im \Sigma_k^{R/A}(\omega,\boldsymbol{p}) = \pm 2\Im R_k^{R/A}(\omega,\boldsymbol{p}) \, . \label{spectralDensHB}
\end{align}
This spectral density must thus be an odd function of frequency as well, $J_k(-\omega,\boldsymbol{p}) = -J_k(\omega,\boldsymbol{p})$. Since we have assumed that the imaginary part of $ R_k^{R/A}(\omega,\boldsymbol{p})$ vanishes for $\omega \to 0$, the integration limit of the spectral integral (\ref{sigmaRADef}) in the infrared exists. One can furthermore see explicitly %in (\ref{sigmaRADef}) 
that the full spectral part of the regulator vanishes for  $\omega \to 0 $, i.e.\ $ \Sigma_k^{R/A}(0,\boldsymbol{p})= 0$, as it must by construction. 

In order to include such a spectral part in a regulator for real-time FRG applications on the CTP one can therefore start with first devising a suitable  imaginary part, i.e.~a scale and frequency-dependent spectral density $J_k(\omega,\boldsymbol{p})$ which may be thought of as modelling an artificial external bath, and then compute the corresponding real part from the (subtracted) Kramers-Kronig relation.    

 For the physical interpretation of such a non-vanishing spectral density, we reiterate 
 %notice that \eqref{sigmaSpectralRep} 
 that such a self energy could have also been obtained by  coupling our system to an external heat bath modelled as an ensemble of harmonic oscillators  after Caldeira and Leggett \cite{CALDEIRA1983587} or, equivalently, as a Gaussian ensemble of bosonic degrees of freedom, upon Gaussian  integration of this ensemble. 
 % see e.g.~Chapter 3.2 of Ref.~\cite{kamenev_2011}.
 Therefore, a causal regulator with a non-vanishing spectral part \eqref{regulatorSPDef}, \eqref{sigmaRADef} can always be interpreted as arising from interactions with a fictitious  external heat bath, specified by the scale dependent spectral density \eqref{spectralDensHB}, which  motivated the name `heat-bath regulator' in earlier work~\cite{Roth:2021nrd}.

If we translate our general expression \eqref{sigmaRADef} for the spectral part of a causal regulator to the Euclidean domain, we obtain the corresponding Euclidean regulator as
\begin{align}
    R_k^E(\omega_E,\boldsymbol{p}) &= - R_k^R(i|\omega_E|,\boldsymbol{p}) = \Delta M_k^2(\boldsymbol{p}) + \int_0^\infty \frac{d\omega'}{\pi} \frac{\omega_E^2 \, J_k(\omega',\boldsymbol{p})}{\omega' (\omega_E^2 + \omega'^2)}
\end{align}
for real Euclidean frequencies $\omega_E$.
Notably, the spectral part, for a positive spectral density, always adds a positive contribution to the positive  mass shift $\Delta M_k^2(\boldsymbol{p})$.
In particular, for  $\omega_E \to \infty $, this contradicts the UV finiteness of the Euclidean regulator in the frequency argument, as we discuss next.

\paragraph{UV finiteness:}

If the spectral density of the regulating bath vanishes for $\omega \to 0$ as we have assumed here, then we can also take the limit $\omega\to\infty $ in (\ref{sigmaRADef}) to compute the ultraviolet limit of the regulator from 
\begin{equation}
\lim_{\omega \to \pm \infty}   R^{R/A}(\omega, \va{p}) = - \Delta M_k^2(\boldsymbol{p})  - \int_0^\infty \frac{d\omega'}{\pi} \frac{J_k(\omega',\va{p})}{\omega' } \, .
  \end{equation}
This shows that a semi-positive spectral density $J_{k}(\omega,\va{p}) \ge 0 $, corresponding to any kind of physically motivated external bath, together with a positive frequency-independent $  \Delta M_k^2(\va{p}) >  0$ is necessarily inconsistent with $ R^{R/A}(\omega, \va{p}) \to 0 $ for $\omega\to\infty $.
In other words, in such a physics motivated setup it is not possible to simultaneously cut off frequency integrals \emph{and} respect the causal structure of the Keldysh action at all intermediate FRG scales~$k$ during the flow.
Moreover, this result does not depend on our substraction at $\omega = 0$ here. In fact the same conclusion is obtained with substraction at complex infinity in Appendix~\ref{sct:KramersKronigRels}.

\paragraph{Lorentz invariance:}

Another issue frequently discussed in the literature \cite{Braun:2022mgx} for real-time applications of the FRG in vacuum concerns Lorentz invariance of regulator and flow equations. Of course, a literal heat bath can at best be Lorentz covariant because it defines a preferred frame.
In the context of our causal regulator \eqref{regRA} Lorentz invariance can be implemented assuming that the Gaussian ensemble of bosonic degrees of freedom is represented by local systems  of Klein-Gordon fields in vacuum quantum field theory, which one can think of as a relativistic version of the Caldeira-Leggett model. This implies that the corresponding spectral density can be expressed as
\begin{align}
	J_k(\omega,\va{p}) = 2\pi \, \mathrm{sgn}(\omega) ~ \theta(p^2) ~ \widetilde{J}_k(p^2)
\end{align}
in terms of the  \emph{invariant} spectral distribution $\widetilde{J}_k(p^2)$ which is a function of the squared invariant mass $s = p^2 = \omega^2 - \va{p}^2$ alone.
In this case, the spectral part~\eqref{sigmaRADef} assumes the form of a standard (subtracted) K\"all\'en–Lehmann spectral representation,
\begin{align}
    \label{bathGeneralSelfEn-li}
    \Sigma_k^{R/A}( \omega,\va{p} ) = -\int_0^\infty \!\! ds \; \frac{ (\omega^2 - \va{p}^2)\, \widetilde{J}_k( s ) }{s \, ((\omega \pm i \varepsilon)^2 - \va{p}^2 - s)} \,.
\end{align}
Moreover, the frequency independent $\Delta M_k(\va{p}) \equiv \Delta M_k$ must then also be independent of the spatial momentum $\va{p}$ in order to be consistent with Lorentz invariance. The $\omega = 0$ substraction has become a light-cone substraction which is allowed, as long as $\widetilde{J}_k(s)$ does not contain massless single-particle contributions (continuous contributions are allowed as long as $\widetilde{J}_k(s) \to 0 $ for $s\to 0$).

As a practical example, consider the problem of respecting the $O(D)$ invariance of $D$-dimensional Euclidean spacetime in a causality-preserving way in the standard formulation of the FRG \cite{Floerchinger:2011sc,Pawlowski:2015mia,Pawlowski:2017gxj}.
Assuming the causal structure of the corresponding Keldysh formulation in thermal equilibrium, a straightforward analytic continuation of \eqref{bathGeneralSelfEn-li} requires the Euclidean regulator to be of the form
\begin{align}
    R_k^E(\omega_E,\va{p}) = \Delta M_k^2  + \int_0^\infty\!\!ds  \;  \frac{ (\omega_E^2+\va{p}^2) \, \widetilde{J}_k( s) }{s\,(\omega_E^2+\va{p}^2 + s)} \label{regEuclSpecRep}
\end{align}
with real Euclidean frequency $\omega_E$, i.e.~to consist of a Callan-Symanzik regulator plus some positive  shift which depends on the Euclidean $O(D)$ invariant squared momentum variable $p_E^2 = \omega_E^2+\va{p}^2$.
This form of the Euclidean regulator then guarantees the existence of a spectral representation of the propagator at all FRG scales $k$, which is not generally the case for an arbitrary regulator and requires extra attention~\cite{Pawlowski:2015mia,Pawlowski:2017gxj}.
Moreover, note that the spectral integral in the momentum-dependent part of the regulator in 
 \eqref{regEuclSpecRep} is ultraviolet finite only as long as 
$ \widetilde{J}_k(s)/s \to 0$  for $ s\to\infty $.
While it is thus not intrinsically UV finite, for an arbitrary spectral density,  this is not a severe restriction  for a regulator, for which  we intuitively expect that $\widetilde{J}_k(s) \to 0 $ for $s \gg k^2  $, anyway. On the other hand, and maybe more importantly however, without the Callan-Symanzik mass shift,  i.e.\ for $\Delta M_k^2 = 0 $,
this expectation would  in turn  imply for $p_E^2 \gg s \sim k^2 $ a  \emph{subtracted superconvergence relation} for the artificial regulator spectral density here,
\begin{equation}
\int_0^\infty\!\!ds  \;  \frac{  \widetilde{J}_k( s) }{s} = 0\, ,
\end{equation}
in order to have
\begin{equation}
R^E_k(p_E^2) \to 0 \, \;\; \mbox{for} \;\; p_E^2 \gg k^2 \, , 
\end{equation}
for our $O(D)$-invariant regulator with spectral representation. In particular, this also shows that it is not possible to maintain the positivity of the regulator spectral distribution $\widetilde{J}_k(s) $, \emph{and} to have the Wilsonian realization of integrating out momentum shell by momentum shell in the FRG flow, at the same time. Instead, a flow generated by \eqref{regEuclSpecRep} with a positive spectral distribution $\widetilde{J}_k(s) \ge 0 $ then necessarily has to be interpreted in the sense of Callan-Symanzik flows, shifting all squared masses uniformly by the square of the FRG scale~$k^2$, and it  thus represents a flow through `theory space' \cite{Braun:2022mgx}.

%Recently, a regulator like \eqref{regEuclSpecRep} with a vanishing spectral density was employed in the context of calculating the phase diagram of the quark-meson model \cite{Otto:2022jzl}, where it was found that it is indeed able to resolve the problem of unphysical negative entropy densities, compared to the more common choice of the purely spatial optimized regulator~\cite{Litim:2001up}.

\paragraph{Causality, Lorentz invariance, UV and IR finiteness:}

It is now straightforward to put together these individual requirements and discuss the consequences.
Lorentz invariance entails that the regulator only depends on the invariant momentum, $R_k^{R/A}= R_k^{R/A}(p^2)$. Causality then requires it to be an analytic function in the cut-complex $p^2$ plane, with the only singularities at the discontinuity along the timelike real axis. The causally regulated theory corresponds to an open system where the self-energy regulator \eqref{regSk} has become the result of integrating a Gaussian ensemble of bosonic degrees of freedom which we may call the \emph{environment} (with Lorentz invariance it is not a heat-bath, of course, but some local field system). Analytic continuation to the Euclidean domain and the existence of a spectral representation are then guaranteed. Ultraviolet finiteness of the Euclidean FRG flows demands that $R^{R/A}_k(p^2) = 0  $ for $-p^2 \gg k^2$. For timelike $p^2>0$ the imaginary part is given by the invariant spectral distribution $\widetilde{J}_k(p^2) $ which describes the interactions of the system with states in the environment of total momentum $p$. The causal extension of the Wetterich equation leads to the conclusion that such artificial interactions in the regulated theory should not occur for $p^2 \gg k^2$, implying that $\widetilde{J}_k(p^2) = 0$, for  $p^2 \gg k^2$. The self-energy regulator therefore then vanishes in all directions at complex infinity of the $p^2$-plane, and no substraction in its spectral representation is required in this case, in the first place. Instead of \eqref{regulatorSPDef}, \eqref{sigmaRADef} we can then simply write
\begin{align}
    R_k^{R/A}(p^2) =  -\int_0^\infty \!\! ds \; \frac{  \widetilde{J}_k( s ) }{(\omega \pm i \varepsilon)^2 - \va{p}^2 - s} \,, 
\end{align}
see Appendix \ref{sct:KramersKronigRels}. Without massless excitations introduced by the regulator, we can furthermore take the limit $p^2\to 0 $ here and obtain for the mass shift
\begin{align}
    \Delta M_k^2 = R_k^E(0)  = - \int_0^\infty \!\! ds \; \frac{  \widetilde{J}_k( s ) }{s} \stackrel{!}{\ge} 0 \, , 
\end{align}
which must be positive to avoid tachyonic regulator singularities. This shows that causality, Lorentz invariance plus UV and IR finiteness together require that the spectral density of the artificial environment cannot be positive. The  bosonic fields in the Gaussian ensemble of the environment thus necessarily violate positivity just as BRST quartets do in covariant gauge theory.\footnote{The closed total system including the artificial environment can still represent a local field system with causality, but there is no spin-statistics theorem for which one needs positivity, in addition.} In fact, the Keldysh action does have a BRST invariance \cite{Canet:2011wf,Marguet:2021gab,Crossley:2015evo,Glorioso:2017fpd,Gao:2018bxz}, expressing the fact that the partition function for vanishing sources of the response fields, $j^q=0$ with  $Z[j^c,0]=1 $, defines a topological quantum field theory. There are of course physical fields with positivity on the CTP, and hence with positive spectral distributions, but in  presence of  interactions the latter are usually ultraviolet divergent. In order to represent an
FRG regulator contribution $\Delta S_k[\Phi]$ as in (\ref{regSk}) in terms of local field systems, so that the regulated partition function and effective average action define a local quantum field theory at  
all FRG scales $k$ during the flow, we can either have positivity or ultraviolet (and infrared) finiteness, but not both at the same time. In particular, we conclude that  representing infrared and ultraviolet finite effective average actions in terms of polynomial algebras of local field systems, with spacelike (anti-)commutativity to maintain causality, necessarily requires a detour into indefinite metric spaces with BRST cohomology construction (of BRST closed modulo BRST exact states) of a physical state space during the flow.  

%
%This might have been an expected consequence of the BRST invariance of the regulated partition function \cite{XX} which for $j^q=0$ defines a Witten-type topological quantum field theory,  with  $Z_k[j^c,0]=1 $. What we have done here is to require this manifestation of unitary time evolution on the CTP to be maintained at all FRG scales $k$ during the flow. The regulator contribution $\Delta S_k[\Phi]$ in (\ref{regSk}) must therefore then be itself BRST exact. 

%we need some references here for the BRST on the CTP.

%\paragraph{Conservation laws}

We end this subsection with a brief mentioning of the implications that causal regulators with non-vanishing spectral parts \eqref{sigmaRADef}  can have on the dynamics.
Coupling any theory to an artificial environment, be it positive and hence physical or not, will  evidently in general violate conservation laws. A causal heat-bath regulator, for example, induces artificial dissipation and therefore also possibly affects the conserved quantities.
This can  potentially lead to a  change of the dynamic universality class in the Halperin-Hohenberg classification~\cite{RevModPhys.49.435} during the flow, when the regulator is present.
Most obviously, due to the auxiliary coupling to a regulating heat bath, energy is not conserved anymore, at least for excitations with frequencies in the support of $J_k(\omega,\va{p})$. An Ohmic heat bath, with $J(\omega,\va{p}) \sim \gamma  \omega $, in the infrared, for example, yields the prototypical realization of Model A dynamics, for which heat-bath regulators are thus particularly well suited.
Special attention is needed, on the other hand, when studying systems with conserved energy (e.g.~theories that classify as Model~C, such as for example realized in Ref.~\cite{Schweitzer:2020noq}), or systems with continuous symmetries (e.g.~Model~B or Model~G as recently studied in \cite{Schweitzer:2021iqk} and \cite{Florio:2021jlx}, for example). 
%cite us in prep as well?
In fact, since imaginary and real parts of our causal regulators are always linked by Kramers-Kronig relations, and a non-vanishing imaginary part always corresponds to a coupling to some external environment, we conclude that the arguably simplest 
way to maintain conservation laws is to assume that the regulator does not depend on $\omega$, 
 implying that the regulator spectral density vanishes, $J_k(\omega,\va{p}) \equiv 0$, and the regulator is thus purely spatial, i.e.
\begin{align}
    R^{R/A}_k(\omega, \va{p}) = -\Delta M_k^2(\va{p})  \, .
\end{align}
This seems particularly well justified when one is interested in the critical dynamics near thermal fixed points, where (a) Lorentz invariance is no issue in the first place,  and (b) one is predominantly interested in an effective theory for the critical long-range infrared modes and ultraviolet finiteness is less of a concern.

\subsection{Truncations for real-time applications}
\label{sct:truncationsForRealTimeApplications}
In order to systematically truncate the effective average action for real-time applications, we start with a general expansion in the 1PI vertex functions of the form
\begin{align}
    \Gamma_k[\phi] &= \sum_{n=0}^Q \frac{1}{n!} \int_{x_1 \ldots x_n} \Gamma_k^{\alpha_1 \ldots \alpha_n}(x_1,\ldots,x_n) \left( \phi^{\alpha_1}(x_1) - \phi_{0,k}^{\alpha_1}(x_1) \right) \cdots  \left( \phi^{\alpha_n}(x_n) - \phi_{0,k}^{\alpha_n}(x_n) \right) \label{eq:generalVertexExp}
\end{align}
around background field expectation values $\phi_{0,k}(x) = (\phi_{0,k}^c(x), \phi_{0,k}^q(x))$ which may depend on the FRG scale~$k$, with the CTP indices $\alpha_1,\ldots,\alpha_n = c,q$, and with the corresponding 1PI vertex functions $\Gamma_k^{\alpha_1 \ldots \alpha_n}(x_1,\ldots,x_n)$ defined at the expansion point and scale-dependent, likewise.
The expansion is truncated at some order $n=Q$, and the $Q$-point vertices are hence the highest ones taken into account.

In general, flowing full $n$-point functions with their complete kinematics included gets prohibitively expensive rather quickly with increasing $n$.
It is, therefore, necessary to develop suitable approximation schemes to the full vertex expansion as written in \eqref{eq:generalVertexExp} in order to reduce the numerical complexity.
One such approximation scheme is to expand in the number of loop structures appearing in the diagrams on the right-hand sides of the flow equations~\cite{Roth:2021nrd}.
In order to do this systematically, one assumes that the $Q$-point vertex is effectively given by a frequency and momentum-independent but scale-dependent constant, and then successively increases the number of loop structures that are taken into account in the lower-order $n$-point functions. Using this procedure, every $n$-point vertex will contain at most $(Q-n)$-loop structures.\footnote{In the  important special case where one expands around $\phi=0$ and the effective average action is invariant under $Z_2$ transformations $\phi \to -\phi$, the loop order in every vertex is of course reduced to $(Q-n)/2$, since all $n$-point vertices with odd $n$ vanish.}
In our quantum mechanical benchmark study of the dissipative anharmonic oscillator in~\cite{Roth:2021nrd} such a combined vertex and loop expansion to the order $Q=6$ has proven to be numerically tractable while producing results for the spectral functions that agree very well with the corresponding ab-initio classical-statistical simulations in the classical regime, for example, even at the quantitative level. We will therefore further pursue this well-approved  expansion scheme  also in our present study of critical dynamics.

Up to this point, we have not specified the point in field space $\phi_{0,k}(x)$ used in the vertex expansion. Although it is not relevant for the general structure of our combined loop and vertex expansion, the right choice of the field expansion point can make a quantitative difference at any finite order.
For our purposes, there are arguably at least two natural choices, namely,
\begin{itemize}
\item the scale-dependent minimum $\phi_{0,k}$ of the effective average action $\Gamma_k$, which satisfies the quantum equations of motion \[ \frac{\delta \Gamma_k[\phi^c,\phi^q]}{\delta \phi^c(x)} \bigg\rvert_{\phi_{0,k}} = 0 \, , \quad \frac{\delta \Gamma_k[\phi^c,\phi^q]}{\delta \phi^q(x)} \bigg\rvert_{\phi_{0,k}} = 0 \] at every FRG scale $k$, and is thus also $k$-dependent, and
\item the fixed infrared minimum obtained in the limit $k \to 0$, \[ \frac{\delta}{\delta \phi^c(x)} \lim_{k \to 0} \Gamma_k[\phi^c,\phi^q] \bigg\rvert_{\phi_0} = 0 \, , \quad \frac{\delta}{\delta \phi^q(x)} \lim_{k \to 0} \Gamma_k[\phi^c,\phi^q] \bigg\rvert_{\phi_0} = 0 \, , \] so that the expansion point $\phi_{0,k}(x) \equiv \phi_0(x)$ is $k$-independent in this case.
\end{itemize}
We will explore both possibilities, and explain them in detail in the following two subsections.
Because the part of the quantum equations of motion obtained from $\delta \Gamma_k/\delta\phi^c  $, is always solved by $\phi_{0,k}^q(x) = 0$ as a consequence of the causality structure \cite{kamenev_2011}, it is understood throughout that we always expand around vanishing response field, $\phi^q_{0,k}(x) = 0$.

\subsubsection{Comoving expansion}
\label{sct:ExpAroundScaleDepMin}

In the `comoving' scheme the vertex expansion \eqref{eq:generalVertexExp} is performed around the scale-dependent minimum $\phi_{0,k}(x)$, which lets the 1PI vertex functions $\Gamma_k^{(n)}$ also implicitly dependent on the minimum $\phi_{0,k}(x)$.
For instance, in the symmetry-broken phase the minimum $\phi_{0,k}(x)$ of the effective average action is indeed $k$-dependent, so the flow of $\Gamma_k^{(n)}$ acquires an additional contribution through the functional chain rule,
\begin{align}
    \partial_k \Gamma_k^{\alpha_1 \dots \alpha_n}(x_1,\dots,x_n) &= \frac{\delta^n \, \partial_k \Gamma_k[\phi]}{\delta \phi^{\alpha_1}(x_1) \dots \delta \phi^{\alpha_n}(x_n)} \bigg\rvert_{\phi_{0,k}} + \nonumber \\ 
    &\hspace{-2cm} \int_y \Gamma_k^{\alpha_1 \dots \alpha_n \beta}(x_1,\dots,x_n,y) \, \partial_k \phi_{0,k}^\beta(y) \, ,
	\label{nPointFncChainRule}
\end{align}
where the first term on the right-hand side is understood as a $k$-derivative at fixed background field configuration $\phi_{0,k}=\phi $ and corresponds to the $n^\text{th}$ functional derivative of the Wetterich equation \eqref{wetterichEq} evaluated at $\phi_{0,k}$.
The second term on the right-hand side reflects an `interaction' with the $k$-derivative of the mean-field expectation value, as shown e.g.~for the 2-point function in the last diagram of Fig.~\ref{gamma2-general-flow}.

For the purpose of obtaining the non-trivial infrared power-law behavior in the critical spectral functions, it turns out that the lowest suitable order in our combined vertex and loop expansion around the scale-dependent minimum  combines the 1-loop structures in the 2-point functions with scale dependent but frequency and momentum-independent higher-order $n$-point vertices (i.e.~for  $n>2$).
%%JR: Hier Ansatz für effective average action einfügen.

Before we discuss the corresponding flow equations, we introduce some convenient notations.
For the 1-loop diagrams shown in Fig.~\ref{gamma2-general-flow}, we define the loop functions
\begin{align}
    I_k^K(x) &\equiv B_k^K(x,x) \, , \\
    J_k^{XY}(x,x') &\equiv \frac{1}{1+\delta_{XY}} \left( B_k^X(x,x') G_k^Y(x,x') + B_k^Y(x,x') G_k^X(x,x') \right) \, ,
\end{align}
with the indices $X$ and $ Y$ denoting either of $R$, $A$ or $K$ for the retarded, advanced, and Keldysh components of the propagators, respectively.
The combinatoric prefactor $1/(1+\delta_{XY})$ is $1/2$, if $X=Y$, and $1$, if $X \neq Y$. It is included here for convenience, to avoid double counting.
The loop functions $J_k$
are related by the exchange symmetry $J_k^{XY}(x,x') = J_k^{YX}(x,x')$.
In a spacetime-translation invariant system, their Fourier-transformed counterparts are given by convolutions, 
\begin{align}
    I_k^K &= \int_{q} B_k^K(q) \, , \label{ILoopDef} \\
    J_k^{XY}(p) &= \frac{1}{1+\delta_{XY}} \int_q \left( B_k^X(q) G_k^Y(p-q) + B_k^Y(q) G_k^X(p-q) \right) \, , \label{JLoopDef}
\end{align}
where $p,q$ denote the energy-momentum vectors in $D$-dimensional spacetime.
%$\delta_{XY}$ denotes the Kronecker-$\delta$ as usual.
In thermal equilibrium, the FDR implies an additional relation between the various $J$'s,
\begin{align}
    \coth\left(\frac{\omega}{2T}\right) \left( J_k^{KR}(\omega,\va{p}) - J_k^{KA}(\omega,\va{p}) \right) = J_k^{KK}(\omega,\va{p}) + J_k^{RR}(\omega,\va{p}) + J_k^{AA}(\omega,\va{p}) \, , \label{JRelFDR}
\end{align}
which can be proven from the following identity relating different distribution functions,
\begin{align}
    \coth(a) \coth(b) + 1 = \coth(a+b) \left( \coth(a) + \coth(b) \right) \, .
\end{align}
In the classical limit $\omega \ll T$ the last two terms on the right-hand side of \eqref{JRelFDR} are of the order $\Ord{(\omega/T)^0}$ and hence subleading  as compared to the others which are of the order $\Ord{(\omega/T)^{-2}}$. In the classical limit, Eq.~\eqref{JRelFDR} therefore reduces to
\begin{align}
    \frac{2T}{\omega} \left( J_k^{KR}(\omega,\va{p}) - J_k^{KA}(\omega,\va{p}) \right) = J_k^{KK}(\omega,\va{p}) \, .\label{JRelFDRClass}
\end{align}
With these notations, we can compactly express the flow equations for the advanced, retarded, and Keldysh components of the full 2-point function as
\begin{subequations}
\begin{align}
    \label{gam2A-flow-full}
	\partial_k \Gamma_{k}^{cq}(\omega,\va{p}) &= -\frac{i\kappa_{k}^2}{2} \, J_k^{KA}(\omega,\va{p}) + \frac{i\lambda_k}{4} \, I_k^K + \frac{\kappa_{k}}{\sqrt{2}} \, \partial_k \phi_{0,k}^c \, , \\
    \label{gam2R-flow-full}
	\partial_k \Gamma_{k}^{qc}(\omega,\va{p}) &= -\frac{i\kappa_{k}^2}{2} \, J_k^{KR}(\omega,\va{p}) + \frac{i\lambda_k}{4} \, I_k^K + \frac{\kappa_{k}}{\sqrt{2}} \, \partial_k \phi_{0,k}^c \, , \\
    \label{gam2K-flow-full}
	\partial_k \Gamma_{k}^{qq}(\omega,\va{p}) &= -\frac{i\kappa_{k}^2}{2} \left( J_k^{KK}(\omega,\va{p}) + J_k^{RR}(\omega,\va{p}) + J_k^{AA}(\omega,\va{p}) \right)  \, ,
\end{align} \label{gamma2-flow-full}
\end{subequations}
where $\kappa_k$ and $\lambda_k$ denote the scale-dependent coupling constants in the 3-point and 4-point vertices, respectively. These formal expressions are represented by the diagrams in Fig.~\ref{gamma2-general-flow}.

\begin{figure*}[t]
\centering
\begin{align*}
    \partial_k \Gamma_k^{cq}(x,x') &= -i
    \left\{\;
	\begin{tikzpicture}[baseline=-0.5ex]
	 	\centerarc[pblue](0,0)(0:45:0.5)
	 	\centerarc[pblue](0,0)(45:90:0.5)
	 	\centerarc[pblue](0,0)(90:135:0.5)
	 	\centerarc[pblue](0,0)(135:180:0.5)
	 	\centerarc[pred](0,0)(180:270:0.5)
	 	\centerarc[pblue](0,0)(270:360:0.5)
		% legs
	 	\draw[pblue] (-0.5,0) -- (-0.5-0.4,0)  node[anchor=north,black] {\LegFontSize $x$};
	 	\draw[pred] (0.5,0) -- (0.5+0.4,0)  node[anchor=north,black] {\LegFontSize $x'$};
	 	% full vertices
	 	\fill[black] (-0.5,0) circle (0.05);
	 	\fill[black] (+0.5,0) circle (0.05);
		% regulator insertion
	 	\fill[black] (0-0.1,0.5-0.1) rectangle ++(0.2,0.2);
	\end{tikzpicture}
	+
	\begin{tikzpicture}[baseline=-0.5ex]
	 	\centerarc[pblue](0,0)(0:45:0.5)
	 	\centerarc[pblue](0,0)(45:90:0.5)
	 	\centerarc[pred](0,0)(90:135:0.5)
	 	\centerarc[pred](0,0)(135:180:0.5)
	 	\centerarc[pblue](0,0)(180:270:0.5)
	 	\centerarc[pblue](0,0)(270:360:0.5)
		% legs
	 	\draw[pblue] (-0.5,0) -- (-0.5-0.4,0)  node[anchor=north,black] {\LegFontSize $x$};
	 	\draw[pred] (0.5,0) -- (0.5+0.4,0)  node[anchor=north,black] {\LegFontSize $x'$};
	 	% full vertices
	 	\fill[black] (-0.5,0) circle (0.05);
	 	\fill[black] (+0.5,0) circle (0.05);
		% regulator insertion
	 	\fill[black] (0-0.1,0.5-0.1) rectangle ++(0.2,0.2);
	\end{tikzpicture}
	+
    \frac{1}{2}\;
	\begin{tikzpicture}[baseline=-0.5ex]
	 	\centerarc[pblue](0,0)(0:45:0.5)
	 	\centerarc[pblue](0,0)(45:90:0.5)
	 	\centerarc[pblue](0,0)(90:135:0.5)
	 	\centerarc[pblue](0,0)(135:180:0.5)
	 	\centerarc[pblue](0,0)(180:270:0.5)
	 	\centerarc[pblue](0,0)(270:360:0.5)
		% legs
	 	\draw[pblue] (0,-0.5) -- (-0.5,-0.5)  node[anchor=north,black] {\LegFontSize $x$};
	 	\draw[pred] (0,-0.5) -- (+0.5,-0.5)  node[anchor=north,black] {\LegFontSize $x'$};
	 	% full vertices
	 	\fill[black] (0,-0.5) circle (0.05);
		% regulator insertion
	 	\fill[black] (0-0.1,0.5-0.1) rectangle ++(0.2,0.2);
	\end{tikzpicture}
    \;\right\}
    +
    \begin{tikzpicture}[baseline=-0.5ex]
        \draw[pblue] (0,0) -- (0,+0.5);
        % legs
        \draw[pblue] (0,0) -- (-0.5,-0.3)  node[anchor=north,black] {\LegFontSize $x$};
        \draw[pred]  (0,0) -- (+0.5,-0.3)  node[anchor=north,black] {\LegFontSize $x'$};
        % full vertices
        \fill[black] (0,0) circle (0.05);
        % mean-field insertion
        \fill[white] (0,0.5) circle (0.1414);
        \draw[black] (-0.1,0.5-0.1) -- (0.1,0.5+0.1);
        \draw[black] (-0.1,0.5+0.1) -- (+0.1,0.5-0.1);
        \draw[black] (0,0.5) circle (0.1414);
    \end{tikzpicture} \\
    \partial_k \Gamma_k^{qq}(x,x') &= -i
    \left\{\;
	\begin{tikzpicture}[baseline=-0.5ex]
	 	\centerarc[pblue](0,0)(0:45:0.5)
	 	\centerarc[pblue](0,0)(45:90:0.5)
	 	\centerarc[pblue](0,0)(90:135:0.5)
	 	\centerarc[pblue](0,0)(135:180:0.5)
	 	\centerarc[pblue](0,0)(180:270:0.5)
	 	\centerarc[pblue](0,0)(270:360:0.5)
		% legs
	 	\draw[pred] (-0.5,0) -- (-0.5-0.4,0)  node[anchor=north,black] {\LegFontSize $x$};
	 	\draw[pred] (0.5,0) -- (0.5+0.4,0)  node[anchor=north,black] {\LegFontSize $x'$};
	 	% full vertices
	 	\fill[black] (-0.5,0) circle (0.05);
	 	\fill[black] (+0.5,0) circle (0.05);
		% regulator insertion
	 	\fill[black] (0-0.1,0.5-0.1) rectangle ++(0.2,0.2);
	\end{tikzpicture}
	+
	\begin{tikzpicture}[baseline=-0.5ex]
	 	\centerarc[pred](0,0)(0:45:0.5)
	 	\centerarc[pred](0,0)(45:90:0.5)
	 	\centerarc[pblue](0,0)(90:135:0.5)
	 	\centerarc[pblue](0,0)(135:180:0.5)
	 	\centerarc[pblue](0,0)(180:270:0.5)
	 	\centerarc[pred](0,0)(270:360:0.5)
		% legs
	 	\draw[pred] (-0.5,0) -- (-0.5-0.4,0)  node[anchor=north,black] {\LegFontSize $x$};
	 	\draw[pred] (0.5,0) -- (0.5+0.4,0)  node[anchor=north,black] {\LegFontSize $x'$};
	 	% full vertices
	 	\fill[black] (-0.5,0) circle (0.05);
	 	\fill[black] (+0.5,0) circle (0.05);
		% regulator insertion
	 	\fill[black] (0-0.1,0.5-0.1) rectangle ++(0.2,0.2);
	\end{tikzpicture}
	+
	\begin{tikzpicture}[baseline=-0.5ex]
	 	\centerarc[pblue](0,0)(0:45:0.5)
	 	\centerarc[pblue](0,0)(45:90:0.5)
	 	\centerarc[pred](0,0)(90:135:0.5)
	 	\centerarc[pred](0,0)(135:180:0.5)
	 	\centerarc[pred](0,0)(180:270:0.5)
	 	\centerarc[pblue](0,0)(270:360:0.5)
		% legs
	 	\draw[pred] (-0.5,0) -- (-0.5-0.4,0)  node[anchor=north,black] {\LegFontSize $x$};
	 	\draw[pred] (0.5,0) -- (0.5+0.4,0)  node[anchor=north,black] {\LegFontSize $x'$};
	 	% full vertices
	 	\fill[black] (-0.5,0) circle (0.05);
	 	\fill[black] (+0.5,0) circle (0.05);
		% regulator insertion
	 	\fill[black] (0-0.1,0.5-0.1) rectangle ++(0.2,0.2);
	\end{tikzpicture} \;\right\}
\end{align*}
\caption{Diagrammatic representation of the flow equations for the advanced and Keldysh 2-point functions $\Gamma_k^{cq}(x,x')$ and $\Gamma_k^{qq}(x,x')$, evaluated at the scale-dependent minimum. The retarded 2-point function $\Gamma_k^{qc}(x,x')$ readily follows from the advanced one by symmetry. They consist of contributions from non-local 1-loop diagrams which, for example, generate a non-trivial dependence of the spectral function $\rho(\omega,\va{p})$ on external frequency $\omega$ and spatial momentum $\va{p}$. 
The advanced 2-point function in the first row in addition acquires 
a frequency and momentum-independent shift in the squared mass
generated from the tadpole diagram in the third term and the interaction with the $k$-dependent field expectation value $\partial_k \phi_{0,k}$ in the fourth. For the Keldysh 2-point function in the second row, only the first diagram survives in the classical limit, because the second and third diagrams both contain a quantum $\phi^q \phi^q \phi^q$ vertex without substructure.} \label{gamma2-general-flow}
\end{figure*}

For spacetime-translation invariant systems, we can utilize the scale-dependent effective potential $V_k(\varphi)$ to effectively encode the flow equations for all higher-order $n$-point vertices (here with $n > 2$), which  we assume  to be given by scale-dependent coupling constants (without substructure).
The effective potential is thereby conveniently obtained from the scale-dependent force,
\begin{align}
    V_k'(\varphi) &\equiv - \frac{1}{\sqrt{2}} \frac{\delta \Gamma_k[\phi]}{\delta \phi^q(x)} \bigg\rvert_{\substack{\phi^c(x) = \sqrt{2}\,\varphi \\ \phi^q(x) =\, 0 \phantom{~~\varphi} }} \, , \label{effPotDef}
\end{align}
where the prime denotes the ordinary derivative with respect to the constant classical field variable $\varphi = \phi^c/\sqrt{2}$. 
This determines $V_k(\varphi)$ up to some (irrelevant) $k$-dependent constant.
The effective potential on the other hand then determines the scale-dependent $n$-point coupling constants  without substructure (here for $n > 2$) via its Taylor expansion around the scale-dependent minimum $\varphi_{0,k} \equiv \phi_{0,k}^c/\sqrt{2}$.
Specifically, the 3 and 4-point coupling constants in Eqs.~\eqref{gamma2-flow-full} are obtained from
\begin{align}
    \kappa_{k} &= V_k^{(3)}( \varphi_{0,k} ) \,, \hspace{1.0cm} \lambda_{k} = V_k^{(4)}( \varphi_{0,k} ) \,.
\end{align}
We can now easily derive the flow equation for the effective potential by differentiating the Wetterich equation~\eqref{wetterichEq} once with respect to the response field $\phi^q(x)$, and setting the response field to zero afterwards, $\phi^q(x) = 0$, but keeping a constant expectation value for the classical field, $\phi^c(x) = \sqrt{2}\, \varphi$ \cite{Tan:2021zid,Roth:2021nrd}.
In the following, we introduce an additional index ${}_\varphi$ to denote an explicit dependence  on such a background field configuration in quantities like $I_{\varphi,k}^K$ or $\Gamma_{\varphi,k}^{(3)}$.
We can then express the flow equation for the effective potential as 
\begin{align}
    \label{flow-eff-pot}
    \partial_k V_k'(\varphi) &=
    -\frac{i}{\sqrt{8}}\;
	\begin{tikzpicture}[baseline=-0.5ex]
	 	\centerarc[pblue](0,0)(-90:90:0.5)
	 	\centerarc[pblue](0,0)(90:270:0.5)
		\draw[pred] (0,-0.5) -- (0,-0.5-0.5);
	 	% bare vertices
	 	\fill[black] (0,-0.5) circle (0.05);
		% regulator insertion
		\fill[black] (0-0.1,0.5-0.1) rectangle ++(0.2,0.2);
	\end{tikzpicture}
    = - \frac{i}{4} V_k^{(3)}(\varphi) I_{\varphi,k}^K \, ,
\end{align}
where we approximated the full 3-point function $\Gamma_{\varphi,k}^{(3)}$ by the third derivative $V_k^{(3)}(\varphi)$ of our effective potential. 
Eq.~\eqref{flow-eff-pot} thus represents a generator for the flow equations of all higher-order $n$-point coupling constants (here for $n>2$) in the approximation where the corresponding vertex functions are assumed to be independent of frequency and momentum.
These flow equations are obtained via differentiation of \eqref{flow-eff-pot} with respect to the classical field expectation value $\varphi$, where one can moreover employ recursion relations between the $\varphi$-derivatives of the various propagators.
For more technical details of this approach, we refer to Sec.~V of Ref.~\cite{Roth:2021nrd}.

\subsubsection{Expansion around vanishing field expectation values}
\label{sct:ExpAroundIRMin}

Because of destabilizing back-coupling terms of higher order vertices arising from the chain rule in \eqref{nPointFncChainRule}, the expansion around the scale-dependent minimum $\phi_{0,k}$ of the previous subsection can sometimes be numerically challenging (see, e.g., the discussion in Chapter~3.4 of Ref.~\cite{Rennecke:2015lur}).
It is therefore valuable to have alternative truncation schemes based on fixed  expansion points in field space for comparison, here especially of our results for the critical spectral functions in Section \ref{sct:results} below. 
When using a fixed classical field configuration as the reference point in the vertex expansion, the most natural first choice is the origin in field space.
The alternative possibility that we have explored here therefore is
the expansion around vanishing field expectation values $\phi_{0,k}(x) = 0$.

Such a symmetric expansion has the added bonus that all odd 1PI vertex functions vanish. At the same time this implies, however, that we need to maintain at least explicit two-loop structures in the 2-point functions to generate non-trivial frequency and momentum dependencies. To achieve this
 one can draw from  existing 2-loop expansion schemes, e.g., as employed in Refs.~\cite{Huelsmann:2020xcy,Tan:2021zid,Roth:2021nrd}.
Here, we build on our combined vertex and loop expansion that we have first introduced in Ref.~\cite{Roth:2021nrd}, as explained in more detail at the beginning of this section. Specifically, we use the expansion order $Q=6$.  
This means that we effectively take into account 2-loop structures in the 2-point functions, and one-loop structures in the 4-point functions, which we split into $s$, $t$, and $u$ channels (cf.~Figs.~\ref{fig:gam4FlowSymm} and \ref{fig:gam2FlowSymm}), whereas we approximate the 6-point function by a frequency and momentum-independent and hence structureless but scale-dependent vertex.

The corresponding flow equations can be obtained as follows:
For the 6-point coupling, we can do a straightforward Taylor expansion of the flow \eqref{flow-eff-pot} of the effective potential, which is explained in detail in Sec.~V of Ref.~\cite{Roth:2021nrd}. 
The flow equation for the 4-point function is written in `local-vertex approximation', i.e.~all occurring 4-point and 6-point functions are replaced by effective local coupling constants, see Fig.~\ref{fig:gam4FlowSymm}.
Consistent with our combined vertex and loop expansion to order $Q=6$, this scheme renders its perturbative structure 1-loop complete and naturally separates the diagrams appearing on the right-hand side of its flow equation into the three different channels.
For instance, the flow equations for advanced, retarded, and anomalous components of the $s$-channel can be compactly expressed as
\begin{subequations}
\begin{align}
    \partial_k V_k^{cl,A}(\omega,\va{p}) &= -\frac{i\lambda_k^2}{4} J_k^{KA}(\omega,\va{p}) + \frac{i\lambda_{6,k}}{24} I_k^K
    \, , \label{vClAFlowEqSymm} \\
    \partial_k V_k^{cl,R}(\omega,\va{p}) &= -\frac{i\lambda_k^2}{4} J_k^{KR}(\omega,\va{p}) + \frac{i\lambda_{6,k}}{24} I_k^K
    \, , \label{vClRFlowEqSymm} \\
    \partial_k V_k^{an}(\omega,\va{p}) &= -\frac{i\lambda_k^2}{4} \left( J_k^{KK}(\omega,\va{p}) + J_k^{RR}(\omega,\va{p}) + J_k^{AA}(\omega,\va{p}) \right) \, . \label{vAnFlowEqSymm}
\end{align}
\end{subequations}
On the right-hand side, these contain (a) the effective 4-point coupling constant already introduced in the previous subsection,
\begin{align}
    \lambda_k = -6 \, V_k^{cl,R/A}(p=0) \, , \label{localVertexApprox}
\end{align}
which here is self-consistently obtained from the full 4-point function on the left, and (b)~the 6-point coupling constant $\lambda_{6,k}$ defined by the corresponding derivative of the effective potential,
\begin{align}
    \lambda_{6,k} = \frac{\partial^6 V_k(\varphi)}{\partial \varphi^6} \bigg\rvert_{\varphi=0} \,.
\end{align}

\begin{figure*}[t]
	\centering
	\begin{align*}
		\partial_k V_k^{cl,A}(x,x') &=
		-i \int\displaylimits_{\substack{x-y,\\x'-y'}}
		\hspace{-1ex}  \Bigg\{\;
		\begin{tikzpicture}[baseline=-0.5ex]
            \centerarc[pblue](0,0)(0:45:0.5)
            \centerarc[pblue](0,0)(45:90:0.5)
            \centerarc[pblue](0,0)(90:135:0.5)
            \centerarc[pblue](0,0)(135:180:0.5)
            \centerarc[pred](0,0)(180:270:0.5)
            \centerarc[pblue](0,0)(270:360:0.5)
			% legs
		 	\draw[pblue] (-0.5,0) -- (-0.5-0.3,0.3) node[anchor=south,black] {\LegFontSize $y$};
		 	\draw[pblue] (-0.5,0) -- (-0.5-0.3,-0.3) node[anchor=north,black] {\LegFontSize $x$};
		 	\draw[pblue] (0.5,0) -- (0.5+0.3,0.3) node[anchor=south,black] {\LegFontSize $y'$};
		 	\draw[pred] (0.5,0) -- (0.5+0.3,-0.3) node[anchor=north,black] {\LegFontSize $x'$};
		 	% full vertices
		 	\fill[black] (-0.5,0) circle (0.05);
		 	\fill[black] (+0.5,0) circle (0.05);
			% regulator insertion
            \fill[black] (0-0.1,0.5-0.1) rectangle ++(0.2,0.2);
		\end{tikzpicture}
		+
		\begin{tikzpicture}[baseline=-0.5ex]
            \centerarc[pblue](0,0)(0:45:0.5)
            \centerarc[pblue](0,0)(45:90:0.5)
            \centerarc[pred](0,0)(90:135:0.5)
            \centerarc[pred](0,0)(135:180:0.5)
            \centerarc[pblue](0,0)(180:270:0.5)
            \centerarc[pblue](0,0)(270:360:0.5)
			% legs
		 	\draw[pblue] (-0.5,0) -- (-0.5-0.3,0.3) node[anchor=south,black] {\LegFontSize $y$};
		 	\draw[pblue] (-0.5,0) -- (-0.5-0.3,-0.3) node[anchor=north,black] {\LegFontSize $x$};
		 	\draw[pblue] (0.5,0) -- (0.5+0.3,0.3) node[anchor=south,black] {\LegFontSize $y'$};
		 	\draw[pred] (0.5,0) -- (0.5+0.3,-0.3) node[anchor=north,black] {\LegFontSize $x'$};
		 	% full vertices
		 	\fill[black] (-0.5,0) circle (0.05);
		 	\fill[black] (+0.5,0) circle (0.05);
			% regulator insertion
            \fill[black] (0-0.1,0.5-0.1) rectangle ++(0.2,0.2);
		\end{tikzpicture} \;\Bigg\} 
          -\frac{i}{6}
        	\int\displaylimits_{\substack{x-y,\\x'-y'}}
        	\begin{tikzpicture}[baseline=-0.5ex]
        	 	\centerarc[pblue](0,0)(-90:90:0.5)
        	 	\centerarc[pblue](0,0)(90:270:0.5)
        		% legs
        	 	%\draw[black] (0,0.5) -- (-0.3,0.5+0.3) node[anchor=south east] {$y$};
        	 	%\draw[black] (0,0.5) -- (+0.3,0.5+0.3) node[anchor=south west] {$y'$};
        	 	%
        	 	\draw[pblue] (0,-0.5) -- (-0.15,-0.5-0.3);
        	 	\draw[pblue] (0,-0.5) -- (+0.15,-0.5-0.3);
        	 	\draw[pblue] (0,-0.5) -- (-0.45,-0.5-0.3);
        	 	\draw[pred] (0,-0.5) -- (+0.45,-0.5-0.3);
        	 	%\node at (0,-1.1) {\LegFontSize $x\,y\,y'\,x'$};
        	 	% bare vertices
        	 	\fill[black] (0,-0.5) circle (0.05);
        		% regulator insertion
            \fill[black] (0-0.1,0.5-0.1) rectangle ++(0.2,0.2);
        	\end{tikzpicture} \\
		\partial_k V_k^{an}(x,x')  &=
		-i \int\displaylimits_{\substack{x-y,\\x'-y'}}
		\hspace{-1ex} \Bigg\{\;
		\begin{tikzpicture}[baseline=-0.5ex]
            \centerarc[pblue](0,0)(0:45:0.5)
            \centerarc[pblue](0,0)(45:90:0.5)
            \centerarc[pblue](0,0)(90:135:0.5)
            \centerarc[pblue](0,0)(135:180:0.5)
            \centerarc[pblue](0,0)(180:270:0.5)
            \centerarc[pblue](0,0)(270:360:0.5)
			% legs
		 	\draw[pblue] (-0.5,0) -- (-0.5-0.3,0.3) node[anchor=south,black] {\LegFontSize $y$};
		 	\draw[pred] (-0.5,0) -- (-0.5-0.3,-0.3) node[anchor=north,black] {\LegFontSize $x$};
		 	\draw[pblue] (0.5,0) -- (0.5+0.3,0.3) node[anchor=south,black] {\LegFontSize $y'$};
		 	\draw[pred] (0.5,0) -- (0.5+0.3,-0.3) node[anchor=north,black] {\LegFontSize $x'$};
		 	% full vertices
		 	\fill[black] (-0.5,0) circle (0.05);
		 	\fill[black] (+0.5,0) circle (0.05);
			% regulator insertion
            \fill[black] (0-0.1,0.5-0.1) rectangle ++(0.2,0.2);
		\end{tikzpicture}
		+
		\begin{tikzpicture}[baseline=-0.5ex]
            \centerarc[pred](0,0)(0:45:0.5)
            \centerarc[pred](0,0)(45:90:0.5)
            \centerarc[pblue](0,0)(90:135:0.5)
            \centerarc[pblue](0,0)(135:180:0.5)
            \centerarc[pblue](0,0)(180:270:0.5)
            \centerarc[pred](0,0)(270:360:0.5)
			% legs
		 	\draw[pblue] (-0.5,0) -- (-0.5-0.3,0.3) node[anchor=south,black] {\LegFontSize $y$};
		 	\draw[pred] (-0.5,0) -- (-0.5-0.3,-0.3) node[anchor=north,black] {\LegFontSize $x$};
		 	\draw[pblue] (0.5,0) -- (0.5+0.3,0.3) node[anchor=south,black] {\LegFontSize $y'$};
		 	\draw[pred] (0.5,0) -- (0.5+0.3,-0.3) node[anchor=north,black] {\LegFontSize $x'$};
		 	% full vertices
		 	\fill[black] (-0.5,0) circle (0.05);
		 	\fill[black] (+0.5,0) circle (0.05);
			% regulator insertion
            \fill[black] (0-0.1,0.5-0.1) rectangle ++(0.2,0.2);
		\end{tikzpicture}
		+
		\begin{tikzpicture}[baseline=-0.5ex]
            \centerarc[pblue](0,0)(0:45:0.5)
            \centerarc[pblue](0,0)(45:90:0.5)
            \centerarc[pred](0,0)(90:135:0.5)
            \centerarc[pred](0,0)(135:180:0.5)
            \centerarc[pred](0,0)(180:270:0.5)
            \centerarc[pblue](0,0)(270:360:0.5)
			% legs
		 	\draw[pblue] (-0.5,0) -- (-0.5-0.3,0.3) node[anchor=south,black] {\LegFontSize $y$};
		 	\draw[pred] (-0.5,0) -- (-0.5-0.3,-0.3) node[anchor=north,black] {\LegFontSize $x$};
		 	\draw[pblue] (0.5,0) -- (0.5+0.3,0.3) node[anchor=south,black] {\LegFontSize $y'$};
		 	\draw[pred] (0.5,0) -- (0.5+0.3,-0.3) node[anchor=north,black] {\LegFontSize $x'$};
		 	% full vertices
		 	\fill[black] (-0.5,0) circle (0.05);
		 	\fill[black] (+0.5,0) circle (0.05);
			% regulator insertion
            \fill[black] (0-0.1,0.5-0.1) rectangle ++(0.2,0.2);
		\end{tikzpicture} \;\Bigg\}
	\end{align*}
	\caption{Diagrammatic representation of the flow equation for the $s$-channel contribution to the 4-point function in our self-consistent `local-vertex approximation'. The flow equations for the $t$ and $u$-channels follow readily from exchange and crossing symmetries. The `local-vertex approximation' manifests itself in the fact that effective local coupling constants derived from the full 4-point and 6-point functions enter the right-hand sides, which makes this particular flow equation 1-loop exact, consistent with our combined vertex and loop expansion to order $Q=6$. \label{fig:gam4FlowSymm}}
\end{figure*}

\medskip\noindent
Finally, the 2-loop complete flow equations  for the two independent 2-point functions are given by
\begin{align}
	\partial_k \Gamma^{cq}_{k}(p) &=
	\frac{i \lambda_k}{12} I_k^K - i \int_q \Big( B_k^{K}(q) V_k^{cl,A}(p-q)+ B_k^{A}(q) V_k^{an}(p-q) \Big) \, , \label{gamma2CQFlowSymm} \\
	\partial_k \Gamma^{qq}_{k}(p) &= - i \int_q \Big( B_k^{K}(q) V_k^{an}(p-q) + B_k^{R}(q) V_k^{qu,R}(p-q) + B_k^{A}(q) V_k^{qu,A}(p-q) \Big) \, . \label{gamma2QQFlowSymm}
\end{align}
These are visualized diagrammatically in Fig.~\ref{fig:gam2FlowSymm}.
In particular, one can see explicitly in this figure how the 1-loop complete 4-point function enters as a decomposition into $s$, $t$, and $u$-channel contributions, which makes our truncation for the 2-point function two-loop exact, including all perturbative two-loop structures, as desired above.
\begin{figure*}[t]
    \centering
    \begin{align*}
        \partial_k \Gamma_k^{cq}(x,x') &= %-i
%        \left\{\;
        -\frac{i}{2}\;
        \begin{tikzpicture}[baseline=-0.5ex]
            \centerarc[pgreen](0,0)(0:45:0.5)
            \centerarc[pgreen](0,0)(45:90:0.5)
            \centerarc[pgreen](0,0)(90:135:0.5)
            \centerarc[pgreen](0,0)(135:180:0.5)
            \centerarc[pgreen](0,0)(180:270:0.5)
            \centerarc[pgreen](0,0)(270:360:0.5)
            % legs
            \draw[pblue] (0,-0.5) -- (-0.5,-0.5)  node[anchor=north,black] {\LegFontSize $x$};
            \draw[pred] (0,-0.5) -- (+0.5,-0.5)  node[anchor=north,black] {\LegFontSize $x'$};
            % vertices
            \fill[black] (0,-0.5) circle (0.1);
            % regulator insertion
            \fill[black] (0-0.1,0.5-0.1) rectangle ++(0.2,0.2);
        \end{tikzpicture} = -\frac{i}{2} \Bigg\{
		\begin{tikzpicture}[baseline=-2ex]
		 	\centerarc[pgreen](0,0)(0:360:0.5)
			% legs
		 	\draw[pblue] (0,-0.8) -- (-0.5,-0.8) node[black,anchor=north] {\LegFontSize $x$};
		 	\draw[pred] (0,-0.8) -- (+0.5,-0.8) node[black,anchor=north] {\LegFontSize $x'$};
		 	% s-, t-, u- vertices
		 	%\draw[black,dotted] (0,-0.5) -- (0,-0.8);
			\draw[fill=black] (0,-0.65) ellipse (0.15/4 and 0.15);
		 	% bare vertices
		 	%\fill[black] (0,-0.5) circle (0.1);
			% regulator insertion
            \fill[black] (0-0.1,0.5-0.1) rectangle ++(0.2,0.2);
		\end{tikzpicture}
		+
		\begin{tikzpicture}[baseline=-1ex]
		 	\centerarc[pgreen](0,0)(-45:90:0.5)
		 	\centerarc[pgreen](0,0)(90:225:0.5)
			% legs
		 	\draw[pblue] (-0.354,-0.354) -- (-0.3-0.354,-0.354-0.3) node[black,anchor=north] {\LegFontSize $x$};
		 	\draw[pred] (0.354,-0.354) -- (+0.3+0.354,-0.354-0.3) node[black,anchor=north] {\LegFontSize $x'$};
		 	% s-, t-, u- vertices
		 	%\draw[black,dotted] (-0.354,-0.354) -- (0.354,-0.354);
			\draw[fill=black] (0,-0.354) ellipse (0.354 and 0.354/4);
		 	% bare vertices
		 	%\fill[black] (0,-0.5) circle (0.1);
			% regulator insertion
            \fill[black] (0-0.1,0.5-0.1) rectangle ++(0.2,0.2);
		\end{tikzpicture}
		+
		\begin{tikzpicture}[baseline=-1ex]
		 	\centerarc[pgreen](0,0)(-45:90:0.5)
		 	\centerarc[pgreen](0,0)(90:225:0.5)
			% legs
		 	\centerarc[pblue](-0.354+0.1,-0.354)(0:-90:0.608)
		 	\centerarc[pred](+0.354-0.1,-0.354)(180:240:0.608)
		 	\centerarc[pred](+0.354-0.1,-0.354)(250:270:0.608)
		 	\node at (-0.708,-0.354-0.608) {\LegFontSize $x$};
		 	\node at (+0.708,-0.354-0.608) {\LegFontSize $x'$};
		 	% s-, t-, u- vertices
		 	%\draw[black,dotted] (-0.354,-0.354) -- (0.354,-0.354);
			\draw[fill=black] (0,-0.354) ellipse (0.354 and 0.354/4);
		 	% bare vertices
		 	%\fill[black] (0,-0.5) circle (0.1);
			% regulator insertion
            \fill[black] (0-0.1,0.5-0.1) rectangle ++(0.2,0.2);
		\end{tikzpicture} \Bigg\} \\
        \partial_k \Gamma_k^{qq}(x,x') &= %-i
%        \left\{\;
        -\frac{i}{2}\;
        \begin{tikzpicture}[baseline=-0.5ex]
            \centerarc[pgreen](0,0)(0:45:0.5)
            \centerarc[pgreen](0,0)(45:90:0.5)
            \centerarc[pgreen](0,0)(90:135:0.5)
            \centerarc[pgreen](0,0)(135:180:0.5)
            \centerarc[pgreen](0,0)(180:270:0.5)
            \centerarc[pgreen](0,0)(270:360:0.5)
            % legs
            \draw[pred] (0,-0.5) -- (-0.5,-0.5)  node[anchor=north,black] {\LegFontSize $x$};
            \draw[pred] (0,-0.5) -- (+0.5,-0.5)  node[anchor=north,black] {\LegFontSize $x'$};
            % vertices
            \fill[black] (0,-0.5) circle (0.1);
            % regulator insertion
            \fill[black] (0-0.1,0.5-0.1) rectangle ++(0.2,0.2);
        \end{tikzpicture} = -\frac{i}{2} \Bigg\{
		\begin{tikzpicture}[baseline=-1ex]
		 	\centerarc[pgreen](0,0)(-45:90:0.5)
		 	\centerarc[pgreen](0,0)(90:225:0.5)
			% legs
		 	\draw[pred] (-0.354,-0.354) -- (-0.3-0.354,-0.354-0.3) node[black,anchor=north] {\LegFontSize $x$};
		 	\draw[pred] (0.354,-0.354) -- (+0.3+0.354,-0.354-0.3) node[black,anchor=north] {\LegFontSize $x'$};
		 	% s-, t-, u- vertices
		 	%\draw[black,dotted] (-0.354,-0.354) -- (0.354,-0.354);
			\draw[fill=black] (0,-0.354) ellipse (0.354 and 0.354/4);
		 	% bare vertices
		 	%\fill[black] (0,-0.5) circle (0.1);
			% regulator insertion
            \fill[black] (0-0.1,0.5-0.1) rectangle ++(0.2,0.2);
		\end{tikzpicture}
		+
		\begin{tikzpicture}[baseline=-1ex]
		 	\centerarc[pgreen](0,0)(-45:90:0.5)
		 	\centerarc[pgreen](0,0)(90:225:0.5)
			% legs
		 	\centerarc[pred](-0.354+0.1,-0.354)(0:-90:0.608)
		 	\centerarc[pred](+0.354-0.1,-0.354)(180:240:0.608)
		 	\centerarc[pred](+0.354-0.1,-0.354)(250:270:0.608)
		 	\node at (-0.708,-0.354-0.608) {\LegFontSize $x$};
		 	\node at (+0.708,-0.354-0.608) {\LegFontSize $x'$};
		 	% s-, t-, u- vertices
		 	%\draw[black,dotted] (-0.354,-0.354) -- (0.354,-0.354);
			\draw[fill=black] (0,-0.354) ellipse (0.354 and 0.354/4);
		 	% bare vertices
		 	%\fill[black] (0,-0.5) circle (0.1);
			% regulator insertion
            \fill[black] (0-0.1,0.5-0.1) rectangle ++(0.2,0.2);
		\end{tikzpicture} \Bigg\}
    \end{align*}
    \caption{Diagrammatic representation of the flow equation for the advanced and Keldysh 2-point functions $\Gamma_k^{cq}(x,x')$, $\Gamma_k^{qq}(x,x')$, respectively, in the symmetric phase, evaluated at vanishing field expectation values. In the second equality, we see the splitting into $s$, $t$, and $u$-channels of the 4-point function, which are diagrammatically denoted by an ellipse. \label{fig:gam2FlowSymm}}
\end{figure*}

\section{Dynamic Models}

\subsection{Model A---Critical dynamics without conserved quantities}
\label{sct:critDynamicsNoConservedQuant}

We consider the Landau-Ginzburg-Wilson  Hamiltonian for a single-component real scalar field $\varphi$ in $d$ spatial dimensions of the form
\begin{align}
    F = \int d^d x \left\{ \frac{1}{2}(\vec{\nabla}\varphi)^2 + \frac{m^2}{2}\varphi^2 + \frac{\lambda}{4!} \varphi^4 \right\} \, , \label{freeEnA}
\end{align}
with $m^2 < 0$ to spontaneously break the $Z_2$ symmetry, and a positive quartic coupling $\lambda > 0$.
The dissipative equation of motion for the non-conserved order parameter field $\varphi(x)$ assumes the form of a stochastic Langevin equation,
\begin{align}
    \partial_t^2 \varphi + \gamma \partial_t \varphi = -\frac{\delta F}{\delta \varphi} + \xi
\end{align}
with damping $\gamma$, and fluctuating white Gaussian noise $\xi$ with zero mean, and variance
\begin{align}
    \langle \xi(t,\va{x}) \xi(t',\va{x}')\rangle_\beta &= 2\gamma T \delta(t-t')\delta(\va{x}-\va{x}') \, ,
\end{align}
to model an external heat bath at temperature $T=1/\beta$.
Since there is no conservation law in this system, its critical point lies in the dynamic universality class of Model~A in the classification scheme of Halperin and Hohenberg~\cite{RevModPhys.49.435}, predicting the dynamic critical exponent $z=2+c\eta^\perp$, where $\eta^\perp$ denotes the spatial anomalous scaling dimension, and  $c$ is a constant of order one which depends on the number of spatial dimensions~$d$.

Model~A can be formulated using a classical-statistical path integral \cite{PhysRevA.8.423,Hertz_2016} with Martin-Siggia-Rose (MSR) action\footnote{The unconventional factor of $2$ in front of the force $\delta F/\delta \phi^c$ originates from defining the Keldysh rotation \eqref{keldyshRot} to be measure-preserving. Hence, we have $\phi^c(x)=\sqrt{2}\,\varphi(x)$.}
\begin{align}
    S &= \int_x \Bigg[ -\phi^q \left( \partial_t^2 \phi^c + \gamma \partial_t \phi^c + 2\frac{\delta F}{\delta \phi^c} \right) +  2i\gamma T (\phi^q)^2 \Bigg] \, . \label{bare-action}
\end{align}
Concerning the truncation of the effective average action, which is necessary for practical calculations within the FRG, we follow the two different vertex expansion schemes, around two different expansion points, as introduced in Section \ref{sct:truncationsForRealTimeApplications}. This is explained more explicitly in the following two subsections.

\subsubsection{Comoving expansion}
\label{sct:ExpAroundScaleDepMinforA}

%%JR: Den folgenden Text habe ich auskommentiert, da er so jetzt in Kapitel 2 steht.
%In the `comoving' expansion the vertex expansion  is performed around the scale-dependent minimum $\phi_{0,k}$.
%This makes the 1PI vertex functions $\Gamma_k^{(n)}$ also implicitly dependent on the minimum $\phi_{0,k}$.
%Since in the symmetry-broken phase the minimum $\phi_{0,k}$ of the effective average action is itself $k$-dependent, the flow of $\Gamma_k^{(n)}$ acquires an additional contribution through the functional chain rule,
%\begin{align}
%    \partial_k \Gamma_k^{\alpha_1 \dots \alpha_n}(x_1,\dots,x_n) &= \frac{\delta^n \, \partial_k \Gamma_k[\phi]}{\delta \phi^{\alpha_1}(x_1) \dots \delta \phi^{\alpha_n}(x_n)} \bigg\rvert_{\phi_{0,k}} + \nonumber \\ 
%    &\hspace{-2cm} \int_y \Gamma_k^{\alpha_1 \dots \alpha_n \beta}(x_1,\dots,x_n,y) \, \partial_k \phi_{0,k}^\beta(y) \, ,
%	\label{nPointFncChainRule}
%\end{align}
%where the first term on the r.h.s.~is understood as a $k$-derivative at fixed background field configuration $\phi = \phi_{0,k}$ and corresponds to the $n^\text{th}$ functional derivative of the Wetterich equation \eqref{wetterichEq} evaluated at $\phi_{0,k}$.
%The second term on the r.h.s.~reflects an `interaction' with the $k$-derivative of the mean-field expectation value, as shown in the last diagram in Fig.~\ref{gamma2-general-flow}, in the case of the 2-point function.

Our goal is to capture non-trivial energy and momentum-dependent effects in the spectral function, specifically the scaling behavior $\rho(\omega) \sim \omega^{-\sigma}$ close to the critical point.
This implies that we have to keep at least some of the full frequency dependence in the 2-point function.
For simplicity, we approximate higher $n$-point functions with $n \geq 3$ to be local, which here means by frequency and momentum-independent coupling constants.
These effective couplings from $n$-point functions at vanishing external momenta can in principle be maintained to arbitrarily high order in the vertex expansion, which results in the introduction of a scale-dependent effective potential $V_k( \varphi )$ (see below). Although higher-order local $n$-point vertices are thus readily incorporated, in this work, we restrict the expansion order to $n_\mathrm{max} =4$, however, which corresponds to the lowest applicable order of our combined vertex and loop expansion around a scale-dependent minimum explained in Sec.~\ref{sct:ExpAroundScaleDepMin}. 
We can then compactly express this setup as the following truncation for the effective average action,
\begin{align}
    \label{vertexExpScaleDepMin}
    \Gamma_k &=
    \frac{1}{2} \int_{xx'} \left( \phi^c(x) - \phi^c_{0,k}, \phi^q(x) \right) \begin{pmatrix}
        0 & \Gamma_k^{cq}(x-x') \\
        \Gamma_k^{qc}(x-x') & \Gamma_k^{qq}(x-x')
    \end{pmatrix} 
    \begin{pmatrix}
        \phi^c(x') - \phi^c_{0,k} \\
        \phi^q(x')
    \end{pmatrix} \\ \nonumber
    &\hspace{0.3cm} - \frac{\kappa_k}{\sqrt{8}} \int_{x} \left( \phi^c(x) - \phi^c_{0,k} \right)^2 \phi^q(x) - \frac{\lambda_k}{12} \int_{x} \left( \phi^c(x) - \phi^c_{0,k} \right)^3 \phi^q(x) \, ,
\end{align}
where we have introduced the effective 3-point coupling constant $\kappa_k$ defined at the non-vanishing field expectation value $\phi_{0,k}^c$.
Using the fact that $\phi^c_{0,k}$ minimizes the effective average action, and employing the $Z_2$ symmetry $\Gamma_k[\phi^c \to -\phi^c] = \Gamma_k$, we find the relations
\begin{align}
    \kappa_k = \frac{\lambda_k \phi^c_{0,k}}{\sqrt{2}} \hspace{0.3cm}\text{and}\hspace{0.3cm} \phi_{0,k}^c = \sqrt{\frac{6m_k^2}{\lambda_k}} \label{kappaScalDepMinRelations}
\end{align}
with $ m_k^2 \equiv -\Gamma_k^{qc}(p=0) $.

\paragraph{2-Point function:}
We assume invariance under spacetime translations and expand the Fourier-transformed  2-point functions in powers of spatial $\va{p}^2$, but keep the full frequency dependence only in the zeroth-order coefficient,
\begin{subequations}
\begin{align}
    \Gamma_k^{qc}(\omega, \va{p}) &= \Gamma_{0,k}^{qc}(\omega) - Z_k^\perp  \va{p}^2 + \cdots \, , \\
    \Gamma_k^{cq}(\omega, \va{p}) &= \Gamma_{0,k}^{cq}(\omega) - Z_k^\perp \va{p}^2 + \cdots \, , \\
    \Gamma_k^{qq}(\omega,\va{p}) &= \frac{2T}{\omega} \left( \Gamma_k^{qc}(\omega, \va{p})-\Gamma_k^{cq}(\omega, \va{p}) \right) \, .
\end{align}
\end{subequations}
We will neglect the $\mathcal O(\va{p}^4) $ terms and furthermore  
 assume that the $\omega$-dependence of the spatial wave-function renormalization factor occuring  at $\mathcal O(\va{p}^2) $  is weak, i.e.~that $Z_k^\perp(\omega) \approx Z_k^\perp$ is approximately frequency independent. Moreover, because we now have  $m_k^2 =  -\Gamma_{0,k}^{qc}(0) $, it is convenient to separate the mass term  from the frequency-dependent part of order zero in the spatial momentum, writing  $\Gamma_{0,k}^{qc}(\omega) = -m_k^2 + \Delta\Gamma_{0,k}^{qc}(\omega)$ so that  $\Delta\Gamma_{0,k}^{qc}(0) = 0 $ (we will discuss the flow equation for the squared mass $m_k^2$ in the context of the effective potential in more detail below).
The scale-dependent retarded, advanced, and Keldysh propagators in the background  of a constant (classical) field expectation value $\varphi$, with $\phi^c(x) \equiv \sqrt{2}\, \varphi$ and $ \phi^q(x) \equiv 0 $, and in thermal equilibrium are then given explicitly by
\begin{subequations}
\begin{align}
    G_{\varphi,k}^{R}( \omega,\va{p} ) &= - \frac{1}{  \Delta\Gamma_{0,k}^{qc}(\omega)  - Z_k^\perp \, \va{p}^2 - \lambda_k \varphi^2/2 + m_k^2/2 + R_k^{R}(\va{p} ) } \, , \label{ret-propagator} \\
    G_{\varphi,k}^{A}( \omega,\va{p} ) &= - \frac{1}{  \Delta \Gamma_{0,k}^{cq}(\omega) - Z_k^\perp \, \va{p}^2 - \lambda_k \varphi^2/2 + m_k^2/2 + R_k^{A}(\va{p} ) } \, , \label{adv-propagator} \\
    G_{\varphi,k}^{K}( \omega,\va{p} ) &= \frac{2T}{\omega} \left( G_{\varphi,k}^{R}( \omega,\va{p} ) - G_{\varphi,k}^{A}( \omega,\va{p} ) \right) \, , \label{kel-propagator}
\end{align} \label{propagators}
\end{subequations}

\noindent
where $\Gamma_k^{qc}(0,\boldsymbol{0}) = \Gamma_k^{cq}(0,\boldsymbol{0}) = -m_k^2 $ holds with $\lambda_k \varphi^2/2 = 3 m_k^2/2 $ at the minimum field value.
For $R^{R/A}_k(\va p) $ in this work we use the optimized regulator of the form given in Eq.~\eqref{eq:litimReg}.
Having the full scale-dependent propagators set up, we can calculate explicit expressions for the loop functions \eqref{ILoopDef}, \eqref{JLoopDef}. 
Analytic results are provided in Appendix~\ref{sct:LoopFncsClosedExpr} whenever available.
Otherwise, the frequency and momentum integrals are solved numerically with the methods outlined in Appendix~\ref{sct:numerical-impl}.
We can now also expand the flow equation, e.g., for the retarded 2-point function in powers of $\va{p}^2$, to obtain
\begin{align}
    \label{gam2R-flow}
    \partial_k \Gamma_{0,k}^{qc}(\omega) &= -\frac{i\kappa_{k}^2}{2} \, J_k^{KR}(\omega,\VecZero) + \frac{i\lambda_k}{4} \, I_k^K + \frac{\kappa_{k}}{\sqrt{2}} \, \partial_k \phi_{0,k}^c \, , \\
    \label{Zperp-flow}
    \partial_k Z_k^\perp &= \frac{i\kappa_{k}^2}{2} \lim_{\va{p}^2 \to 0} \, \frac{\partial }{\partial \va{p}^2} \, J_k^{KR}(0,\va{p}) \, , \;\text{etc.}
\end{align}
%Moreover, one can separate the frequency-dependent part from the zeroth order ($\va{p}=0$) term \eqref{gam2R-flow} according to $\Gamma_{0,k}^{qc}(\omega) = -m_k^2 + \Delta\Gamma_{0,k}^{qc}(\omega)$ for convenience, with $\Delta\Gamma_{0,k}^{qc}(0) = 0$ set to zero to avoid double counting.
Finally, inserting the explicit expressions for the $J$'s from Appendix~\ref{sct:LoopFncsClosedExpr} into~\eqref{Zperp-flow}, we can convert this flow equation into an expression for the spatial anomalous scaling dimension $\eta_k^\perp$, which in this truncation hence reads
\begin{align}
    \label{flow-eta-perp}
    \eta_k^\perp &= -k \partial_k \log Z_k^\perp =  \frac{\Omega_d k^{d+2} T Z_k^\perp}{ (2\pi)^d } \frac{\kappa_{k}^2}{\left( Z_k^\perp k^2 + m_k^2 \right)^4} \, ,
\end{align}
with $\Omega_d$ denoting the volume of the $d$-dimensional ball with unit radius, e.g.~$\Omega_3 = 4\pi/3$.

\paragraph{Effective potential:}
We start with the general flow equation \eqref{flow-eff-pot} for the effective potential in spacetime-translation invariant systems. 
Using that $B_{\varphi,k}^K(x,x) = I_{\varphi,k}^K$ with the expression for $I_{\varphi,k}^K$ from Appendix~\ref{sct:LoopFncsClosedExpr}, and doing a formal integration with respect to $\varphi$, we obtain
\begin{align}
    \label{flow-eff-pot-rho-cl-limit}
    \partial_k V_k(\varphi) &= \left( 1 - \frac{\eta_k^\perp}{2 + d} \right) \frac{\Omega_d k^{d+1} T Z_k^\perp}{(2\pi)^{d}} \frac{ 1 }{Z_k^\perp k^2 + V_k''(\varphi)}
\end{align}
for the flow of the effective potential, which remarkably is the same as in the static $d$-dimensional Euclidean case.
Because of the $Z_2$ symmetry, the effective potential only depends on $\varphi^2$, which suggests the usual substitution $\rho \equiv \varphi^2$, and effectively replaces $V_k''(\varphi) \to 2 V_k'(\rho) + 4 \rho V_k''(\rho)$ in the flow equation.
Subsequently, a common approximation scheme is to expand the effective potential $V_k(\rho)$ in a power series in $\rho$ around the square of the scale-dependent minimum~\cite{Berges:2000ew},
\begin{align}
    \label{eff-pot-taylor-expansion}
    V_k(\rho) = \sum_{n=2}^{n_{\text{max}}/2} \frac{v_{n,k}}{n!} (\rho - \rho_{0,k})^n
\end{align}
up to some cutoff order $n_{\text{max}}/2$.
The flow equations for the coefficients $v_n$ are then easily obtained by differentiating \eqref{flow-eff-pot-rho-cl-limit} $n$ times with respect to $\rho$ and setting $\rho=\rho_{0,k}$ to the field expansion point afterward.
The flow of the minimum $\rho_{0,k}$ can be projected by requiring that it stays a minimum during the flow \cite{Wetterich:1992yh,Sinner:2007ws}, which immediately entails
\begin{align}
	\partial_k \rho_{0,k} = - \frac{(\partial_k V_k')(\rho_{0,k})}{V_k''(\rho_{0,k})}
\end{align}
for the flow of $\rho_{0,k}$.
The lowest non-trivial cutoff order is given by $n_{\text{max}} = 4$.
In this case, all third derivatives of the effective average potential with respect to $\rho$ vanish, and the final flow equations read
\begin{align}
    \partial_k \rho_{0,k} &= \left( 1 - \frac{\eta_k^\perp}{2 + d} \right) \frac{6\Omega_d Z_k^\perp k^{d+1} T}{(2\pi)^{d}} \frac{1}{\left(\lambda_k \rho_{0,k}/3 + Z_k^\perp k^2 \right)^2} \, , \label{eq:flowRho0} \\
    \partial_k \lambda_k
    &= \left( 1 - \frac{\eta_k^\perp}{2 + d} \right) \frac{6 \Omega_d Z_k^\perp k^{d+1} T }{(2\pi)^{d}} \frac{\lambda_k^2}{\left( \lambda_k \rho_{0,k}/3 + Z_k^\perp k^2 \right)^3} \, , \label{eq:flowLambdaFiniteFieldExpVal}
\end{align}
if $\rho_{0,k} > 0$. When reaching $\rho_{0,k} = 0$, on the other hand, we dynamically switch to
\begin{align}
    \partial_k m_k^2 &= - \left( 1 - \frac{\eta_k^\perp}{2 + d} \right) \frac{\Omega_d Z_k^\perp k^{d+1} T}{(2\pi)^{d}}  \frac{\lambda_k}{\left( m_k^2 + Z_k^\perp k^2 \right)^2} \, , \label{eq:flowM2ZeroField} \\
    \partial_k \lambda_k &= \left( 1 - \frac{\eta_k^\perp}{2 + d} \right) \frac{6 \Omega_d Z_k^\perp k^{d+1} T }{(2\pi)^{d}} \frac{\lambda_k^2}{\left( m_k^2 + Z_k^\perp k^2 \right)^3}\, , \label{eq:flowLambdaZeroField}
\end{align}
with the squared mass $m_k^2 = 2V_k'(\rho=0)$.
Concerning the critical scaling of the quartic coupling $\lambda_k$ and the squared mass $m_k^2$, we find the expected result
\begin{align}
    m_k^2 \sim k^{2-\eta^\perp} \,, \hspace{0.5cm} \lambda_k \sim k^{4-d-2\eta^\perp} \, , \label{m2LamCritScal}
\end{align}
which can be readily verified by inserting power-law ans\"atze into the flow equations listed above.

%Moreover, the 3-point coupling constant, which occurs in many of the flow equations, is then in the Taylor expansion \eqref{eff-pot-taylor-expansion} just given by $\kappa_k = V_k'''(\varphi_{0,k}) = \sqrt{ \rho_{0,k} } \lambda_k$ for $\rho_{0,k} > 0$ and $\kappa_k = V_k'''(0) = 0$ for $\rho_{0,k} = 0$.

\paragraph{Damping:}
In presence of the non-local one-loop diagrams in the flow equation in Fig.~\ref{gamma2-general-flow}, which occur because of the non-vanishing 3-point vertex at the scale-dependent minimum in the broken phase, there is a contribution to the damping during the flow even in a truncation that only contains one-loop structures as used here. This is in contrast to the symmetric phase, where the damping only receives a contribution at two-loop level~\cite{Huelsmann:2020xcy,Tan:2021zid,Roth:2021nrd}.
Its flow equation can be straightforwardly derived by virtue of the FDR \eqref{FDRForGam2}, and with noting that $ \partial_k\Gamma^{qq}_k(0, \va{0}) = 4i T\partial_k \gamma_k $. 
In the classical limit,
it is thus given by the first diagram in the second line of Fig.~\ref{gamma2-general-flow}, which yields
\begin{align}
    \label{gamGeneralFlow}
    \partial_k \gamma_k
    &= -\frac{\kappa_{k}^2}{8T} \, J_k^{KK}(0,\VecZero) = -\frac{\kappa_{k}^2}{8T} \int_p B_k^K(p) G_k^K(p) \, .
\end{align}

\subsubsection{Expansion around the IR minimum}
\label{sct:ExpAroundIRMinforA}

Another possibility that we have explored is to use our combined vertex and loop expansion around vanishing field expectation values $\phi_{0}(x) = 0$ from Sec.~\ref{sct:ExpAroundIRMin}, which corresponds to the minimum of the symmetric phase. More precisely, we make the spacetime-translation invariant ansatz
\begin{align}
    \Gamma_k = &
    \frac{1}{2} \int_{xx'} \left( \phi^c(x), \phi^q(x) \right)
    \begin{pmatrix}
        0 & \Gamma_k^{cq}(x-x') \\
        \Gamma_k^{qc}(x-x') & \Gamma_k^{qq}(x-x')
    \end{pmatrix}
    \begin{pmatrix}
        \phi^c(x') \\
        \phi^q(x')
    \end{pmatrix} + \nonumber \\
    & \frac{3\cdot 2^2}{4!} \int_{xx'} \phi^q(x) \phi^c(x) V_k^{an}(x-x') \phi^q(x') \phi^c(x') + \nonumber \\
    & \frac{3\cdot 2}{4!} \int_{xx'} \phi^q(x) \phi^c(x) V_k^{cl,R}(x-x') \phi^c(x') \phi^c(x') + \nonumber \\
    & \frac{3\cdot 2}{4!} \int_{xx'} \phi^c(x) \phi^c(x) V_k^{cl,A}(x-x') \phi^q(x') \phi^c(x') + \nonumber \\
    & \text{6-point function} \label{vertexExpSymm}
\end{align}
for the effective average action.
The powers of two in the prefactors of the vertices in \eqref{vertexExpSymm} count the number of ways one can permute the Keldysh indices in the 4-point $s$, $t$ and $u$-channels, without affecting the value of the integral, and we have an overall factor of three which expresses the number of channels.
We will neglect a possible 6-point function in the last line in \eqref{vertexExpSymm} in order to ensure that the truncation is comparable with the comoving expansion \eqref{vertexExpScaleDepMin} of the previous Sec.~\ref{sct:ExpAroundScaleDepMin} in terms of the highest order of vertices that is still taken into account.

To generalize our earlier truncation of Ref.~\cite{Roth:2021nrd} to $d > 0$ spatial dimensions, we equip the combined vertex and loop expansion with a simultaneous expansion in spatial gradients (analogous to what we did for the 2-point function in the comoving expansion in  Subsection~\ref{sct:ExpAroundScaleDepMinforA}).
For the scope of this work, we again restrict ourselves to include only the first order corrections, i.e.~the  $\Ord{\va{p}^2}$ terms.
Then the vertices are also expanded according to
\begin{subequations}
\label{eq:4vertex}
\begin{align}
	V_k^{cl,A}(\omega,\va{p}) &= -\frac{\lambda_k}{6} + \Delta V_{0,k}^{cl,A}(\omega) + V_{1,k}^{cl,A}(0) \va{p}^2 + \cdots \, , \label{eq:4vertexR} \\
    V_k^{cl,R}(\omega,\va{p}) &= -\frac{\lambda_k}{6} + \Delta V_{0,k}^{cl,R}(\omega) + V_{1,k}^{cl,R}(0) \va{p}^2 + \cdots \, , \label{eq:4vertexA}\\
    V_k^{an}(\omega,\va{p}) &= \frac{2T}{\omega} \left( V_k^{cl,R}(\omega,\va{p}) - V_k^{cl,A}(\omega,\va{p}) \right) \, ,
\end{align} 
\end{subequations}
where we have separated off the constant $-\lambda_k/6$ term (in momentum space), and we have set $ \Delta V_{0,k}^{cl,R/A}(0) = 0 $, $ \Delta V_{0,k}^{an}(0) = 0 $, accordingly, to avoid double counting.
Analogous to the case of the spatial wave function renormalization factor $Z_k^\perp$, we employ the approximation that the first order coefficients $V_{1,k}^{cl,R}(\omega)$ only depend weakly on the frequency $\omega$, and can hence be approximated by $ V_{1,k}^{cl,R}(\omega) \approx V_{1,k}^{cl,R}(0) $.

The propagators at vanishing field expectation value $\varphi=0$ are also given by the expressions \eqref{ret-propagator} -- \eqref{kel-propagator} as in the comoving expansion, with the exception that the  squared mass $m_k^2$ is here, of course, also understood at vanishing field expectation value.

\paragraph{2-Point function:}
Our two-loop complete flow equation \eqref{gamma2CQFlowSymm} for the 2-point function can be readily expanded in powers of $\va{p}^2$ to project the flow equation of the advanced 2-point function onto the expansion coefficients $\Gamma_{0,k}^{qc}(\omega)$, $Z_k^\perp$, \dots.
Like in the comoving expansion it is both numerically and analytically convenient to separate off the frequency-dependent part $ \Delta \Gamma_{0,k}^{qc}(\omega) $ of the full advanced 2-point function. Correspondingly, setting $\omega = 0$ in \eqref{gamma2CQFlowSymm} yields a flow equation for the squared mass,
\begin{align}
	-\partial_k m_k^2
	&= \frac{i \lambda_k}{4} I_k^K - \frac{4 \, \Omega_d k^{d+3} T Z_k^\perp}{(2\pi)^d} \frac{d}{d+2} \left( 1-\frac{\eta_k^\perp}{4+d} \right) \frac{ V_{1,k}^{cl,R}(0) }{\left( m_k^2 + Z_k^\perp k^2 \right)^2} \, , \label{m2FlowSymmFinalResult}
\end{align}
which we have obtained by solving the frequency and momentum integrals analytically.
Moreover, the flow equation for the spatial wave function renormalization factor $Z_k^\perp$ is obtained by differentiating \eqref{gamma2CQFlowSymm} straightforwardly with respect to $\va{p}^2$,
\begin{align}
    \partial_k Z_k^\perp &= i V_{1,k}^{cl,R}(0) \, I_k^K \, , \label{flowZPerpSymm}
\end{align}
which can be evaluated in closed form and rearranged into the explicit expression, 
\begin{align}
	\eta^\perp_k 	&= \left( \frac{1}{2+d} - \frac{ (2\pi)^d }{ 4 \, \Omega_d k^{d+2} T } \frac{ \left( m_k^2 + Z_k^\perp k^2 \right)^2 }{ V_{1,k}^{cl,R}(0) } \right)^{-1} \, , \label{spatAnomDimSymm}
\end{align}
for the spatial anomalous dimension.

\paragraph{4-Point function:}
We split the full 4-point function $\Gamma_k^{(4)}$ into $s$, $t$, and $u$ channels, as already mentioned above, and follow our loop expansion of Ref.~\cite{Roth:2021nrd} to rewrite the right-hand side of its full flow equation in our `local-vertex approximation'.
In this approximation, we replace all occurring 4-point functions with effective local coupling constants, cf.~Fig.~\ref{fig:gam4FlowSymm}.
The flow equation for the 4-point function expanded in powers of $\va{p}^2$ can hence be compactly expressed as
\begin{subequations}
\begin{align}
    \label{vClR0FlowEq}
    \partial_k V_{0,k}^{cl,R}(\omega) &=  -\frac{i \lambda_k^2}{4} J_k^{KR}(\omega,\VecZero) \, , \\
    \label{vClR1FlowEq}
    \partial_k V_{1,k}^{cl,R}(0) &= -\frac{i \lambda_k^2}{4} \lim_{\va{p}\to 0} \frac{\partial}{\partial \va{p}^2} J_k^{KR}(0,\va{p}) \, ,\;\text{etc.},
\end{align}
\end{subequations}
here just for the retarded part, since the others (advanced and Keldysh) follow readily by symmetry, and also with the aforementioned self-consistently determined 4-point coupling $\lambda_k$ given in Eq.~\eqref{localVertexApprox}.
We can now set $\omega=0$ in \eqref{vClR0FlowEq} and solve the remaining integral analytically to find a flow equation for the quartic coupling constant,
\begin{align}
    \label{flowLambdaSymm}
	\partial_k \lambda_k %&=  \frac{3i \lambda_k^2}{2} J_k^{KA}(0,\VecZero)
	= \frac{6\Omega_d k^{d+1} T Z_k^\perp}{(2\pi)^d} \left( 1-\frac{\eta_k^\perp}{2+d} \right) \frac{\lambda_k^2}{\left( m_k^2 + Z_k^\perp k^2 \right)^3} \, ,
\end{align}
and correspondingly for the first-order momentum-dependent correction in \eqref{vClR1FlowEq},
\begin{align}
    \partial_k V_{1,k}^{cl,A}(0) &=  \frac{\Omega_d k^{d+1}  T\, ( Z_k^\perp )^2}{2 \, (2 \pi)^d} \frac{ \lambda_k^2 }{\left( m_k^2 + Z_k^\perp k^2 \right)^4} \, , \label{vClA1FlowEqClass}
\end{align}
see Appendix~\ref{sct:LoopFncsClosedExpr}.

As a final remark concerning our truncation scheme, we emphasize that the vertex expansion around $\phi_{0,k}(x) =0$ is valid in the symmetric phase for temperatures $T \geq T_c$, i.e.~when the $Z_2$ symmetry is restored at some scale~$k$ during the FRG flow.
Otherwise, the effective mass parameter $m_k^2 + Z_k^\perp k^2$ crosses zero at some scale~$k$ which leads to divergences in the corresponding propagators and hence  invalidates parts of the flow equations. 
We thus have to require $m_k^2 + Z_k^\perp k^2 > 0$ at all FRG scales~$k>0$, which for our applications is equivalent to restricting to temperatures $T \geq T_c$ in the symmetric phase.

\subsection{Model B---Linear coupling to a conserved density}
\label{sct:ModB}

Motivated by the work of Son and Stephanov \cite{Son:2004iv} on the dynamic universality class of the critical endpoint in the phase diagram of QCD, we now introduce a linear coupling of the order parameter field $\varphi(x)$ to a conserved `baryon' number density $n(x)$.
The slow infrared mode is then determined by the diffusive 
density fluctuations, 
and one expects to find critical Model B dynamics instead of Model A above. More specifically, we modify our free energy \eqref{freeEnA} according to
\begin{align}
    F = \int d^d x \left\{ \frac{1}{2} (\vec{\nabla}\varphi)^2 + \frac{m^2}{2} \varphi^2 + \frac{\lambda}{4!} \varphi^4 + B \varphi n + \frac{n^2}{2\chi_0}  \right\} \, , \label{freeEnB}
\end{align}
with the coupling $B$ between the order parameter $\varphi(x)$ and the conserved density $n(x)$, and with the baryon susceptibility $\chi_0$.
The stochastic hydrodynamic equations of motion can be derived using standard rules \cite{RevModPhys.49.435,Son:1999pa,Son:2002ci,Son:2004iv,Fujii:2004za,Nakano:2011re} and assume the form
\begin{subequations}
\begin{align}
    \partial_t^2 \varphi + \gamma \partial_t \varphi &= - \frac{\delta F}{\delta \varphi} + \xi \, , \label{hydroBPhi} \\
    \tau_R \partial_t^2 n + \partial_t n &= 
    \lambda_n \vec{\nabla}^2 
    \frac{\delta F}{\delta n} + \vec{\nabla} \cdot \vec{\zeta} \, , \label{hydroBn}
\end{align}
\end{subequations}
where we have introduced the baryon conductivity $\lambda_n$ (not to be confused with the quartic coupling $\lambda$), a finite Israel-Stewart-type relaxation time $\tau_R$ to exclude propagation at superluminal velocities \cite{Jeon:2015dfa}, and a white Gaussian noise vector $\vec{\zeta}$, with 
\begin{align}
    \langle \zeta_i(t,\va{x}) \zeta_j(t',\va{x}')\rangle_\beta &= 2\lambda_n  T \delta_{ij} \delta(t-t')\delta(\va{x}-\va{x}') \, ,
\end{align}
for the diffusive density fluctuations. 
The equation of motion \eqref{hydroBn} for the conserved density $n(x)$ represents the relaxation equation for the associated baryon current $\vec j(x)$,
\begin{equation}
\partial_t \vec j = -\frac{1}{\tau_R} 
\Big(\vec j 
%- \vec K \big)\, , \;\;\mbox{where} \;\; %\vec K = -
+ \lambda_n \vec{\nabla}
    \frac{\delta F}{\delta n} + \vec{\zeta}\,
    \Big) \, ,
\end{equation}
so that $\partial_t n  +\vec\nabla \!\cdot\!\vec j = 0$
and the total baryon number $Q= \int d^d x \, n(\va{x})$ in the system
is conserved.
To prepare for our real-time FRG formulation, we can now continue writing down the corresponding MSR action $S$,
\begin{align}
	S = \int_x \bigg[
	&- \phi^q \left( \partial_t^2 \phi^c + \gamma \partial_t \phi^c + 2\frac{\delta F}{\delta \phi^c} \right) + 2i \gamma T \, \left( \phi^q \right)^2 + \nonumber \\ 
	&\hspace{-1.0cm} - n^q \left( \tau_R \partial_t^2 n^c + \partial_t n^c - 2\lambda_n \vec{\nabla}^2 \frac{\delta F}{\delta n^c} \right) - 2i \lambda_n T \, n^q\vec{\nabla}^2 n^q \bigg] \, , \label{generalMSRAction}
\end{align}
where we have also used our measure-preserving definition of the Keldysh classical and quantum components for the conserved-density field on the closed-time path,
\begin{align}
	n^{c,q}(x) &\equiv \frac{1}{\sqrt{2}} \left( n^{+}(x) \pm n^{-}(x) \right) \, , \label{keldyshRotN}
\end{align}
which we state explicitly here for completeness.

Since our conserved-density fields $n^c(x)$, $n^q(x)$ both enter only quadratically in the MSR action \eqref{generalMSRAction}, we can perform the corresponding Gaussian integrations  which amounts to modifying the bare retarded/advanced 2-point functions of the order parameter field  at the UV initial scale ($k=\Lambda$) according to
\begin{align}
    \label{bareS2B}
    %{S^{(2)}}^{R/A}
    \Gamma_\Lambda^{R/A}
    (\omega,\va{p}) &= \omega^2 \pm i\gamma\omega - m^2 - \va{p}^2 -   \frac{\lambda_n B^2 \va{p}^2}{\tau_R \omega^2 \pm i\omega - D_0 \va{p}^2} \, ,
\end{align}
where  $D_0 \equiv \lambda_n/\chi_0$ is the baryon  diffusion constant, and the Keldysh component is again determined by the FDR.
Because of the diffusive character of the conserved density, the limits $\omega \to 0$ and $\va{p} \to 0$ of the inverse propagator \eqref{bareS2B} no longer commute.
This has the unavoidable consequence that we have to distinguish between the \emph{static screening mass} $m_s$ (or just `static mass'),
\begin{align}
    m_s^2 &\equiv - \lim_{\va{p}\to 0} \lim_{\omega \to 0} %{S^{(2)}}
    \Gamma_\Lambda^{R/A}(\omega,\va{p}) = m^2 - B^2 \chi_0 \, , \label{modBStatMass}
\end{align}
and the \emph{plasmon} mass $m_p$,
\begin{align}
    m_p^2 &\equiv -\lim_{\omega \to 0} \lim_{\va{p}\to 0} %{S^{(2)}}
    \Gamma_\Lambda^{R/A}(\omega,\va{p}) = m^2 \, , \label{modBPlasmonMass}
\end{align}
in our calculations.
It is important to note here that it is the \emph{static} mass that is inversely proportional to the correlation length, $\xi = \sqrt{Z^\perp}/m_s$, and hence vanishes at the critical point. Compared to that, the plasmon mass $m_p$
contains the strictly positive offset $B^2 \chi_0$ and hence  stays finite at criticality.
As we will show in the following subsection, the product $B^2 \chi_0$  does in fact  remain unchanged under the FRG flow.

\subsubsection{Truncation}

In order to truncate the effective average action for Model~B we employ our vertex expansion from Sec.~\ref{sct:ExpAroundScaleDepMin} around the scale-dependent minimum.
This is well motivated by the fact that the only essential difference compared to the corresponding Model~A calculation concerns the one-loop $J$-functions, where we now have to use a form for the Model~B 2-point functions that is modified appropriately.  
Based on the initial form \eqref{bareS2B}, a truncation analogous to Section~\ref{sct:ExpAroundScaleDepMinforA}, e.g.~for the retarded 
2-point function then reads
\begin{align}
    \Gamma_k^{qc}(\omega,\va{p}) &= - m_{s,k}^2 + \Delta \Gamma_{0,k}^{qc}(\omega) - Z_k^\perp \va{p}^2 - B^2 \chi_{0} \, \left( 1 + \frac{D_{0} \va{p}^2}{\tau_{R} \omega^2 + i\omega - D_{0} \va{p}^2} \right) \, . \label{gam2ModBAnsatz} 
\end{align}
The additional non-trivial $\va{p}$-dependence in the propagator directly reflects the diffusive nature of the conserved density fluctuations that mix with those of the order parameter for any non-zero value of the coupling $B$ between the two.
We again require $ \Delta \Gamma_{0,k}^{qc}(0) = 0 $ assuming all frequency and momentum-independent contributions to be subsumed in
the mass term (split here into  static  mass and offset).
The full expressions for the corresponding retarded/advanced propagators $G_{\varphi,k}^{R/A}(\omega,\va{p})$ analogous to Eqs.~\eqref{propagators} then follow readily from the ansatz \eqref{gam2ModBAnsatz}. Because they are is rather lengthy
and not particularly illuminating, they are not  explicitly repeated here.

The flow equation for the effective potential stays the same as in Model~A, cf.~Sec.~\ref{sct:ExpAroundScaleDepMin}, and is given by Eq.~\eqref{flow-eff-pot-rho-cl-limit}.
Thereby one only has to be careful about the quadratic term in the effective potential since it involves the \emph{static} mass, i.e.~$V_k(\rho) = m_{s,k}^2 \rho/2 + \cdots $, because it originates from the static limit of \eqref{gam2ModBAnsatz}.
Likewise, the flow of the spatial wave function renormalization factor $Z_k^\perp$ is also the same as in Model~A and given by Eq.~\eqref{flow-eta-perp}, wherein the mass term $m_k^2$ is also again understood as the static screening mass $m_{s,k}^2$.

It is now straightforward to see that the flows of the new parameters $B^2 \chi_{0}$, $D_{0}$ and $\tau_{R}$ which model the coupling to the conserved density all vanish:
If we assume that we did not integrate out the conserved density $(n^c,n^q)$, then the effective average action would be a functional of all four variables $(\phi^c,\phi^q,n^c,n^q)$ (but without a regulator term for the conserved density).
On the right-hand side of the Wetterich equation \eqref{wetterichEq} the second functional derivative $\Gamma_k^{(2)}[\phi^c,\phi^q,n^c,n^q]$ enters, which is however independent of $(n^c,n^q)$ since $\Gamma_k$ is only quadratic in $(n^c,n^q)$ and the coupling of $n$ to the order parameter $\varphi$ is linear.
Therefore every functional derivative with respect to $n^c$ and $n^q$ will annihilate the right-hand side of \eqref{wetterichEq}.
This is not the case in Model~C, where the coupling between $\varphi$ and $n$ is \emph{non-linear} and hence this argument no longer holds~\cite{Mesterhazy:2013naa} (as we demonstrate in detail below).
Consequently, the difference $ m_{p,k}^2 - m_{s,k}^2 = B^2 \chi_0 $ is independent of the FRG scale~$k$.
This implies that while the static screening mass $m_{s}$ vanishes at the critical point (due to the divergent correlation length $\xi \sim m_{s}^{-1}$ at our second-order phase transition), the plasmon mass $m_{p,k}$ remains finite.
This is closely related to prior observations that the sigma-meson mode stays massive in large-$N$ NJL models \cite{Scavenius:2000qd,Fujii:2003bz,Fujii:2004za}, and will be reflected in the critical spectral functions shown in Sec.~\ref{sct:results} below.

\subsubsection{Critical modes}
\label{sct:ModBCritModes}

In the long-wavelength limit $|\va{p}| \ll k$ at some finite FRG scale~$k$, the two low-frequency poles of the retarded propagator (i.e.~the roots of \eqref{gam2ModBAnsatz} with the optimized regulator of Eq.~\eqref{eq:litimReg} added), can be expanded in a power series in $\va{p}^2$ around $\va{p}=0$,
\begin{subequations}
\begin{align}
    \omega_{1,k}(\va{p}) &= -i D_{\text{eff},k} \va{p}^2 + \Ord{|\va{p}|^4} \, , \label{modBMode1Diff} \\
    \omega_{2,k}(\va{p}) &= -\frac{i}{\gamma_k}\left( m_{p,k}^2 + Z_k^\perp k^2 \right) + \Ord{\va{p}^2} \, , \label{modBMode1Relax}
\end{align}
\end{subequations}
where we have introduced a  scale-dependent effective diffusion constant $D_{\text{eff},k}$ defined as
\begin{align}
    \label{critModeDiffConst}
    D_{\text{eff},k} &\equiv  D_0\left( 1 - \frac{\chi_0 B^2}{m_{s,k}^2 + Z_k^\perp k^2 + \chi_0 B^2} \right) \,.
\end{align}
It is associated with the diffusive mode in \eqref{modBMode1Diff}.  We have furthermore 
taken the low-frequency limit $\omega \ll \gamma_k,\tau_R^{-1}$, such that quadratic and higher-order terms in $\omega$ can effectively be neglected, in particular, those due to  the small but finite Israel-Stewart relaxation time $\tau_R$. Also note that the scale-dependent damping constant $\gamma_k$ in our truncation, obtained from   Eq.~\eqref{gam2ModBAnsatz}, agrees with the general definition  
%$ \gamma_k \equiv -i \lim_{\omega \to 0}  \Delta\Gamma_{0,k}^{qc}(\omega)/ \omega $.
$ \gamma_k \equiv \lim_{\omega \to 0}  \Im\Gamma_{k}^{qc}(\omega,\boldsymbol{0})/ \omega $.

The first mode \eqref{modBMode1Diff} is purely diffusive and represents fluctuations along a linear combination of $\varphi$ and $n$ in the 2-dimensional $(\varphi,n)$ field space \cite{Son:2004iv}, while the second one \eqref{modBMode1Relax} is purely relaxational.
Both these modes can reflect the singular behavior near criticality at the second-order phase transition. 
For the diffusive mode, the usual argument is as follows:
For $k\ll |\va{p}| $ we can neglect the influence of the regulator and the critical singularity gradually 
builds up as $\omega_{1,k}(\va{p}) \sim |\va{p}|^z$ \cite{Schweitzer:2021iqk}, where the finite momentum acts as the relevant infrared cutoff to limit the spatial correlation length to $ \xi^{-1}\sim |\va{p}| $. For $|\va{p}| \ll k$, on the other hand, the infrared cutoff is provided by the regulator $k$, and  the dispersion relation becomes regular, see \eqref{modBMode1Diff}. Requiring these two limits to match smoothly at $|\va{p}| \sim k$, one finds $\omega_{1,k}(k) = -iD_{\text{eff},k} k^2 \sim k^z$. Since it is this mode that gives rise to the Model-B dynamics \cite{Son:2004iv}, with dynamic critical exponent $z=4-\eta^\perp$, the critical divergence of the effective diffusion constant $D_{\text{eff},k}$ must thus read $ D_{\text{eff},k} \sim k^{2-\eta^\perp} $. To verify this explicitly from its definition in \eqref{critModeDiffConst}, note that the regularized static screening mass vanishes as $m_{s,k}^2 + Z_k^\perp k^2 \sim k^{2-\eta^\perp}$ at the critical point.

In contrast, the plasmon mass $m_{p,k}^2$ stays finite, and hence the critical behavior of the relaxational mode $\omega_{2,k}$ is fully determined by the critical divergence of the damping constant $\gamma_k \sim k^{-\eta^\gamma}$, with $\eta^\gamma \equiv -k \partial_k \log \gamma_k$ denoting its scaling dimension. 
Together we can thus summarize the behavior of the two modes at criticality as
\begin{align}
    \omega_{1,k} \sim -i D_{\text{eff},k} k^2 \sim k^{4-\eta^\perp} \, , \quad \omega_{2,k} \sim k^{\eta^\gamma} \, .
\end{align}
In order to calculate the critical exponent $\eta^\gamma$ associated with such a critical divergence of the damping constant, we consider its flow equation from Eq.~\eqref{gamGeneralFlow}, which in this form remains valid also for Model~B. Although in the present case of course we have to use the modified propagators in this equation which include the extra diffusive self-energy from \eqref{gam2ModBAnsatz}.
The structure of the retarded propagator (with the regulator \eqref{eq:litimReg})
at small frequencies and spatial momenta is best emphasized  by a partial fraction decomposition,
\begin{align}
    G_k^R(\omega,\va{p}) &= -\frac{1}{\gamma_k} \frac{i\omega - D_0 \va{p}^2}{\omega_{2,k}(\va{p}) - \omega_{1,k}(\va{p})} \bigg( \frac{1}{\omega-\omega_{1,k}(\va{p})} -  \frac{1}{\omega-\omega_{2,k}(\va{p})} \bigg) \,, \label{GRetModBPFD}
\end{align}
where we have only included the leading order $\Delta \Gamma_{0,k}^{qc}(\omega) = i\gamma_k \omega + \cdots$, consistent with our limit of small frequencies $\omega \ll \gamma_k, \tau_R^{-1}$ as explained above.
Taylor expanding the two modes $\omega_{1,k}(\va{p})$ and $\omega_{2,k}(\va{p})$  in spatial momentum $\va{p}$ and keeping only the first term from the diffusive pole \eqref{modBMode1Diff} in the brackets  which represents the dominant low frequency contribution  in \eqref{GRetModBPFD},
one obtains the critical part of the retarded propagator,
\begin{align}
    G_k^R(\omega,\va{p}) = \frac{1}{m_{p,k}^2 + Z_k^\perp k^2} \frac{\omega + i D_0 \va{p}^2}{\omega+iD_{\text{eff},k} \va{p}^2} + \cdots \, . \label{GRetModBExpanded}
\end{align}
First, we observe that it is indeed independent of $\gamma_k$, as expected for Model-B dynamics.
Hence, by inserting the critical parts  of the retarded \eqref{GRetModBExpanded} and advanced propagators into the flow equation for $\gamma_k$ in \eqref{gamGeneralFlow}, we can isolate the critical contribution to the flow of $\gamma_k$ according to
\begin{align}
    \partial_k \gamma_k = f(k) + \text{less singular (for $k\to 0$)} \,, \label{flowGamModBSchematic}
\end{align}
where $f(k)$ is some function independent of $\gamma_k$.
Moreover, in the critical region, we have $f(k) = A\, k^{-x_f}$ with some possibly non-universal prefactor $A$ and a universal critical exponent $x_f$.
For such a functional form of $f(k)$ the flow \eqref{flowGamModBSchematic} can be directly integrated, which yields
\begin{align}
    \gamma_k = \frac{A}{1-x_f} k^{1-x_f} + \text{less singular}
\end{align}
and we hence have $\eta^\gamma = x_f-1$ from the definition of $\eta^\gamma$. 
By solving the integral in \eqref{gamGeneralFlow} at criticality, where only the IR-dominant diffusive mode \eqref{GRetModBExpanded} is needed the function $f(k)$ can be obtained analytically (the frequency integral can be solved with the residue theorem), and explicitly expressed in $d$ spatial dimensions as
\begin{align}
    f(k)
    %&= -\frac{\kappa_k^2}{8T} \frac{\Omega_d k^{d-1}}{(2\pi)^d} \frac{4(D_0-D_k)^2 (3D_0+D_k) (d-\eta_k^\perp) T^2 Z_k^\perp}{(d-2) D_k^4 (m_{p,k}^2 + Z_k^\perp k^2)^3} \\
    &= - \frac{\Omega_d Z_k^\perp k^{d-1} T}{(2\pi)^d} \frac{(d-\eta_k^\perp) (D_0-D_k)^2 (3D_0+D_k)  \kappa_k^2}{2(d-2) D_k^4 (m_{p,k}^2 + Z_k^\perp k^2)^3}  \,. \label{eq:fexpl}
\end{align}
The bare diffusion constant $D_0$ is independent of $k$ and the plasmon mass $m_{p,k}$ has a finite $k\to 0$ limit. Moreover, the first relation in \eqref{kappaScalDepMinRelations} together with the static scaling behavior \eqref{m2LamCritScal} entail that the 3-point coupling constant critically vanishes as
\begin{align}
    \kappa_k^2 \sim k^{6-d-3\eta^\perp} \; .
\end{align}
Inserting this into (\ref{eq:fexpl}), together with the critical singular behavior $D_{\text{eff},k} \sim k^{2-\eta^\perp}$ of the effective diffusion constant as explained earlier, we find
\begin{align}
    f(k) \sim \frac{k^{d-1} \kappa_k^2}{D_k^4} \sim k^{-3}
\end{align}
and hence $x_f = 3$. The corresponding exact result for the critical scaling behavior of the damping constant $\gamma_k$ in Model~B is $\eta^\gamma = 2$.

In the IR limit $k \to 0$, criticality is also manifest in the spectral function. In this case we now therefore use that $|\va{p}| \gg k$ for any finite value of the spatial momentum.
Sending $k \to 0$ in the effective diffusion constant \eqref{critModeDiffConst} of the critical mode, for example, we find
\begin{align}
    D_{\text{eff},\text{IR}}(\va{p}) &= D_0 \left(1-\frac{\chi_0 B^2}{m_{s,\text{IR}}^2 + Z_{\text{IR}}^\perp \va{p}^2 + \chi_0 B^2} \right) = \frac{D_0}{\chi_0 B^2} \left( m_{s,\text{IR}}^2 + Z_{\text{IR}}^\perp \va{p}^2 \right) + \cdots \,. \label{diffConstEffIR}
\end{align}
The critical part of the retarded propagator in the IR limit ($k\to 0$) reads (cf.~Eq.~\eqref{GRetModBExpanded}),
\begin{align}
    G_{\text{IR}}^R(\omega,\va{p}) = \frac{1}{m_{p,\text{IR}}^2} \frac{\omega + i D_0 \va{p}^2}{\omega+iD_{\text{eff},\text{IR}}(\va{p}) \va{p}^2} + \cdots
\end{align}
which immediately translates into the corresponding critical part of the spectral function,
\begin{align}
    \rho_{\text{IR}}(\omega,\va{p}) = \frac{1}{\pi m_{p,\text{IR}}^2} \frac{\omega(D_0-D_{\text{eff},\text{IR}}(\va{p})) \va{p}^2}{\omega^2+\left(D_{\text{eff},\text{IR}}(\va{p}) \va{p}^2\right)^2} + \cdots \,,
\end{align}
where we can directly see that it exhibits a maximum at $\omega = D_{\text{eff},\text{IR}}(\va{p})\va{p}^2$.
From \eqref{diffConstEffIR}, on the other hand, we obtain for  $|\va{p}| \gg \xi^{-1}$  in the critical regime, 
\begin{align}
    D_{\text{eff},\text{IR}}(\va{p}) = \frac{D_0 Z^\perp_{\text{IR}}}{\chi_0 B^2} \va{p}^2 \,,
\end{align}
and therefore the dispersion relation $\omega \sim |\va{p}|^4$ for the critical diffusive mode, yielding the mean-field exponent $z=4$. Such a mean-field scaling in the IR limit ($k\to 0$) is expected for a truncation that relies on an expansion in spatial gradients where the propagator never becomes a non-analytic function of $\va{p}$. A factor $\sim |\va p|^{-\eta^\perp}$
from the momentum dependence of the spatial wave-function renormalization
$Z^\perp_\text{IR}$ is obviously missing in our truncation. This is why the correct dynamic critical exponent can be observed only in the FRG scale $k$ dependence here, for which we only need the scaling $Z_k^\perp \sim k^{-\eta^\perp} $ at the critical point. 

In the opposite limit $|\va{p}| \ll \xi^{-1}$, e.g.~in any finite volume, we have $D_{\text{eff},\text{IR}}(\va{p}) = \text{const.}$, and as such the usual diffusive dispersion relation $\omega \sim \va{p}^2$, as expected. Moreover, precisely at criticality ($\xi^{-1} = 0$), for $|\va p| \ll \omega$, the critical spectral function behaves as  $\rho(\omega,\va p) \sim \va{p}^2/\omega $, while for 
$|\va p| \gg \omega$ we obtain $\rho(\omega,\va p) \sim \omega \va{p}^{-2(z-1)}  $ (here with our mean-field $z$). In particular, at any finite $\omega$, for example,  it first increases with spatial momentum (as long as $|\va p| \ll \omega $) and eventually decreases again (when $|\va p| \gg \omega $), with the maximum in the transition region scaling as  %$\omega \propto |\va p|^z $.
$|\va p | \propto \omega^{1/z} $.
All these properties of the critical spectral function can be derived quite generally from the underlying dynamic scaling functions as discussed in detail for Model B in \cite{Schweitzer:2021iqk}.

\ifModC
\subsection{Model C---Non-linear coupling of a conserved (energy) density}
\label{sct:ModC}

As mentioned in our Introduction, it was historically first argued by Berdnikov and Rajagopal in Ref.~\cite{Berdnikov:1999ph} that the dynamic universality class of the critical endpoint in the phase diagram of QCD should be that of Model~C, when they analyzed critical slowing down and off-equilibrium phenomena in heavy-ion collisions and employed the value $z \approx 2.17$ for the dynamic critical exponent, reflecting the underlying assumption of critical Model-C dynamics.
They derived the latter by exploiting the fact that $z=2+\alpha/\nu$ in the case of Model~C is fully determined by static exponents \cite{RevModPhys.49.435}, and inserted $\alpha \approx 0.11$ and $\nu \approx 0.630$ which were both known for the $3d$ Ising model \cite{Guida:1998bx}. 
The question on the dynamic universality class of the CEP was later revisited by Son and Stephanov in Ref.~\cite{Son:2004iv} where they argued in favor of Model~H, i.e.~that of the liquid-gas transition in a pure fluid, and predicted $z \approx 3$, which is the accepted theory to date.
In the context of the dynamic renormalization group and the $\varepsilon$-expansion, an analysis of Model~C was later performed e.g.~in Ref.~\cite{Nakano:2011re} and a study in the context of the FRG was performed e.g.~in Ref.~\cite{Mesterhazy:2013naa}. For a more general introduction, see especially also Ref.~\cite{tauber}.

In comparison with Model~B from Sec.~\ref{sct:ModB} above, the hydrodynamic equations of motion \eqref{hydroBPhi}, \eqref{hydroBn} stay the same, but the coupling between $\varphi$ and $n$ in the free energy changes,
\begin{align}
    B \, \varphi \, n \to \frac{g}{2} \, \varphi^2 \, n \, ,
\end{align}
i.e.~there is no more mixing between $\varphi$ and $n$, such that the Landau-Ginzburg-Wilson free-energy functional now reads\footnote{We have not included a gradient term $\sim (\vec{\nabla} n)^2$ since it is irrelevant in the renormalization group sense~\cite{Nakano:2011re}.}
\begin{equation}
    F = \int d^d x \left\{ \frac{1}{2} (\vec{\nabla}\varphi)^2 + \frac{m^2}{2} \varphi^2 + \frac{\lambda}{4!} \varphi^4 + \frac{g}{2} \varphi^2 n + \frac{n^2}{2\chi_0}   \right\} \, . \label{freeEnergyC}
\end{equation}
Notably, a mixing between $\varphi$ and $n$ is forbidden in Model~C due to the reflection symmetry $\varphi \to -\varphi$ of the order parameter, which excludes linear coupling terms of the form $ \sim \varphi n $ and $ \sim (\vec{\nabla}\varphi)\cdot \vec{\nabla} n $ \cite{Son:2004iv}.
In contrast, when such linear coupling terms appear in the free energy, as in Eq.~\eqref{freeEnB} above, they fundamentally change the infrared structure of the theory, and subsequently the dynamic universality class to that of Model~B, as we have seen explicitly in Sec.~\ref{sct:ModB}.
Due to the non-linear coupling between $\varphi^2$ and $n$, on the other hand,  the fluctuations of the conserved (energy) density are expected to diverge as $\langle nn \rangle \sim k^{-\alpha/\nu}$ at the critical point, when $\alpha > 0$~\cite{Mesterhazy:2013naa}, where $\alpha$ denotes the critical exponent of the specific heat.
We will indeed find that the susceptibility  then becomes 
FRG-scale $k$ dependent, $\chi_0\to \chi_{0,k}$, and diverges at the critical point with an exponent $x_{\chi_0} = \alpha/\nu$. This divergence thus is directly related to the specific-heat exponent $\alpha \geq 0$, which is positive with Ising universality in $d=2,3$ spatial dimensions.
From the hyperscaling relation $z=2+\alpha/\nu$, we then conclude that the dynamic critical behavior of Model~C is mainly driven by the static critical behavior of the susceptibility $\chi_{0,k}$ of $n$.

The bare MSR action for the classical-statistical equations of motion coincides with Eq.~\eqref{generalMSRAction}, with the difference here of course being that the free energy \eqref{freeEnergyC} of Model~C is used.
For a first overview of the real-time structure of the theory, we explicitly insert the free energy \eqref{freeEnergyC} into the general MSR action \eqref{generalMSRAction}, which then becomes
\begin{align}
    \label{modelCKeldyshAction}
    S &= S_{A} + \int_x \bigg[ -  n^q \left( \tau_R \partial_t^2 + \partial_t - \frac{\lambda_n}{\chi_0} \vec{\nabla}^2 \right) n^c  \\ \nonumber 
    &\hspace{+0.5cm} -\frac{g}{\sqrt{2}}\,\phi^q \phi^c n^c + \frac{g \lambda_n }{\sqrt{8}} \phi^c \phi^c \vec{\nabla}^2 n^q - 2i \lambda_n T n^q\vec{\nabla}^2 n^q \bigg] \, ,
\end{align}
where $S_A$ denotes the MSR action of Model~A in Eq.~\eqref{bare-action}.
Our new MSR action \eqref{modelCKeldyshAction} for Model~C contains two new 3-point vertices: $\phi^c \phi^q n^c$ and $\phi^c \phi^c n^q$.
The latter also contains a factor of $\va{p}^2$, reflecting the diffusive dynamics of the $n$-field.
One of the prominent qualitative features that comes with the diffusive structure of the vertices is that certain $\omega \to 0$ and $\va{p} \to 0$ limits no longer commute.
%We encountered this characteristic feature already in Sec.~\ref{sct:ModB} in the context of  Model~B and the distinction between the static and the plasmon mass.
For the diffusive propagators of the conserved density,  for example, we have
\begin{subequations}
\begin{align}
    \lim_{\va{p} \to 0} \lim_{\omega \to 0} \frac{g^2 \lambda_n  \va{p}^2}{\tau_R \omega^2 + i\omega - D_0 \va{p}^2} &=
    -\frac{g^2 \lambda_n}{D_0} = -g^2 \chi_0 \, , \label{diffVertStaticLimit} \\
    \lim_{\omega \to 0} \lim_{\va{p} \to 0} \frac{g^2 \lambda_n  \va{p}^2}{\tau_R \omega^2 + i\omega - D_0 \va{p}^2} &= 0 \, , \label{diffVertPlasmonLimit}
\end{align}
\end{subequations}
where we used the definition of the diffusion constant $D_0 = \lambda_n/\chi_0$ in the first line.
The first order  of limits  in \eqref{diffVertStaticLimit} corresponds to the static case. It entails that the interaction with the conserved density effectively induces a shift in the 4-point coupling according to
\begin{align}
    \lambda \to \lambda -3g^2 \chi_0 \, . \label{quartCouplEffShift}
\end{align}
We will therefore introduce the FRG scales~$k$ dependent `static' coupling constant defined by this shifted coupling, $ \lambda_{s,k} \equiv \lambda_k - 3g_k^2 \chi_{0,k} $.
The second  order of limits in  \eqref{diffVertPlasmonLimit}  corresponds to the plasmon case.

\subsubsection{Truncation}

In order to construct a systematic truncation scheme, we expand the effective average action around vanishing field expectation values according to the 2-loop expansion scheme explained in Sec.~\ref{sct:ExpAroundIRMin} above.  We thereby allow all additional couplings to the conserved (energy) density to be scale dependent, and assume an arbitrary 2-point function $\Gamma_k^{(2)}(\omega,\va{p})$ to be able to capture an infrared power-law behavior of the spectral function at criticality.
With these steps, our ansatz for the effective average action reads
\begin{align}
    \Gamma_k &= 
    \frac{1}{2} \int_{xx'} \hspace{-0.2cm} \left( \phi^c(x), \phi^q(x) \right)
    \begin{pmatrix}
        0 & \Gamma_k^{cq}(x-x') \\
        \Gamma_k^{qc}(x-x') & \Gamma_k^{qq}(x-x')
    \end{pmatrix}
    \begin{pmatrix}
        \phi^c(x') \\
        \phi^q(x')
    \end{pmatrix} + \nonumber \\
    &\hspace{0.5cm} \frac{1}{2} \int_{p} (\phi^q \phi^c)_{-p} \left( -\frac{\lambda_k}{6} + V_{1,k}^{cl,R}(0) \, \va{p}^2 \right) (\phi^c \phi^c)_{p} + \nonumber \\
    &\hspace{0.5cm} \frac{1}{2}  \int_{xx'} \hspace{-0.2cm} \left( n^c(x), n^q(x) \right)
    \begin{pmatrix}
        0 & -\tau_{R}\partial_t^2 \hspace{-0.08cm} + \hspace{-0.08cm} \partial_t \hspace{-0.08cm} + \hspace{-0.08cm} D_{0,k} \vec{\nabla}^2 \\
        -\tau_{R}\partial_t^2 \hspace{-0.08cm} - \hspace{-0.08cm} \partial_t \hspace{-0.08cm} + \hspace{-0.08cm} D_{0,k} \vec{\nabla}^2 & -4i\lambda_n T \vec{\nabla}^2
    \end{pmatrix}
    \begin{pmatrix}
        n^c(x') \\
        n^q(x')
    \end{pmatrix} +  \nonumber \\ 
    &\hspace{0.5cm}  \int_x \bigg[ -\frac{g_k}{\sqrt{2}}\,\phi^q \phi^c n^c + \frac{g_k \lambda_n }{\sqrt{8}} \phi^c \phi^c \vec{\nabla}^2 n^q \bigg] \, . 
    \label{effAvgActionModC}
\end{align}
Apart from the static first-order momentum correction $V_{1,k}^{cl,R}(0)\va{p}^2$ to the 4-point vertex, which is needed in order to have a non-vanishing anomalous dimension~$\eta^\perp$, we have thereby neglected 
all frequency dependent non-local corrections to the 4-point vertices of our symmetric vertex expansion in Eq.~\eqref{vertexExpSymm}. In particular, compared to the expansion in \eqref{eq:4vertex} the frequency-dependence of the order zero contribution 
$\Delta V_{0,k}^{cl,R}(\omega)$ is neglected here as well,
because the dominant sources of the infrared power law  in the critical  spectral function, from the flow of the 2-point function, now are the non-local one-loop diagrams containing two 3-point $\varphi^2 n$ vertices anyway, see Fig.~\ref{fig:gam2FlowModC}. Hence an additional frequency dependence of the 4-point vertex function is no-longer needed to see the critical dynamics, here.

\begin{figure}[t]
    \centering
    \begin{align*}
        \partial_k \Gamma_k^{qc}(x,x') &= -i
    \Bigg\{\;
	\begin{tikzpicture}[baseline=-0.5ex]
	 	\centerarc[pblue](0,0)(0:45:0.5)
	 	\centerarc[pblue](0,0)(45:90:0.5)
	 	\centerarc[pblue](0,0)(90:135:0.5)
	 	\centerarc[pblue](0,0)(135:180:0.5)
	 	\centerarc[pblue,dashed](0,0)(180:270:0.5)
	 	\centerarc[pred,dashed](0,0)(270:360:0.5)
		% legs
	 	\draw[pred] (-0.5,0) -- (-0.5-0.5,0) node[anchor=north,black] {\LegFontSize $x$};
	 	\draw[pblue] (0.5,0) -- (0.5+0.5,0) node[anchor=north,black] {\LegFontSize $x'$};
	 	% vertices
	 	\fill[black] (-0.5,0) circle (0.05);
	 	\fill[black] (+0.5,0) circle (0.05);
		% regulator insertion
	 	\fill[black] (0-0.1,0.5-0.1) rectangle ++(0.2,0.2);
	 	% labels
	 	\node[black] at (-0.8,+0.3) {\LegFontSize $\phi^q$};
	 	\node[black] at (+0.8,+0.3) {\LegFontSize $\phi^c$};
	 	\node[black] at (0,+0.2) {\LegFontSize $\phi$};
	 	\node[black] at (0,-0.3) {\LegFontSize $n$};
	\end{tikzpicture}
	+
	\begin{tikzpicture}[baseline=-0.5ex]
	 	\centerarc[pred](0,0)(0:45:0.5)
	 	\centerarc[pred](0,0)(45:90:0.5)
	 	\centerarc[pblue](0,0)(90:135:0.5)
	 	\centerarc[pblue](0,0)(135:180:0.5)
	 	\centerarc[pblue,dashed](0,0)(180:360:0.5)
		% legs
	 	\draw[pred] (-0.5,0) -- (-0.5-0.5,0) node[anchor=north,black] {\LegFontSize $x$};
	 	\draw[pblue] (0.5,0) -- (0.5+0.5,0) node[anchor=north,black] {\LegFontSize $x'$};
	 	% vertices
	 	\fill[black] (-0.5,0) circle (0.05);
	 	\fill[black] (+0.5,0) circle (0.05);
		% regulator insertion
	 	\fill[black] (0-0.1,0.5-0.1) rectangle ++(0.2,0.2);
	 	% labels
	 	\node[black] at (-0.8,+0.3) {\LegFontSize $\phi^q$};
	 	\node[black] at (+0.8,+0.3) {\LegFontSize $\phi^c$};
	 	\node[black] at (0,+0.2) {\LegFontSize $\phi$};
	 	\node[black] at (0,-0.3) {\LegFontSize $n$};
	\end{tikzpicture}
	+
    \frac{1}{2}\;
	\begin{tikzpicture}[baseline=-0.5ex]
	 	\centerarc[pblue](0,0)(0:45:0.5)
	 	\centerarc[pblue](0,0)(45:90:0.5)
	 	\centerarc[pblue](0,0)(90:135:0.5)
	 	\centerarc[pblue](0,0)(135:180:0.5)
	 	\centerarc[pblue](0,0)(180:270:0.5)
	 	\centerarc[pblue](0,0)(270:360:0.5)
		% legs
	 	\draw[pred] (0,-0.5) -- (-0.5,-0.5) node[anchor=north,black] {\LegFontSize $x$};
	 	\draw[pblue] (0,-0.5) -- (+0.5,-0.5) node[anchor=north,black] {\LegFontSize $x'$};
	 	% vertices
	 	\fill[black] (0,-0.5) circle (0.1);
		% regulator insertion
	 	\fill[black] (0-0.1,0.5-0.1) rectangle ++(0.2,0.2);
	 	% labels
	 	\node[black] at (-0.67,-0.4) {\LegFontSize $\phi^q$};
	 	\node[black] at (+0.67,-0.4) {\LegFontSize $\phi^c$};
	 	\node[black] at (0,+0.2) {\LegFontSize $\phi$};
	\end{tikzpicture}
    \;\Bigg\}
	\end{align*}
    \caption{Diagrammatic representation of the flow of the retarded 2-point function in Model~C around the symmetric expansion point $\phi = n = 0$. Compared to the corresponding flow in Model~A, shown in Fig.~\ref{fig:gam2FlowSymm}, there are two additional one-loop diagrams due to the non-vanishing $\phi\phi n$ vertex, which generate the critical power-law behavior in the spectral function. In contrast, the third (tadpole) diagram is assumed to be local in frequency here as compared to the first row of Fig.~\ref{fig:gam2FlowSymm}. This is sufficient to obtain the correct dynamic critical behavior in the spectral function, precisely
    because of the additional one-loop diagrams that involve interactions with the conserved energy density and hence dominate the critical dynamics of Model C.
   \label{fig:gam2FlowModC}}
\end{figure}

Because in the static limit ($\omega \to 0$) the only effect of the 
interaction with the conserved (energy) density, which can  be trivially integrated out by Gaussian integration, is the shift in the quartic coupling, cf.~\eqref{quartCouplEffShift}, with the replacement $ \lambda_{k} \to  \lambda_{s,k} = \lambda_k -3g_k^2 \chi_{0,k}$, the flows of the static quantities $m_k^2$, $\lambda_{s,k}$, and $Z_k^\perp$ are otherwise identical to those of Model~A, as given in Eqs.~\eqref{m2FlowSymmFinalResult}, \eqref{spatAnomDimSymm} and \eqref{flowLambdaSymm}.  

The essential new quantities, compared to Model A, are the susceptibility $\chi_{0,k}$ and the static coupling $g_k$ which can be obtained from  the following functional derivatives of the effective average action,
\begin{align}
    -\frac{\lambda_n}{\chi_{0,k}} \delta^{D}(0) &= \lim_{\va{p}\to 0} \frac{1}{\va{p}^2} \frac{\delta^2 \Gamma_k}{\delta n^q(0,-\va{p}) \delta n^c(0,\va{p})} \, , \\
    -\frac{g_k}{\sqrt{2}} \delta^{D}(0) &= \frac{\delta^3 \Gamma_k}{\delta \phi^q(p) \delta \phi^c(p) \delta n^c(p)} \bigg\rvert_{p=0} \, ,
\end{align}
where the infinite spacetime volume factor (in equilibrium $\delta^{D}(0)\propto V/T $) originates from overall momentum conservation and cancels on both sides.
Projecting the Wetterich equation \eqref{wetterichEq} onto the flow of  $g_k$  and $\chi_{0,k}$ gives rise to the diagrams shown in Fig.~\ref{fig:gAndChi0EuclFlow}. In the flow of $\partial_k \chi_{0,k}^{-1}$ we note that the $1/\va{p}^2$ prefactor is precisely canceled by a factor  $\va{p}^2$ from the one diffusive $\phi^c \phi^c n^q$ vertex involved in the diagram.
In practice, we evaluate these diagrams by inserting the FDR between the propagators,
%and primarily obtain
%\begin{align}
%    \partial_k g_k &= -2ig_k^3 \lambda T \int_q  \frac{\va{q}^2}{q^0} \Big[ B_k^R(q) G_k^R(q) D_k^R(q) - \\ \nonumber
%    &\hspace{1cm} B_k^A(q) G_k^A(q) D_k^A(q) \Big] + \frac{i\lambda_{k} g_k}{2} J_k^{KR}(0,\VecZero) \, ,\\
%    \partial_k \chi_{0,k}^{-1} &= \frac{i g_k^2}{2} J_k^{KR}(0,\va{0}) \, ,
%\end{align}
solve the remaining $q^0$-integral with Kramers-Kronig relations, and use the convenient form of the frequency-independent optimized regulator in \eqref{eq:litimReg} for the $\va{q}$-integral, such that we finally arrive at
\begin{align}
	\partial_k g_k &= \frac{\Omega_d k^{d+1} T Z_k^\perp}{(2\pi)^d} \left( 1 - \frac{\eta_k^\perp}{2+d} \right)  \frac{ 2 g_k \left( \chi_{0,k} g_k^2 + \lambda_{s,k} \right) }{ (m_k^2 + Z_k^\perp k^2)^3 }  \, , \label{gFlowEq} \\
    \partial_k \chi_{0,k}^{-1} &= \frac{ \Omega_d k^{d+1} T  Z_k^\perp}{(2\pi)^d} \left( 1 - \frac{\eta_k^\perp}{2+d} \right) \frac{ 2 g_k^2 }{ (m_k^2 + Z_k^\perp k^2)^3 } \, . \label{chi0InvFlowEq}
\end{align}
As before, we have also applied our local-vertex approximation from Sec.~\ref{sct:ExpAroundIRMin} (where we introduced  it for  the flow of the quartic coupling $\lambda_k$) 
again in the flow of the 3-point coupling constant $g_k$. Here, this simply implies that we also neglect the terms involving $V_{1,k}^{cl,R/A}(0)$ in the  flow equation for $g_k$, which is then consistent with the approximation used for the flow of $\lambda_k$.

Last but not least, we have tacitly assumed that any possible  dependence of  $\lambda_n$ and $\tau_R$  on the FRG scale $k$ and hence their flow can be neglected. 
While this might seem to be a rather crude approximation at first, it nevertheless well motivated as long as we are primarily interested in the dynamical critical behavior: 
($i$) the finite relaxation time $\tau_R$ only serves to prevent superluminal signals at \emph{very large} spatial momenta and plays no role for long wavelength infrared modes, and ($ii$) the dynamic critical exponent $z=2+\alpha/\nu$ of Model C is fully determined by static quantities, so we do not expect the real-time kinetic coefficient $\lambda_{n,k}$ (which only enters through the hydrodynamic equations of motion \eqref{hydroBn}) to have an impact on the critical dynamics, in this particular case.
This will of course not generally be true in other Models, such as Model~H for example, where the flow of $\lambda_n$ has to be taken into account \cite{Son:2004iv}.

\begin{figure}[t]
	\centering
	\begin{align*}
		\partial_k g_k &= i\sqrt{2}\Bigg\{
		\begin{tikzpicture}[baseline=-0.5ex]
	        \centerarc[->,pgreen](0,0)(0:45:0.5)
	        \centerarc[pgreen](0,0)(45:90:0.5)
	        \centerarc[->,pgreen](0,0)(90:150:0.5)
	        \centerarc[pgreen](0,0)(150:180:0.5)
	        \centerarc[->,pgreen](0,0)(180:225:0.5)
	        \centerarc[pgreen](0,0)(225:275:0.5)
	        \centerarc[->,pgreen,dashed](0,0)(275:325:0.5)
	        \centerarc[pgreen,dashed](0,0)(325:360:0.5)
			% legs
		 	\draw[dashed,pblue] (-0.5,0) -- (-1.0,0);
		 	\draw[pblue] (0.5,0) -- (1.0,0);
		 	\draw[pred] (0,-0.5) -- (0,-1.0);
		 	% vertices
		 	\fill[black] (-0.5,0) circle (0.05);
		 	\fill[black] (+0.5,0) circle (0.05);
		 	\fill[black] (0,-0.5) circle (0.05);
			% regulator insertion
	 	    \fill[black] (0-0.1,0.5-0.1) rectangle ++(0.2,0.2);
	 	    % labels
	 	    \node[black] at (0.0,0.75) {\LegFontSize $q$};
	 	    \node[black] at (-0.54,-0.54) {\LegFontSize $q$};
	 	    \node[black] at (0.54,-0.54) {\LegFontSize $q$};
		\end{tikzpicture}+
		\begin{tikzpicture}[baseline=-0.5ex]
	        \centerarc[->,pgreen](0,0)(0:45:0.5)
	        \centerarc[pgreen](0,0)(45:90:0.5)
	        \centerarc[->,pgreen](0,0)(90:150:0.5)
	        \centerarc[pgreen](0,0)(150:180:0.5)
	        \centerarc[->,pgreen,dashed](0,0)(180:225:0.5)
	        \centerarc[pgreen,dashed](0,0)(225:275:0.5)
	        \centerarc[->,pgreen](0,0)(275:325:0.5)
	        \centerarc[pgreen](0,0)(325:360:0.5)
			% legs
		 	\draw[pred] (-0.5,0) -- (-1.0,0);
		 	\draw[dashed,pblue] (0.5,0) -- (1.0,0);
		 	\draw[pblue] (0,-0.5) -- (0,-1.0);
		 	% vertices
		 	\fill[black] (-0.5,0) circle (0.05);
		 	\fill[black] (+0.5,0) circle (0.05);
		 	\fill[black] (0,-0.5) circle (0.05);
			% regulator insertion
	 	    \fill[black] (0-0.1,0.5-0.1) rectangle ++(0.2,0.2);
	 	    % labels
	 	    \node[black] at (0.0,0.75) {\LegFontSize $q$};
	 	    \node[black] at (-0.54,-0.54) {\LegFontSize $q$};
	 	    \node[black] at (0.54,-0.54) {\LegFontSize $q$};
		\end{tikzpicture}+
		\begin{tikzpicture}[baseline=-0.5ex]
	        \centerarc[->,pgreen](0,0)(0:45:0.5)
	        \centerarc[pgreen](0,0)(45:90:0.5)
	        \centerarc[->,pgreen](0,0)(90:150:0.5)
	        \centerarc[pgreen](0,0)(150:180:0.5)
	        \centerarc[->,pgreen](0,0)(180:275:0.5)
	        \centerarc[pgreen](0,0)(275:360:0.5)
			% legs
		 	\draw[pblue] (-0.5,0) -- (-0.5-0.3,0.3);
		 	\draw[pred] (-0.5,0) -- (-0.5-0.3,-0.3);
		 	\draw[pblue,dashed] (0.5,0) -- (1.0,0);
		 	% vertices
		 	\fill[black] (-0.5,0) circle (0.05);
		 	\fill[black] (+0.5,0) circle (0.05);
			% regulator insertion
	 	    \fill[black] (0-0.1,0.5-0.1) rectangle ++(0.2,0.2);
	 	    % labels
	 	    \node[black] at (0.0,0.75) {\LegFontSize $q$};
	 	    \node[black] at (0.0,-0.75) {\LegFontSize $q$};
		\end{tikzpicture} \Bigg\} \,,\hspace{0.1cm}
		\partial_k \chi_{0,k}^{-1} = \frac{i}{\lambda_n} \lim_{\va{p}\to 0} \frac{1}{\va{p}^2}\,
		\begin{tikzpicture}[baseline=-0.5ex]
	        \centerarc[->,pgreen](0,0)(0:45:0.5)
	        \centerarc[pgreen](0,0)(45:90:0.5)
	        \centerarc[->,pgreen](0,0)(90:150:0.5)
	        \centerarc[pgreen](0,0)(150:180:0.5)
	        \centerarc[->,pgreen](0,0)(180:275:0.5)
	        \centerarc[pgreen](0,0)(275:360:0.5)
			% legs
		 	\draw[pred,dashed] (-0.5,0) -- (-1.0,0);
		 	\draw[pblue,dashed] (0.5,0) -- (1.0,0);
		 	% vertices
		 	\fill[black] (-0.5,0) circle (0.05);
		 	\fill[black] (+0.5,0) circle (0.05);
			% regulator insertion
	 	    \fill[black] (0-0.1,0.5-0.1) rectangle ++(0.2,0.2);
	 	    % labels
	 	    \node[black] at (0.0,0.75) {\LegFontSize $q+\va{p}$};
	 	    \node[black] at (0.0,-0.75) {\LegFontSize $q$};
	 	    \node[black] at (0.8,0.2) {\LegFontSize $\va{p}$};
	 	    \node[black] at (-0.8,0.2) {\LegFontSize $-\va{p}$};
		\end{tikzpicture}
	\end{align*}
	\caption{Flow of the static quantities $g_k$ and $\chi_{0,k}^{-1}$ in the local-vertex approximation. Arrows indicate the loop-momentum routing. The limit $\va{p} \to 0$ in the equation for  $\partial_k \chi_{0,k}^{-1}$ is well-defined because of the diffusive nature of the $n$ field fluctuations: all diagrams involve one diffusive $\phi^c \phi^c n^q$ vertex proportional to $\va{p}^2$.}
	\label{fig:gAndChi0EuclFlow}
\end{figure}

\fi

\section{Results}
\label{sct:results}

The initial values of the various parameters at the UV scale $k=\Lambda $  used in our FRG flows are listed in Table~\ref{tab:uvParamsFlow}, in units of the appropriate powers of $\Lambda$.
By tuning the temperature $T$ close to its critical value $T_{c}$, while keeping all other initial values fixed, we observe the second-order phase transition where the system restores its spontaneously broken $Z_2$ symmetry. We therefore start with discussing the static critical behavior at the observed second-order phase transition with $Z_2$ universality.  
The most important aspect for our purposes will thereby be  the flow of the  anomalous spatial scaling dimension  $\eta_k^\perp \equiv -k\partial_k \log Z_k^\perp$.

\subsection{Statics}
\label{sct:staticsResults}

The power-law behavior of the static correlator  of the order-parameter field at criticality,
$ \langle \varphi(\va{x}) \varphi(\va{y}) \rangle \sim 1/|\va{x}-\va{y}|^{d-2+\eta^\perp} $, immediately entails that the corresponding two-point function cannot be analytic at $\va p^2 = 0$ in momentum space. 
On the other hand, such a non-analyticity will not be seen at any finite order in an expansion in terms of  spatial derivatives, where the critical propagator \emph{never} becomes a non-analytic function of $\va{p}^2$ in  the IR ($k \to 0$).
We have mentioned this already in the discussion of the 
critical spectral function of Model~B in Sec.~\ref{sct:ModBCritModes}, and it will manifest itself explicitly in the Model-B results in Sec.~\ref{sct:critSFs} below. In particular, to extract 
the corresponding critical exponent $\eta^\perp$,  we therefore follow  the proposition by Berges et al.~in Ref.~\cite{Berges:2000ew} that one can  identify a non-vanishing anomalous scaling dimension with the logarithmic $k$-derivative of the spatial wave function renormalization factor $\eta_k^\perp \equiv -k\partial_k \log Z_k^\perp$, even in an expansion in spatial gradients.
This argument was also used in Refs.~\cite{Canet:2002gs,Canet:2003qd} where it was  moreover confirmed that higher orders, including  $\Ord{\vec{\nabla}^4}$, of the derivative  expansion indeed converge rather quickly towards  exactly or more precisely  known values  from the  literature.

\begin{figure*}[t]
    \centering
    \begin{minipage}{.49\textwidth}
        \centering
        {\footnotesize (a) $d = 2$}
        \includegraphics[width=\textwidth]{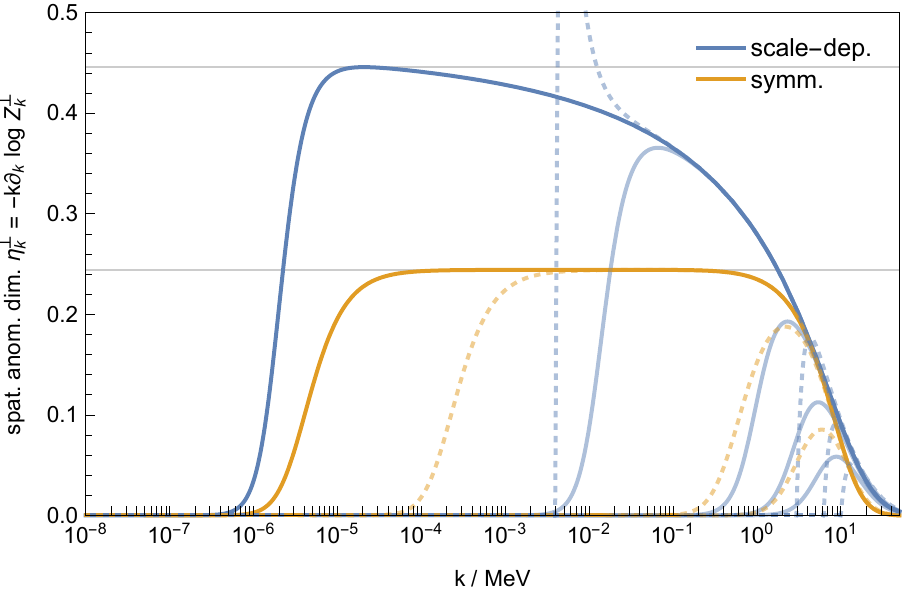}
    \end{minipage}
    \begin{minipage}{.49\textwidth}
        \centering
        {\footnotesize (b) $d = 3$}
        \includegraphics[width=\textwidth]{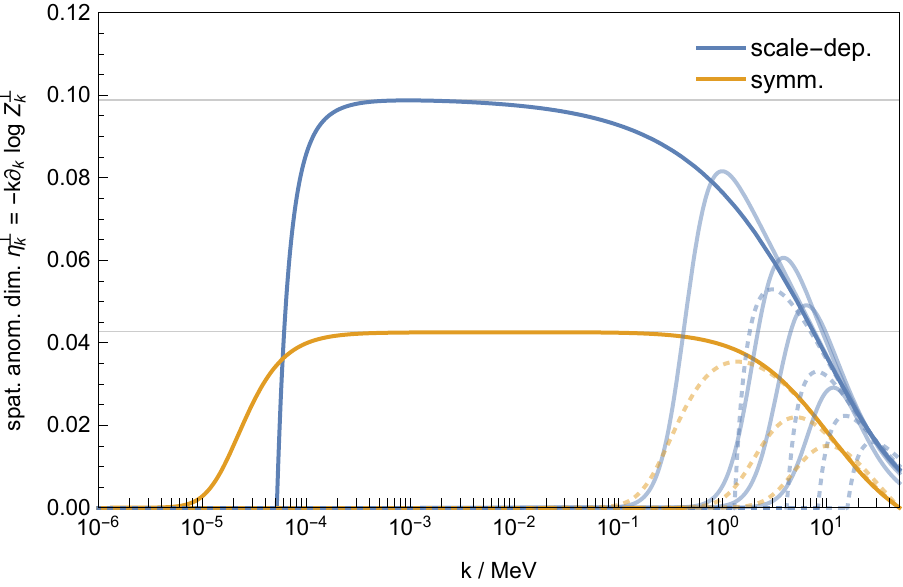}
    \end{minipage}
    \caption{Flow of the  anomalous spatial scaling dimension  $\eta_k^\perp \equiv -k\partial_k \log Z_k^\perp$ for various temperatures around criticality, (a) left in  $d=2$ and (b) right in $d=3$ spatial dimensions. The thick solid lines are closest to the critical temperature $T_c$. Thin solid lines are used for $T<T_c$, and dotted lines for  $T> T_c$. The results from  the comoving  expansion of Sec.~\ref{sct:ExpAroundScaleDepMin} (blue) are labeled `scale-dep.' and the ones from that of  Sec.~\ref{sct:ExpAroundIRMin} around the symmetric IR minimum (yellow) are labeled `symm.' (available only for $T\ge T_c$). 
    (a) In $d=2$ spatial dimensions, the symmetric expansion scheme produces a stable plateau approaching a fixed-point value of $\eta^\perp = 0.244$. The situation is less clear in  the comoving expansion, where no clear fixed point is reached yet, and we hence use the estimate $\eta_k^\perp = 0.446$ from where the flow $\partial_k \eta_k^\perp = 0$ is stationary. (b) In $d=3$ spatial dimensions, both expansion schemes exhibit rather clear plateaus, yielding  $ \eta^\perp = 0.0988 $ (scale-dep.) and $\eta^\perp = 0.0427$ (symm.), respectively. See text for details. \label{fig:etaPerp}}
\end{figure*}

Our results for the flow of the  anomalous spatial scaling dimension  $\eta_k^\perp \equiv -k\partial_k \log Z_k^\perp$ are plotted for various temperatures near criticality in  $d=2$ (left) and $ 3$ (right) spatial dimensions in Fig.~\ref{fig:etaPerp}. For comparison and test of the systematics we have included both results, from the  two different expansion schemes in Secs.~\ref{sct:ExpAroundScaleDepMin} and \ref{sct:ExpAroundIRMin}, the comoving expansion around the scale-dependent minimum and that around the symmetric IR minimum at $\phi=0$ above $T_c$.

\begin{table*}[t]
    \centering
    \scriptsize
    \begin{tabular}[t]{ |c|c|c|c|c|c|c|c|c|c||c| } \hline
              & \shortstack{$m^2_\Lambda$\\$[\Lambda^2]$} & \shortstack{$\lambda_\Lambda$\\$[\Lambda^{3-d}]$} & \shortstack{$Z_\Lambda^\perp$\\$\phantom{[1]}$} & \shortstack{$\gamma_\Lambda$\\$[\Lambda]$} & \shortstack{$\lambda_n$\\$[\Lambda^{d-2}]$} & \shortstack{$\chi_{0,\Lambda}$\\$[\Lambda^{d-1}]$} & \shortstack{$B$\\$[\Lambda^{\frac{3-d}{2}}]$} & \shortstack{$g_\Lambda$\\$[\Lambda^{2-d}]$} & \shortstack{$\tau_{R}$\\$[\Lambda^{-1}]$} & \shortstack{$T_c$\\$[\si{\MeV}]$} \\ \hline
        $d=2$ & $-0.04$ & $0.1$ & $1$ & $0.2$, $0.4^c$ & 1 & 1 & $-$ & 1 & 1 & $141.2^a, 147.9^b$ \\
        $d=3$ & $-0.04$ & $1.0$ & $1$ & $0.2$ & 1 & 1 & 1 & 1 & 1 & $155.3^a, 155.7^b$ \\ \hline
    \end{tabular}
	\caption{Initial conditions used at the UV initial scale $k=\Lambda$ for $d=2$ and 3 spatial dimensions. 
 Arbitrarily assuming $\Lambda = \SI{50}{\MeV}$, the resulting critical temperatures $T_c$ in MeV in the two different truncation schemes (indicated by superscripts) are shown in the last column: ($a$)  from the comoving expansion of Sec.~\ref{sct:ExpAroundScaleDepMin}, and ($b$) from the expansion around the IR minimum in the symmetric phase of Sec.~\ref{sct:ExpAroundIRMin}.
 %We measure the order parameter $\varphi$ in units of $\si{\mega\eV}^{(d-1)/2}$, and the conserved density $n$ in units of $\si{\mega\eV}^d$. 
 The initial value for the mass in Model~B refers to the static screening mass, $m_{s,\Lambda}^2$,  that for the quartic coupling in Model~C to the statically shifted one, $\lambda_{s,\Lambda}$. Moreover, the superscript  $(c)$ refers to the damping constant of Model~C in $d=2$ spatial dimensions for which we use a somewhat larger value  than in the other cases, in order to broaden an otherwise sharp quasi-particle peak that would require an unnecessarily fine $\omega$-grid.}
	\label{tab:uvParamsFlow}
\end{table*}

 Starting in three spatial dimensions ($d=3$), as shown in Fig.~\ref{fig:etaPerp}~(b) on the right, where both expansion schemes show pronounced plateau structures, we obtain two stable but distinct values for the anomalous scaling dimension, as listed in Table~\ref{crit-exp-results} below.
 At least qualitatively, we can therefore conclude that the spatial wave function renormalization factor $Z_k^\perp$ indeed assumes a stable infrared power-law behavior in the scaling regime close to the critical point.
Quantitatively, the value $\eta^\perp = 0.0988$ for the anomalous scaling dimension in the comoving expansion is still almost about a factor of $3$ off compared to the high precision result of $\eta^\perp = 0.036298$ obtained for three spatial dimensions from the conformal bootstrap approach~\cite{Kos:2016ysd,Komargodski:2016auf}.
We attribute this discrepancy predominantly to the missing two-loop structures, which is further supported by the observation that the  value of $\eta^\perp = 0.0427$ from our symmetric expansion, with its momentum-dependent vertex corrections, is already considerably  closer to the high precision result.
It is nevertheless reassuring, that our result of $\eta^\perp = 0.0988$ from the comoving expansion is at least compatible with earlier ones in comparable truncations such as those of Ref.~\cite{Canet_2007}, listed in Table~\ref{crit-exp-results} below, or 
 that of Ref.~\cite{Sinner:2007ws}, where an analogous truncation of the effective average action was used in the Euclidean FRG (in combination with a $4^\text{th}$ order Taylor expansion of the effective potential). The latter can in fact be seen as an important test of the consistency between our real real-time results and those from the standard Euclidean FRG.
The comparatively large difference, here about a factor of $2$,  between our own results for $\eta^\perp$, from the two different expansion schemes, once again demonstrates the importance of such handles on systematic uncertainties. To better understand the underlying systematics here, also  note that our symmetric expansion scheme effectively takes into account higher orders in the derivative expansion (through the order $\Ord{\vec{\nabla}^2}$ terms in  $V_{1,k}^{cl,R/A}(\omega,\va p)$ contained in the 4-point vertex function).
Such higher orders in the derivative expansion are 
known to  be rather important from previous studies within the Euclidean FRG \cite{Canet:2002gs,Canet:2003qd}.
Momentum-dependent vertex corrections of this kind, which represent  genuine two-loop contributions to the two-point function, are not included in the one-loop exact expansion around the scale-dependent minimum, here.
One remaining task for the future will therefore be to also include those momentum-dependent vertex corrections in the truncation of Sec.~\ref{sct:ExpAroundScaleDepMin}, in order to verify that the resulting values for the anomalous scaling dimensions from the two expansions indeed converge towards one another.

In fact, in $d=2$ spatial dimensions, cf.~Fig.~\ref{fig:etaPerp}~(a), we observe 
that \emph{only} the symmetric expansion scheme produces a clear plateau where it  approaches a stable fixed-point value.  The situation in the comoving expansion, where no stable fixed-point value is reached, is even less clear than in $d=3$.
As a rough estimate of $\eta^\perp$ in this case, we simply use the value where $\partial_k \eta_k^\perp = 0$, i.e.~where the flow  is stationary. This yields  $\eta^\perp = 0.446$. 
One has to keep in mind, however,  that no pronounced infrared  power-law behavior of $Z_k^\perp$ emerges in this setup in the first place.
%A first potential reason for this could be that we have to get much closer to the critical point to really see scaling behavior in the spatial wave function renormalization factor in the sense that $\eta_k^\perp$ becomes a constant over many orders of magnitude in~$k$.
%However, this argument is weakened by the fact that the symmetric expansion around the IR minimum already shows a definite plateau structure upon hitting the critical point approximately with the same accuracy.
For comparison, the clear plateau obtained from the expansion around the symmetric IR minimum yields $\eta^\perp = 0.244$ as our best estimate of  the anomalous spatial scaling dimension at criticality. As in $d=3$ dimensions, it is again about a factor of 2 lower than that from the comoving expansion. Luckily, however, in this case we know the exact value, $\eta^\perp = 1/4 = 0.25 $, from Onsager's exact solution of the two-dimensional Ising model~\cite{onsager}.
The close agreement with this exact value of our result from the symmetric expansion scheme with its two-loop contributions to the next-to-leading order terms in the derivative expansion is very reassuring. The momentum-dependent vertex corrections that are taken into account in the symmetric expansion scheme, as explained above,  evidently suffice to produce a stable power-law behavior in the spatial wave function renormalization factor $Z_k^\perp$, with an anomalous scaling dimension very close to the exact result in two spatial dimensions. While the symmetric expansion scheme is  quantitatively thus more reliable than our comoving expansion scheme around the scale-dependent minimum, when it comes to extracting this particular static critical exponent $\eta^\perp$, the larger value from the latter, for $d=2$, is again comparable with that  of Ref.~\cite{Canet_2007}, cf.~Table~\ref{crit-exp-results}. We emphasize once more, however, that the main merit of the comoving expansion is its simplicity when it comes to calculating the dynamic critical exponent $z$ in the next subsection. It is the non-vanishing value of the scale dependent minimum that allows to compute the corresponding critical infrared behavior of the spectral function from this simple one-loop exact scheme. A two-loop extension analogous to that of our symmetric expansion scheme, with the corresponding higher-order corrections to the spatial derivative expansion, should be straightforward (although more tedious) to include for a  quantitative improvement of the static results, especially of the critical exponent $\eta^\perp$  from the spatial wave function renormalization, also in the comoving expansion in the future.

In summary, the flow in our symmetric expansion scheme around $\phi=0$ more clearly shows the fixed-point behavior of  $\eta^\perp$, and also produces quantitatively more accurate results for this particularly difficult static exponent.
%Both could be due to the inclusion of higher order momentum corrections in the vertices.
The general pattern that Taylor expansions of the effective potential around a fixed field value converge more quickly than those around a scale-dependent one has in fact  been observed previously  already~\cite{Rennecke:2015lur}.
To further improve our  expansion schemes systematically in the future one can take into account higher order vertices ($\phi^6$, $\phi^8$, \dots) and/or higher orders in the (spatial)  gradient expansion to include more momentum-dependent structure in the vertices. Both have been found to improve the precision  of the results for $ \eta^\perp $ in the past
\cite{Canet:2002gs,Canet:2003qd}. With the main focus on critical dynamics and the infrared behavior of critical spectral functions in our present study, we leave these quantitavive improvements on the statics for future work.

\subsection{Critical spectral functions}
\label{sct:critSFs}

\begin{figure*}[t]
    \centering
    \rotatebox[origin=t]{90}{\footnotesize spectral function $\rho(\omega)~/~\si{\mega\eV}^{-2}$}
    \begin{minipage}{0.95\textwidth}
        \centering
        \begin{minipage}{.45\textwidth}
            \centering
    		{\footnotesize Model A}
    		\includegraphics[width=\textwidth]{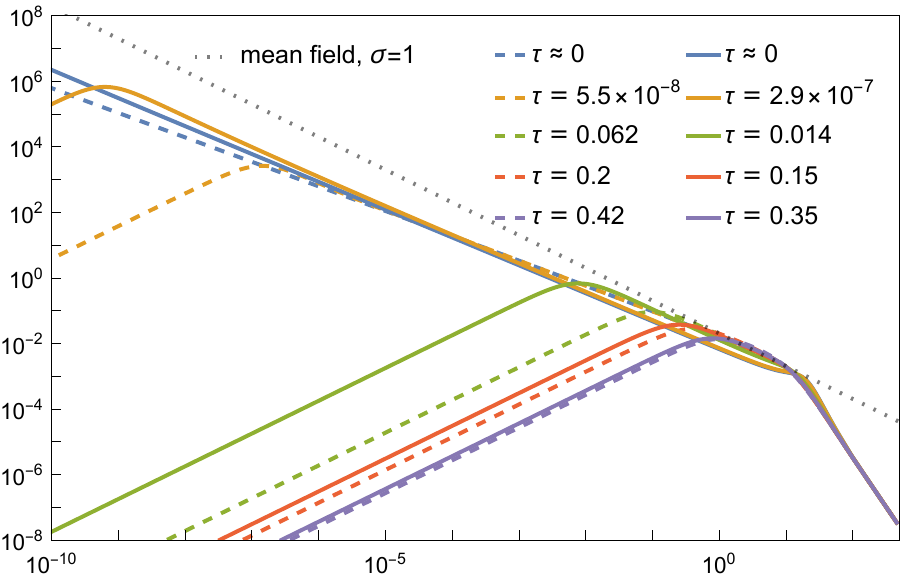}
    	\end{minipage}
%        \begin{minipage}{.32\textwidth}
%            \centering
%    		{\footnotesize Model B}
%    		\includegraphics[width=\textwidth]{{sfs_ModB_d2_aboveTc_loglog}.pdf}
%    	\end{minipage}
        \begin{minipage}{.45\textwidth}
            \centering
    		{\footnotesize Model C}
    		\includegraphics[width=\textwidth]{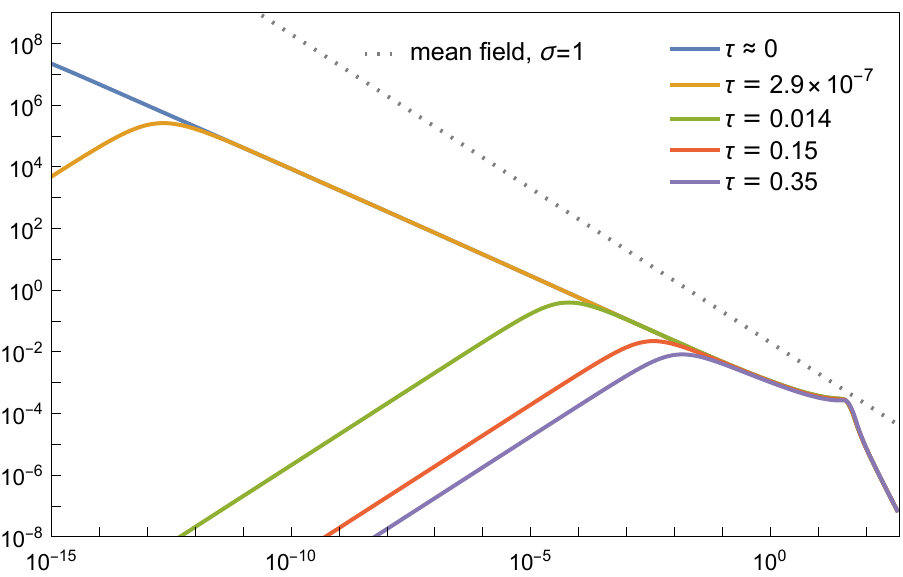}
    	\end{minipage} 
        \rotatebox[origin=t]{-90}{\footnotesize $d=2$} \\
        \begin{minipage}{.45\textwidth}
            \centering
    		%{(d) Model A}
    		\includegraphics[width=\textwidth]{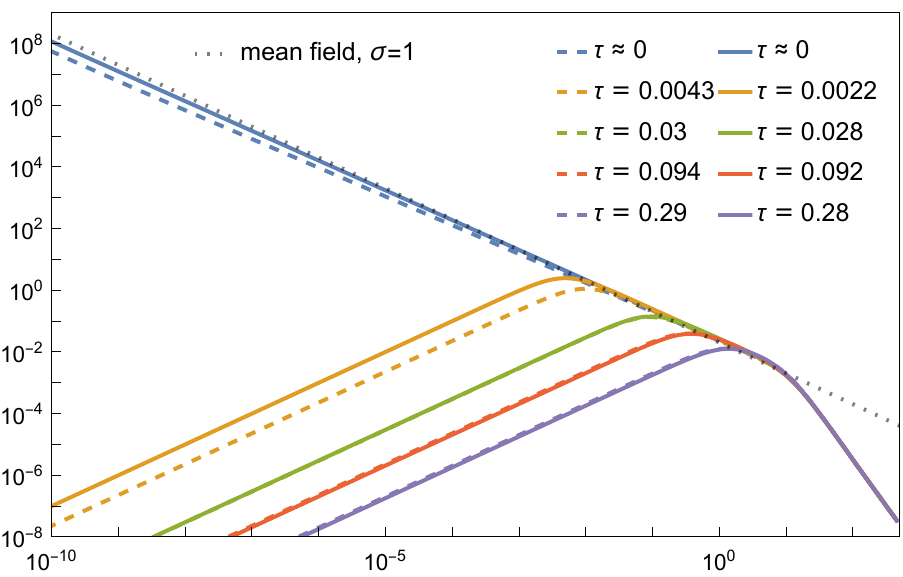}
    	\end{minipage}
 %       \begin{minipage}{.32\textwidth}
 %           \centering
 %   		%{(e) Model B}
 %   		\includegraphics[width=\textwidth]{{sfs_ModB_d3_aboveTc_loglog}.pdf}
 %   	\end{minipage}
        \begin{minipage}{.45\textwidth}
            \centering
    		%{(f) Model C}
    		\includegraphics[width=\textwidth]{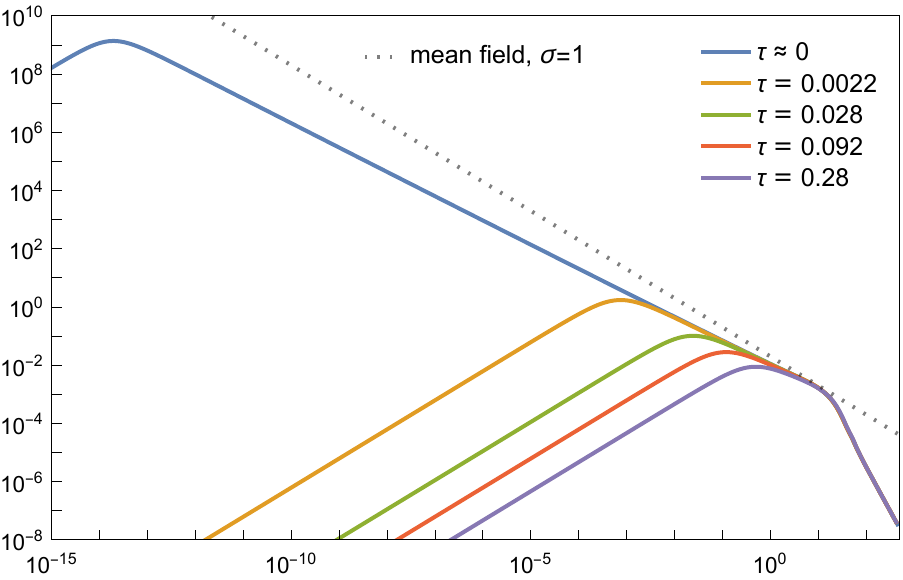}
    	\end{minipage}
        \rotatebox[origin=t]{-90}{\footnotesize $d=3$}
    \end{minipage}\\
	\centering
	{\footnotesize $\omega~/~\si{\mega\eV}$}
    \caption{Critical infrared power-law \eqref{eq:infpow}
    in the spectral functions of Models A (left) and C (right) in $d=2$ (top) and $d=3$ (bottom) spatial dimensions
     at vanishing momentum gradually builds up  for $T \to T_c$,  when approaching the critical temperature from above. Mean-field results are shown as dotted gray lines for comparison. Solid lines represent the results from the symmetric expansion scheme around $\phi=0$ explained in Secs.~\ref{sct:ExpAroundIRMin} and \ref{sct:ExpAroundIRMinforA} for Model~A (left), and Sec.~\ref{sct:ModC} for Model~C (right). Dashed lines represent the results from the comoving expansion around the scale-dependent minimum explained in Secs.~\ref{sct:ExpAroundScaleDepMin} and \ref{sct:ExpAroundScaleDepMinforA} for Model A (left).}
    \label{fig:critSfs}
\end{figure*}

Upon approaching the critical temperature from above we observe scaling behavior gradually building up in the resulting IR ($k \to 0$) spectral functions,  which is shown in Fig.~\ref{fig:critSfs} for Models~A and~C, and in two and three spatial dimensions.
Specifically, the critical spectral function
is expected to show the scaling behavior~\cite{Berges:2009jz},
\begin{align}
    \label{sf-scaling}
    s^{2 - \eta^\perp} \rho_\text{IR}(s^z \omega, s \va{p}, s^{1/\nu} \tau) = \rho_\text{IR}(\omega, \va{p}, \tau)
\end{align}
for  $s > 0$, where the third argument here indicates the dependence on the reduced temperature $\tau \equiv (T-T_c)/T_c$. Introducing dimensionless frequency and momentum variables $\bar\omega = f_t^+ \omega$ and $\bar p = f^+ |\va p|$, where $ f^+$ and $ f_t^+ $ are the non-universal amplitudes of the static correlation length $\xi  $ and the correlation  time $\xi_t$ at $T>T_c$, the critical spectral function can therefore be expressed in terms of universal dynamic scaling functions \cite{Schweitzer:2020noq}, e.g. 
\begin{align}
\rho_\text{IR}(\omega, \va{p}, \tau) &= \bar\omega^{-(2-\eta^\perp)/z} f_\omega( x, y) \label{eq:fomegadef}\\
&=  \bar p^{-(2-\eta^\perp)} f_p (1/x, y/x^{1/\nu z} ) \,, \label{eq:fpdef}
\end{align}
where $x= \bar p^z/\bar\omega$ and $y = \tau/\bar\omega^{1/\nu z}$. These are not independent, of course, but chosen as convenient for the specific purpose and related by $f_\omega(x,y) = x^{-(2-\eta^\perp)/z} f_p (1/x, y/x^{1/\nu z} ) $. 
Whenever the dynamic frequency scaling function $f_\omega(x,y)$ is regular in the origin, with $f_\omega \equiv f_\omega(0,0) >0 $,
we immediately obtain for $\va{p}=0$, $\tau = 0$, i.e.~in the long-wavelength limit  at criticality, the infrared power-law,
\begin{align}
\rho_\text{IR}(\omega, 0, 0) &= f_\omega \, \bar\omega^{-\sigma}  \, , \;\;\mbox{with} \;\; \sigma = (2-\eta^\perp)/z \,. \label{eq:infpow}
\end{align}
This is the case for Model~A and Model~C. It will  not be the case for Model~B, however, where the diffusive dynamics of the critical mode in fact implies that $f_\omega(0,y) = 0$ for all values of  $y$ \cite{Schweitzer:2021iqk}. In particular,  with $y=0$ at criticality, in Model~B, one finds 
\begin{align}
f_\omega(x,0) \sim x^{1-\sigma }\, , \;\; 
f_p(1/x,0) \sim x \, , \;\;
&\mbox{for} \, x\to 0 \,, \label{eq:fsmallx}\\
f_\omega(x,0) \sim x^{-1-\sigma }\, , \;
f_p(1/x,0) \sim x^{-1}\, , \;\; 
&\mbox{for} \, x\to \infty \,,\label{eq:flargex}
\end{align}
with a maximum around  $x\approx 1 $ in either case \cite{Schweitzer:2021iqk}, corresponding to a peak in the critical spectral function at spatial momenta scaling as $|\va p| \propto \omega^{1/z}$. This  also explains the behavior of the critical spectral function discussed at the end of Sec.~\ref{sct:ModBCritModes} for Model~B.

For the infrared power-law $\rho_\text{IR}(\omega) \sim \omega^{ -\sigma }$  of the critical zero-momentum spectral function in Model~A and Model~C, we define  the logarithmic derivative
\begin{equation}
	\sigma(\omega) \equiv -\omega \frac{\partial}{\partial \omega} \log \rho_\text{IR}(\omega) \, , \label{sfLogDeriv}
\end{equation}
which then assumes a constant value $\sigma(\omega) = \sigma = \text{const.}$ in the scaling regime.
%However, due to the Taylor expansion in $\va{p}^2$, we have no direct access to the anomalous scaling dimension $\eta^\perp$ through the obtained IR ($k\to 0$) data, and we hence extract the spatial anomalous scaling dimension $\eta^\perp$ from the scaling of $Z_k^\perp$ at the FRG fixed point \cite{Berges:2000ew,Canet:2002gs,Canet:2003qd}, which is shown in Fig.~\ref{fig:etaPerp}.
It is evident from Fig.~\ref{fig:critSfs}
that we approach such non-trivial fixed-point values of  $\sigma $ close to criticality, indicating that the flow indeed generates this power-law behavior in the correlation functions.
For a first intuitive understanding, remember that at  mean-field level one has $\eta^\perp = 0$ and  $ z=2$ so that the corresponding mean-field exponent $\sigma $ is given by $\sigma_\text{mf} \equiv 1 $.
The resulting mean-field scaling of the critical spectral functions in the infrared is indicated 
in Fig.~\ref{fig:critSfs} by the dotted gray lines.

\subsubsection*{Model A}

We first discuss the two-dimensional result.
Using the scaling relation $z=(2-\sigma)/\eta^\perp$, the value of $\sigma$ obtained from the scaling of the critical spectral function at the plateau of its logarithmic derivative \eqref{sfLogDeriv}, with the respective value of $\eta^\perp$ from the FRG fixed-point as explained in Sec.~\ref{sct:staticsResults} above, yields $z=2.094$ for the comoving expansion of Sec.~\ref{sct:ExpAroundScaleDepMin}, and $z=2.073$ for the symmetric expansion of Sec.~\ref{sct:ExpAroundIRMin}.

There is a slight tension in the literature on Model~A in two spatial dimensions, as different results seem to converge to two slightly different values (although  within about $2.5$ standard deviations, see the discussion in \cite{Schweitzer:2020noq}). The lower one, see e.g.~Refs.~\cite{pft-2d,Schweitzer:2020noq}, is close to the experimental value of $z= 2.09(6)$ from Ref.~\cite{PhysRevB.71.144406}, while most simulations tend to settle around  $z=2.17(3)$, cf.~Ref.~\cite{zhong_critical_2018}, which is close to the original Monte-Carlo estimate of Nightingale and Bl\"ote \cite{PhysRevB.62.1089} and also agrees with other calculations, e.g.~see Refs.~\cite{Canet_2007,Duclut:2016jct,Adzhemyan:2021hvo}.
With all due appreciation of our systematic uncertainties, especially  in two spatial dimensions, our result for this dynamic critical exponent $z$ seems to be in favor of the experimental value of $2.09(6)$ in both expansions. Although higher expansion orders are needed to determine the scaling exponent in two spatial dimensions with certainty (and reliable error estimates)  within the framework of the real-time FRG, it is perhaps worth noting that our rough estimate here is fully consistent  with that from the classical-statistical simulations  in Ref.~\cite{Schweitzer:2020noq} which, in principle, provide ab-inito results for  critical spectral functions near thermal fixed-points.

In three spatial dimensions, the situation is somewhat more settled, and most precision results from the literature collectively find $ z \approx 2.024 $, e.g.~see Refs.~\cite{Canet_2007,Duclut:2016jct,Mesterhazy:2015uja,Adzhemyan:2021hvo,Hasenbusch_2020} and the references therein).
Our results of $z=2.042$ and $z=2.035$ for the comoving and symmetric expansions, respectively, are in reasonable  agreement, given that  the error induced by the finite truncation is at least of the same order as the discrepancy between our two schemes.
The results for the static $\eta^\perp$ and the dynamic critical exponent $z$ of Model~A are summarized in Table~\ref{crit-exp-results}.
%This is confirmed by our use of two different expansion points whose difference in the result for $z$ can be used as an estimate for the truncation error. 

\begin{table*}
    \centering
    \begin{tabular}[t]{ |l|l|l|l| }
    \hline
        $d=2$ & $\eta^\perp$ & $z$ & Reference \\ \hline \hline
        FRG   & $0.446^a$, $0.244^b$         & $2.094^a$, $2.073^b$                           & (This work.)   \\
            & 0.43         & 2.15 (LPA), 2.17 (UZA)         & \cite{Canet_2007}   \\
            & 0.29         & 2.16, 2.15, and 2.14           & \cite{Duclut:2016jct} \\ \hline
        MC    &              & 2.1667(5)                      & \cite{PhysRevB.62.1089} \\ \hline
        PFT   &              & 2.093                          & \cite{pft-2d} \\ \hline
        CS    &              & 2.10(4)                        & \cite{Schweitzer:2020noq} \\ \hline
        $\varepsilon$-expansion & & 2.14(2)                        & \cite{Adzhemyan:2021hvo} \\ \hline
        Experiment  &              & 2.09(6)                        & \cite{PhysRevB.71.144406} \\ \hline
        Exact & 1/4          &                                & \cite{onsager} \\ \hline
    % \end{tabular}
    % \begin{tabular}[t]{ |l|l|l|l| }
    \hline
        $d=3$ & & &   \\ \hline \hline
        %$d=3$ & $\eta^\perp$ & $z$ & Reference  \\ \hline \hline
        FRG   & $0.0988^a$, $0.0427^b$      & $2.042^a$, $2.035^b$                            & (This work.)   \\
            & 0.11         & 2.05 (LPA), 2.14 (UZA)          & \cite{Canet_2007}   \\
            & 0.0443       & 2.024, 2.024, and 2.023         & \cite{Duclut:2016jct} \\
            & 0.055        & 2.025                           & \cite{Mesterhazy:2015uja} \\
            & 0.0443 ($\partial^2$), 0.033 ($\partial^4$) &                                 & \cite{Canet:2002gs,Canet:2003qd} \\
            & 0.0361(11) ($\partial^6$) & & \cite{DePolsi:2020pjk} \\ \hline
        MC    &              & 2.0245(15)                      & \cite{Hasenbusch_2020} \\ \hline
        CS    &              & 1.92(11)                        & \cite{Schweitzer:2020noq} \\ \hline
        $\varepsilon$-expansion & & 2.0235(8)                       & \cite{Adzhemyan:2021hvo} \\ \hline
        CB    & 0.036298     &                                 & \cite{Kos:2016ysd,Komargodski:2016auf} \\ \hline
    \end{tabular}
\caption{Comparison of results for the critical exponents $\eta^\perp$ and $z$ with Ising universality and  Model~A dynamics in $d = 2,3$ spatial dimensions: These include 
Monte-Carlo (MC) simulations~\cite{PhysRevB.62.1089,Hasenbusch_2020}, FRG calculations~\cite{Canet_2007,Duclut:2016jct,Mesterhazy:2015uja}, perturbative field theory (PFT) methods~\cite{pft-2d}, classical-statistical lattice simulations~\cite{Schweitzer:2020noq}, and five-loop $\varepsilon$-expansion \cite{Adzhemyan:2021hvo}. In two spatial dimensions, one furthermore has  the exact value for $\eta^\perp$ from Onsager's solution~\cite{onsager} and
the experimental value for $z$ from Ref.~\cite{PhysRevB.71.144406}. In three spatial dimensions, high-precision results are available  from conformal bootstrap (CB)~\cite{Kos:2016ysd,Komargodski:2016auf}. 
The FRG results of Ref.~\cite{Duclut:2016jct} were obtained for three different `frequency regulators.'  The results from Refs.~\cite{Canet:2002gs,Canet:2003qd} are understood as second $\Ord{\partial^2}$ and fourth $\Ord{\partial^4}$ order in the gradient expansion, and the more recent one of Ref.~\cite{DePolsi:2020pjk} as sixth order $\Ord{\partial^6}$ together with a novel error estimate. The superscripts on our results indicate: ($a$) comoving expansion around scale-dependent minimum $\phi=\phi_{0,k}$, and ($b$) symmetric expansion around  $\phi=0$.}
\label{crit-exp-results}
\end{table*}

\begin{figure*}[t]
    \centering
    \rotatebox[origin=t]{90}{\footnotesize spectral function $\rho(\omega,\boldsymbol{p})~/~\si{\mega\eV}^{-2}$}
    \begin{minipage}{.45\textwidth}
        \centering
        {(a) $\tau = -0.0021$}
		\includegraphics[width=\textwidth]{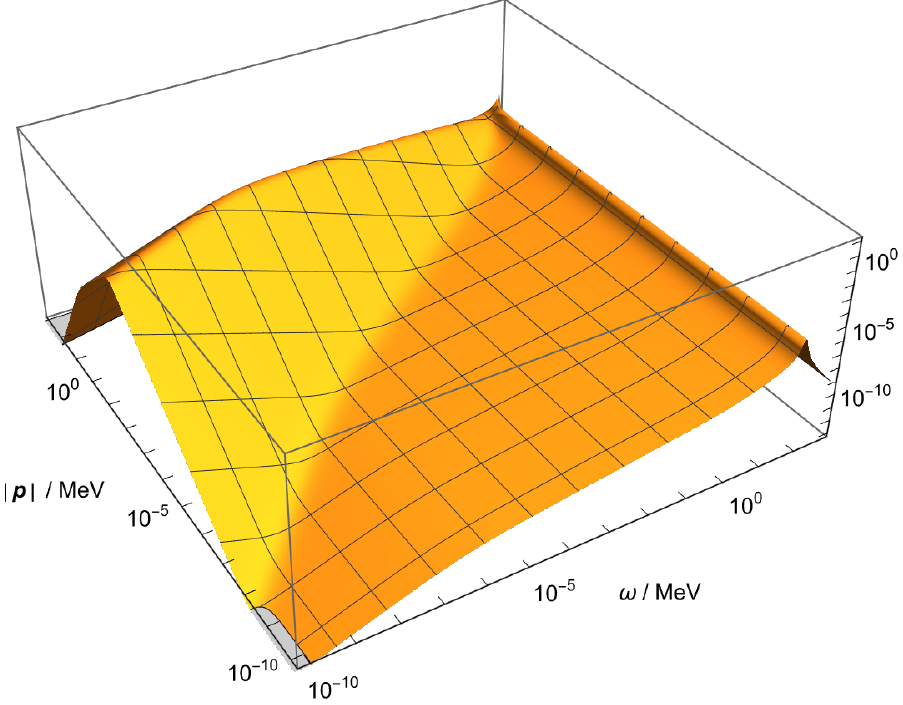}
	\end{minipage} 
    \begin{minipage}{.45\textwidth}
        \centering
        {(b) $\tau \approx 0$}
		\includegraphics[width=\textwidth]{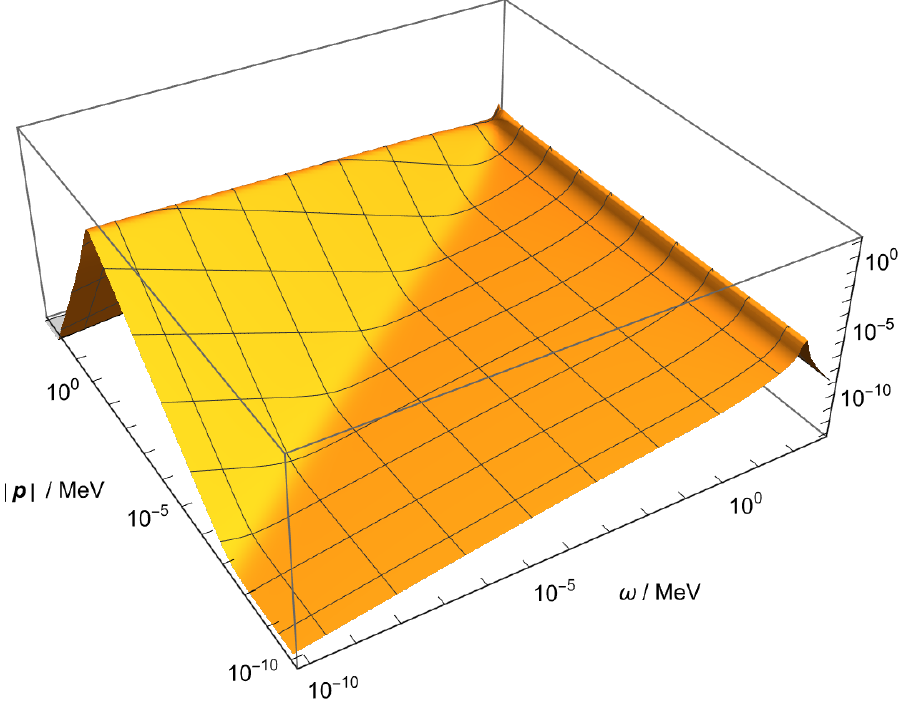}
	\end{minipage} 
%    \begin{minipage}{.3\textwidth}
%        \centering
%        {(c) $\tau = 0.0043$}
%		\includegraphics[width=\textwidth]{{sfs3d_ModB_d3_T156.0000000000}.pdf}
%	\end{minipage} 
	\centering
    \caption{Non-critical (a) and critical (b) spectral functions of Model~B in three spatial dimensions, realized after Son and Stephanov~\cite{Son:2004iv} by coupling a conserved density linearly to the non-conserved order parameter. We see that near the critical temperature in (b) the order parameter (interpreted as fluctuations in the sigma-meson channel) stays massive. Indeed, it is instead the diffusive mode which emerges as a mixture of fluctuations in the (baryon) density and fluctuations in the sigma-meson channel which becomes critical, and gives rise to the critical scaling behavior in the spacelike region, as visible in (b), where the spectral function obeys the scaling hypothesis~\eqref{sf-scaling} with mean-field exponents $\eta^\perp = 0$ and $z=4$. \label{fig:critSfsModB}}
\end{figure*}

\subsubsection*{Model B}

Exemplary spectral functions for Model~B are shown
in Fig.~\ref{fig:critSfsModB} for two  temperatures: somewhat below the critical temperature in Fig.~\ref{fig:critSfsModB}~(a) on the left, and approximately at the criticality in (b) on the right.
The spectral function essentially consists of two modes: First, there is the quasi-particle pole of the non-conserved but massive order parameter fluctuations which is gapped in the long-wavelength limit by the plasmon mass $m_p$.
Secondly, and more interestingly, already in Fig.~\ref{fig:critSfsModB}~(a) there is a  
gapless mode with diffusive dispersion  relation $\omega \sim \va{p}^2$ visible in the spacelike region, which is neither present in Model~A nor~C. This is of course  due to the mixing of the order parameter fluctuations with those of the conserved (baryon) density.

If we now tune the temperature very close to the critical one, as shown in Fig.~\ref{fig:critSfsModB}~(b), we observe that in contrast to Models~A and~C (cf.~Fig.~\ref{fig:critSfs}), the quasi-particle ridge stays gapped here because the plasmon mass remains finite. In particular, 
no power-law divergence builds up for $|\va p|\to 0$, but the spectral function  in the long wavelength limit vanishes for $\omega \to 0$. This is not the part of the critical spectral function that is described by the dynamic scaling functions in Eqs.~\eqref{eq:fomegadef}, \eqref{eq:fpdef}, here.
What actually does becomes critical is the part of the spectral function induced by the mixing with the conserved (baryon) density, as discussed in Sec.~\ref{sct:ModBCritModes} above. We first note that the dispersion relation of the diffusive peak in the spacelike region changes from $\omega \sim |\va p|^2$ in Fig.~\ref{fig:critSfsModB}~(a)
 to $\omega \sim |\va p|^z$ in Fig.~\ref{fig:critSfsModB}~(b), here with the mean-field exponent $z=4$ because of our derivative expansion as explained in Sec.~\ref{sct:ModBCritModes}. Moreover, the $\va p^2 $ dependence of this diffusive peak 
is described by the dynamic scaling functions in \eqref{eq:fomegadef}, \eqref{eq:fpdef}  (here with $\tau\approx 0$), e.g.~by
\begin{align}
\rho_\text{IR}(\omega, \va{p}, 0) &= \bar\omega^{-\sigma}
\,f_\omega( x, 0) \, .
\end{align}
This scaling function was numerically determined in Ref.~\cite{Schweitzer:2021iqk} to be very well described by 
\begin{align}
  f_\omega (x,0) = \rho_0 \, \frac{x^{-\sigma}}{x^{-1} + x} \, , \;\; x = \bar p^z/\bar \omega \, ,
\end{align}
with some non-universal normalization $\rho_0 $, yielding 
\begin{align}
\rho_\text{IR}(\omega=\mathrm{const.}, \va{p}, 0) &\propto    \frac{{\bar p}^{-(2-\eta^\perp) }}{\bar\omega/{\bar p}^z + {\bar p}^z/\bar\omega } \, .
\end{align}
This diffusive part of the spectral function 
at criticality is thus exactly described by the scaling hypothesis \eqref{sf-scaling}, here  with mean-field exponents $\eta^\perp = 0$ and $z=4$,  however. In particular, this also implies the limiting behavior in Eqs.~\eqref{eq:fsmallx} and \eqref{eq:flargex} for $|\va p|\ll \omega $ and $|\va p| \gg\omega $, as already discussed at the end of Sec.~\ref{sct:ModBCritModes}.  

\begin{figure}[t]
    \centering
    \includegraphics[width=0.5\columnwidth]{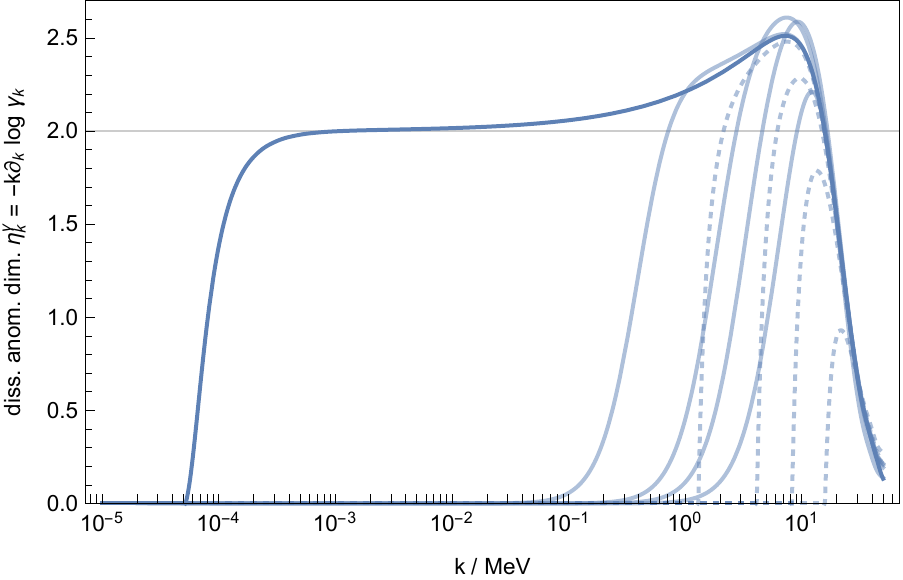}
    \caption{Flow of the scaling dimension $\eta_k^\gamma$ of the damping constant $\gamma_k$ for Model~B in three spatial dimensions at various temperatures $T$ around the critical one, where it approaches a fixed-point value of $\eta^\gamma = 2$. Solid lines indicate $T<T_c$, and dashed lines $T>T_c$, respectively.  \label{fig:etaGam_ModB_d23}}
\end{figure}

To explicitly verify the infrared divergence of the FRG-scale $k$ dependent  damping constant $ \gamma_k \equiv \lim_{\omega \to 0}  \Im\Gamma_{k}^{qc}(\omega,0)/ \omega $  at the critical point, in Fig.~\ref{fig:etaGam_ModB_d23} we show  our results for its logarithmic scale derivative 
$\eta^\gamma_k \equiv -k \partial_k \log \gamma_k$, for temperatures $T$ around the critical temperature $T_c$. 
We clearly observe that the corresponding scale-dependent exponent $\eta^\gamma_k$ 
nicely levels in a plateau, yielding  $\gamma_k \sim  k^{-{\eta^\gamma}} $ with $\eta^\gamma = 2$ as $T$ approaches $T_c$ at criticality. This is the numerical confirmation of  our explicit analytic derivation in Sec.~\ref{sct:ModBCritModes}. 

In summary, our Model~B spectral function in Fig.~\ref{fig:critSfsModB} nicely illustrates how both features discussed by Son and Stephanov in \cite{Son:2004iv} as a result of the mixing of the chiral order parameter field with the baryon density fluctuations near the critical endpoint in the phase diagram of QCD, after Gaussian integration of the conserved density, are combined in a single spectral function: The quasi-particle peak in the timelike region, which represents the fast and gapped fluctuations of the chiral condensate, and which practically instantaneously adjusts to the slow diffusive critical mode representing the mixture between chiral condensate and baryon-density fluctuations that gives rise to Model~B dynamics (without the coupling to the shear modes in the energy-momentum tensor in QCD).

\ifModC
\subsubsection*{Model C}

The critical spectral functions of Model~C are shown in the right column in Fig.~\ref{fig:critSfs}.
With the same procedure outlined for Model~A above, we obtain $z=2.56$ in two spatial dimensions, and $z=2.31$ in three spatial dimensions, respectively.
This is to be compared with the exact result $z=2$ in two spatial dimensions (from Onsager's solution \cite{onsager}), and $z=2+\alpha/\nu\approx 2.17$ in three spatial dimensions, where for  the latter we inserted the known numerical result for $\nu = 0.629971(4) $ \cite{Kos:2016ysd,Komargodski:2016auf} and used the hyperscaling relation $\alpha=2-\nu d$ for the specific heat exponent.
While the three-dimensional result has the qualitatively correct property of being larger than the Model-A value of $z = 2+c\eta^\perp$ \cite{RevModPhys.49.435}, such that the ordering of the two Models is captured correctly within our truncations, this is not the case in two spatial dimensions.
The rather large discrepancy from the exact  value arises from  the specific heat exponent $\alpha $ which vanishes in the two-dimensional Ising model, whereas our result corresponds to  $\alpha/\nu \approx 0.56$.
The reason for not reproducing $\alpha=0$ in our truncation lies in the crude approximation we have made concerning the conserved density: its fluctuations 
diverge with some  power of $k$ at the critical point
in our truncation, whereas 
$\langle nn\rangle \sim k^{-\alpha/\nu}$ should diverge logarithmically, for $\alpha=0$. To cancel this artificial power-law divergence, we expect that 
higher-order non-Gaussian correlations $\langle n \cdots n \rangle$ are needed which are not included in our truncation.
%Incidentally, our results for $z$ are surprisingly close to the values one obtains in lowest order $\varepsilon$-expansion, where 
This is analogous to the $\varepsilon$-expansion (with $\varepsilon = 4-d$), where at 
first order one has $z=2+\varepsilon/3 + \Ord{\varepsilon^2}$ for Model~C \cite{RevModPhys.49.435}. In fact, it is precisely what one also obtains from linearizing the static flow equations \eqref{m2FlowSymmFinalResult}, \eqref{spatAnomDimSymm}, \eqref{flowLambdaSymm}, \eqref{gFlowEq}, and \eqref{chi0InvFlowEq} around the Wilson-Fisher fixed point (up to small corrections of order $O(\eta^\perp)$). This shows that the present truncation does not capture enough orders in the $\varepsilon$-expansion needed for a reliable  estimate of the specific heat exponent~$\alpha$.
We hence expect substantial improvements when higher expansion orders of the combined effective potential $U_k(\varphi,n)$ are included.\footnote{We have also neglected a possible flow of the baryon conductivity $\lambda_n$ and the relaxation time $\tau_{R}$. Although the inclusion of flowing $\lambda_n$ and $\tau_{R}$ would be beneficial for the overall quantitative correctness, it would not improve the value of the dynamical critical exponent $z$, as the latter is fully determined by static quantities in Model~C~\cite{RevModPhys.49.435}.}
Instead of relying on a (static) flow equation for $U_k(\varphi,n)$, it might be interesting for future work to  reformulate it as a hydrodynamic advection-diffusion equation, and then solve it with the methods described in e.g.~Refs.~\cite{Grossi:2019urj,Koenigstein:2021syz,Koenigstein:2021rxj,Steil:2021cbu,Ihssen:2023qaq} in the two-dimensional $(\varphi,n)$ field space.
%It would be indeed an interesting topic for future work to instead rely on a flow equation  for the combined effective potential $U_k(\varphi,n)$, possibly reformulate it as a hydrodynamic (continuity) equation, and then solve it via methods described in e.g.~Refs.~\cite{Grossi:2019urj,Koenigstein:2021syz,Koenigstein:2021rxj,Steil:2021cbu} in the two-dimensional $(\varphi,n)$ field space.
%One can then extract the critical exponent $\alpha/\nu$ from the flow of the effective potential in the vicinity of the fixed point solution, and correspondingly extract a value for the dynamical critical exponent $z=2+\alpha/\nu$, without invoking the real-time FRG at all.
While the FRG-calculation of the heat capacity exponent $\alpha$ is a known challenge of the static critical behavior,   
we are nevertheless able to successfully describe the dynamic critical behavior, at least qualitatively also in Model~C. Our real-time FRG calculations clearly produce
a  stable power-law behavior in the correlation functions.
Systematic extensions of the truncation scheme to improve the numerical accuracy of all relevant critical exponents (including the specific heat exponent $\alpha$) will be left for future studies.
%This issue of calculating the exponent $\alpha/\nu$ to a high precision is fully pushed into static quantities, and hence fully within the scope of the static Euclidean FRG.
\fi

\section{Conclusion \& Outlook}
\label{sct:conclusion}

We have developed a real-time formulation of the functional renormalization group (FRG) for studying critical dynamics in the vicinity of continuous phase transitions.
This includes a general prescription to construct causal regulators, which are bilinear in the fields, from Kramers-Kronig relations based on spectral distributions of unphysical states in an indefinite metric space representing an artificial environment. As another important prerequisite for FRG calculations on the Schwinger-Keldysh closed time path,  based on combined vertex, loop and derivative expansions, we have devised truncations suitable to capture the relevant critical dynamics. 

As a first application of our real-time FRG formulation, we have calculated critical spectral functions for the relaxational Models~A,~B, and~C. 
We have thereby employed two distinct expansion schemes of the effective average action in terms of 1PI vertex functions, to assess the influence of truncation uncertainties on our results.
Dynamically generating critical infrared power-law behavior in the correlation functions (here specifically in the two and 4-point functions) was possible due to incorporating non-local loop diagrams self-consistently in the flow equations. 
Our results for the dynamic critical exponents of the purely dissipative Model A dynamics (with static Ising universality) in two and three spatial dimensions are very promising and compare quantitatively well with the literature. 

The diffusive dynamics of Model~B produces the expected scaling behavior as described by dynamic scaling functions. 
In addition, the spectral function of our realization of the Model~B dynamics nicely combines the gapped non-critical fluctuations of the order parameter field with the diffusive critical mode. Translated to the QCD phase-diagram language
after Son and Stephanov~\cite{Son:2004iv}, the former are the fast fluctuations of the chiral condensate that adjust practically instantaneously to the slow diffusive critical mode that mixes with the fluctuations of the conserved baryon density in the vicinity of the critical endpoint. 

A natural next step would be to include the two conserved shear modes in the fluctuations of the energy-momentum tensor, to describe the universal critical behavior of Model~H as expected to be relevant for critical endpoint in QCD. The corresponding critical dynamics and the universal behavior of non-equilibrium phase transitions along protocols with time-dependent variations of the system parameters could then eventually bring the phenomenology of relativistic heavy-ion collisions and the static properties of the QCD phase diagram closer to one another. In general, however, more conserved fields bring along additional difficulties such as non-vanishing Poisson brackets or commutator relations between order parameter and/or conserved fields corresponding to reversible  mode-mode couplings, which we did not encounter in this work.

%In the future, they might be useful in comparison with experimental observations, for example in the search of the 
%For Model~B we included a linear coupling to a conserved density, and hence confirmed the prediction of  Son and Stephanov~\cite{Son:2004iv} that the linear coupling of a conserved density to a non-conserved order parameter indeed changes the critical dynamics to be diffusive.

\ifModC
For Model~C we formulated a symmetric two-loop expansion scheme around vanishing field expectation value, along the lines of our general truncation strategy,
in order to include a non-linear cubic coupling between the square of the order parameter field and a  conserved energy density.
While this suffices to generate the expected  infrared power-law in the critical spectral function, just as with Model~A dynamics, 
the resulting values for the dynamic critical exponent $z= 2+\alpha/\nu $ are off in this case, mainly because of the known  difficulty in reliably calculating the static critical exponent $\alpha$ of the specific heat. We expect that this requires non-Gaussian moments in the higher-order fluctuations of the conserved energy density, which we did not include here. Considering the flow of the full effective potential in the two-dimensional field space $(\varphi,n)$ might be the remedy needed to achieve this in the future.
\fi

Another future  extension of our present 
study would be to include finite spatial momenta beyond the derivative expansions used here, and to systematically 
determine the complete dynamic scaling functions  introduced in Refs.~\cite{Schweitzer:2020noq,Schweitzer:2021iqk} that describe the critical spectral functions in all three dynamic models. Such dynamic scaling functions could then again be used to make predictions for the universal part of QCD spectral functions in the vicinity of the $Z_2$-critical endpoint, for example.

\section*{Acknowledgements}

We thank S\"oren Schlichting, Dominik Schweitzer, Leon Sieke, Arno Tripolt, and Yunxin Ye for stimulating discussions and collaborations on closely related work. We also acknowledge instructive conversations with Jan Horak, Jan Pawlowski, and Nicolas Wink. 
This work was supported by the Deutsche Forschungsgemeinschaft (DFG) through the grant CRC-TR 211 ``Strong-interaction matter under extreme conditions.'' JVR is supported by the Studienstiftung des deutschen Volkes.
Computational resources were provided by the HPC Core Facility of Justus-Liebig University Giessen.

\appendix

\section{Causality and Kramers-Kronig relations}
\label{sct:KramersKronigRels}

In Section \ref{sct:causalReg}, we started with assuming  that the regulator $R_k^{R/A}(\omega,\boldsymbol{p})$  is analytic at $\omega = 0$ for infrared finiteness. One can then write down subtracted Kramers-Kronig relations based on the analytic properties of $(R_k^{R/A}(\omega,\boldsymbol{p}) -  R_k^{R/A}(0,\boldsymbol{p}))/\omega $,
\begin{subequations}
\begin{align}
    \Re R^{R/A}(\omega, \va{p}) &= \Re R^{R/A}(0, \va{p})  \pm P\int_{-\infty}^\infty \frac{d\omega'}{2\pi} \frac{2 \omega^2 \Im R^{R/A}(\omega', \va{p})}{\omega'(\omega'^2 - \omega^2)} \, , \label{kramersKronig1s} \\
    \Im R^{R/A}(\omega, \va{p}) 
    %\mp \frac{C_k(\boldsymbol p)}{\omega}  
    &=  \mp P\int_{-\infty}^\infty \frac{d\omega'}{2\pi} \frac{2 \omega\,\Re R^{R/A}(\omega', \va{p})}{\omega'^2 - \omega^2} \, , \label{kramersKronig2s}
\end{align}
\end{subequations}
where $P$ denotes the Cauchy principal value of the integral. Here, we have furthermore used that $\Im R^{R/A}(0, \va{p}) = 0 $ for an analytic function with odd imaginary part, so that Eq.~(\ref{kramersKronig2s}) stays formally the same as in the unsubtracted case in Eq.~(\ref{kramersKronig2}) below. 
The subtracted Kramers-Kronig relations (\ref{kramersKronig1s}) and (\ref{kramersKronig2s}) are combined in  the subtracted spectral representation,
\begin{equation}
  R^{R/A}(\omega, \va{p}) = R^{R/A}(0, \va{p})  - \int_{0}^\infty \frac{d\omega'}{2\pi} \frac{2 \omega^2 J(\omega', \va{p})}{\omega'((\omega \pm i\epsilon)^2 - \omega'^2)} \, ,
\end{equation}
corresponding to Eqs.~\eqref{regulatorSPDef}, \eqref{sigmaRADef} of Section~\ref{sct:causalReg}.

Alternatively to first assuming analyticity of the regulator at $\omega=0$, as described in Section~\ref{sct:causalReg}, we might as well start with assuming that the regulator is analytic at complex infinity, and perform the subtraction of the frequency-independent mass shift there. Then, for any finite value of $k$ the limit 
\begin{align}
    \Delta M_{\infty,k}^2(\boldsymbol{p}) \equiv -\lim_{\omega \to \infty} R_k^{R/A}(\omega,\boldsymbol{p}) \label{massShiftUV}
\end{align}
exists, is real, and unique in the sense that it can be taken in any direction and there is no singularity at infinity in the complex $\omega$-plane. We can then formally separate this part from the regulator and use this as an alternative definition of the  spectral part of the self-energy regulator,  
\begin{align}
    \Sigma_k^{R/A}(\omega,\boldsymbol{p}) \equiv R_k^{R/A}(\omega,\boldsymbol{p}) + \Delta M_{\infty,k}^2(\boldsymbol{p})\, , 
\end{align}
replacing (\ref{regulatorSPDef}) by the one subtracted at infinity. Then, by definition we have in this case,
\begin{align}
    \Sigma_k^{R/A}(\omega,\boldsymbol{p}) \to 0 \hspace{0.2cm} \text{for $\omega \to \infty$}\,. \label{sigmaRALimitWInfty}
\end{align}
%Moreover, the causal structure demands that the retarded and advanced parts are connected by complex conjugation. Together with the requirement that they are real in the time domain for real scalar fields, we thus have $\Sigma_k^R(-\omega,\boldsymbol{p}) = {\Sigma_k^R}^*(\omega,\boldsymbol{p}) = \Sigma_k^A(\omega,\boldsymbol{p})$. 
Because it vanishes at infinity, its real and imaginary parts are now related by unsubtracted Kramers-Kronig relations,
for the retarded/advanced components $\Sigma^{R/A}(\omega,\va{p})$ of the spectral part of the regulator,
\begin{subequations}
\begin{align}
    \Re \Sigma^{R/A}(\omega, \va{p}) &= \pm P\int_{-\infty}^\infty \frac{d\omega'}{2\pi} \frac{2 \omega'\,\Im \Sigma^{R/A}(\omega', \va{p})}{\omega'^2 - \omega^2} \, , \label{kramersKronig1} \\
    \Im \Sigma^{R/A}(\omega, \va{p}) 
    %\mp \frac{C_k(\boldsymbol p)}{\omega}  
    &=  \mp P\int_{-\infty}^\infty \frac{d\omega'}{2\pi} \frac{2 \omega\,\Re \Sigma^{R/A}(\omega', \va{p})}{\omega'^2 - \omega^2} \, , \label{kramersKronig2}
\end{align}
\end{subequations}
Together, these imply the unsubtracted (standard) spectral representation replacing (\ref{sigmaRADef}),
\begin{align}
    \Sigma_k^{R/A}(\omega,\boldsymbol{p}) = -\int_0^\infty \frac{d\omega'}{2\pi} \frac{2\omega' J_k(\omega',\boldsymbol{p})}{(\omega \pm i\varepsilon)^2-\omega'^2} \, , \label{sigmaSpectralRepA}
\end{align}
where the FRG-scale dependent spectral density $J_k(\omega,\va{p})$ is again given by the imaginary part of the regulator,
\begin{align}
    J_k(\omega,\boldsymbol{p}) = \pm 2\Im \Sigma_k^{R/A}(\omega,\boldsymbol{p}) = \pm 2\Im R_k^{R/A}(\omega,\boldsymbol{p}) \, . \label{spectralDensHBA}
\end{align}
Eqs.~\eqref{kramersKronig1} and \eqref{kramersKronig2} use the analyticity of $\Sigma^{R(A)}(\omega,\va{p})$ in the upper (lower) half plane.
They are thus valid for any retarded (advanced) function $f(\omega)$ in the complex plane, as long as it falls off at complex infinity, $f(\omega) \to 0$ for  $\omega\to \infty $. In particular, analyticity at complex infinity  implies that 
\begin{align}
    \Sigma_k^{R}(\omega,\boldsymbol{p}) \to \frac{i C_k(\boldsymbol p)}{\omega} \hspace{0.2cm} \text{or faster for $\omega \to \pm\infty$}\,,  \label{sigmaRALimitWInftys}
\end{align}
with possibly non-vanishing but real $C_k(\boldsymbol p)$. Contour integration then yields
\begin{align}
    \int_{-\infty}^\infty \frac{d\omega}{2\pi} ~ \Big( \Sigma_k^{R}(\omega,\boldsymbol{p}) + \Sigma_k^{A}(\omega,\boldsymbol{p})\Big) = C_k(\boldsymbol{p}) \, . \label{anasigma}
\end{align}
This shows explicitly that, when the spectral part $\Sigma_k^{R/A}(\omega,\boldsymbol{p})$ tends to zero faster than $1/\omega$ for $\omega \to \infty$, i.e.~for $C_k(\boldsymbol  p) \equiv 0$, we obtain the typical causality sum rule
\begin{align}
    0 &= \int_{-\infty}^\infty \frac{d\omega}{2\pi} \left( \Sigma_k^R(\omega,\boldsymbol{p}) + \Sigma_k^A(\omega,\boldsymbol{p}) \right) = \int_0^\infty \frac{d\omega}{2\pi}~4 \Re \Sigma_k^{R}(\omega,\boldsymbol{p}) \, .
\end{align}
The real part $\Re \Sigma_k^{R/A}(\omega,\boldsymbol{p})$ of our causal regulator then necessarily has a zero crossing along the frequency axis.
As a manifestation of this zero crossing  we furthermore observe that a positive spectral density $J_{k}(\omega,\va{p})$, which also vanishes when $\omega\to 0$ for IR finiteness, then
necessarily leads to a \emph{negative} shift in the squared mass caused by the spectral part \eqref{sigmaSpectralRepA} of the regulator,
\begin{align}
    \Delta m_k^2(\va{p}) \equiv -\Sigma_k^{R/A}(0, \va{p}) = - \int_0^\infty \frac{d\omega'}{\pi} \frac{J_k(\omega',\va{p})}{\omega' } \,. \label{bathMassShift}
\end{align}
%Adding \eqref{massShiftUV} and \eqref{bathMassShift} yields the total shift of the squared mass that our scalar field obtains due to the regulator, which must necessarily be positive at all FRG scales~$k$ in order to properly regulate IR modes, 
%\begin{align}
%    0 \stackrel{!}{\leq} \Delta M_{\infty,k}^2(\va{p}) + \Delta m_k^2(\va{p}) \,. \label{positiveMassInIR}
%\end{align}
Assuming that the regulator is IR and UV finite with respect to frequencies, i.e.\ that 
the large-frequency limit \eqref{massShiftUV} vanishes as well, and hence $\Delta M_{\infty,k}^2(\va{p}) \equiv 0$, we obtain the same result  as in Section \ref{sct:causalReg} again,
%By then combining \eqref{positiveMassInIR} with \eqref{bathMassShift} we get
\begin{align}
    0 \stackrel{!}{\leq} \Delta m_k^2(\va{p}) = R^E_k(0,\va p) = - \int_0^\infty \frac{d\omega'}{\pi} \frac{J_k(\omega',\va{p})}{\omega' } \, ,
\end{align}
to avoid acausal regulator singularities. And this contradicts the positivity of the regulator spectral density $J_k(\omega,\va{p})$ in exactly the same way as in Section~\ref{sct:causalReg}, when infrared and ultraviolet finiteness are required for the regulator at the same time.

\section{Explicit expressions for loop functions}
\label{sct:LoopFncsClosedExpr}

In this appendix we list closed form expressions for the loop functions $I_k^K$ and $J_k^{KR}(0,\va{0})$.
The corresponding frequency integrals are solved by virtue of Kramers-Kronig relations, and the spatial momentum integrals using the convenient form of the optimized (Litim) regulator,
\begin{align}
    I_k^K &= -\frac{ 4i \, \Omega_d k^{d+1} T Z_k^\perp}{(2\pi)^d} \left( 1\hspace{-0.05cm}-\hspace{-0.05cm}\frac{\eta_k^\perp}{2+d} \right)  \frac{ 1}{\left( m_k^2 + Z_k^\perp k^2 \right)^2} \, , \label{IKClass} \\
	J_k^{KR}(0,\VecZero) &= -\frac{4i\Omega_d k^{d+1} T Z_k^\perp}{(2\pi)^d} \left( 1\hspace{-0.05cm}-\hspace{-0.05cm}\frac{\eta_k^\perp}{d+2} \right) \frac{1}{\left( m_k^2 + Z_k^\perp k^2 \right)^3} \, , \label{JKRAtZeroClass}
\end{align}
and the analytic $\va{p}^2$ derivative,
\begin{align}
    \lim_{\va{p} \to 0} \frac{\partial}{\partial \va{p}^2} J_k^{KA}(0,\va{p})
    &=  \frac{2i \, \Omega_d k^{d+1}  T\, ( Z_k^\perp )^2}{(2 \pi)^d} \frac{1}{\left( m_k^2 + Z_k^\perp k^2 \right)^4} \, . \label{JKAP2DerivQuarticClass}
\end{align}
Note that no further approximations are involved in deriving these expressions.
Hence they are valid even for an arbitrary frequency dependence $\Gamma_{0,k}^{qc}(\omega)$ of the 2-point function, with the identification $m_k^2 = -\Gamma_{0,k}^{qc}(0)$.
Furthermore, all expressions given in this appendix are valid both for the comoving expansion scheme of Sec.~\ref{sct:ExpAroundScaleDepMin} and the symmetric expansion scheme of Sec.~\ref{sct:ExpAroundIRMin}.

\section{Numerical implementation}
\label{sct:numerical-impl}

We numerically discretize functions $f(\omega)$ which depend on frequency by putting them on a linearly spaced grid $\{ u_0, \dots, u_{N_g-1} \}$ in logarithmic $u\equiv \log(\omega/\omega_{\text{max}})$ space, where the two grid points on the boundary are given by $u_0 = \log(\omega_\text{min} / \omega_\text{max})$ and $u_{N_g} = 0$.

The frequency parts of the (loop-)integrals (convolutions) are calculated via simple lattice sums.
Thereby 2- and 4-point functions are extrapolated for large $\omega$ by their UV ($k=\Lambda$) values, and are linearly interpolated between grid points.

Integrals over the spatial momenta $\va{p}$ can generally be reduced to integrals over the magnitude $|\va{p}|$, which is then calculated using a simple lattice sum on an equally spaced grid in $ \log(|\va{p}|/k) $-space from 0 to some finite lower bound (here taken as coinciding with $ u_0 $), where each of the subintervals are evaluated using an $N_{p,GL}$-point Gauss-Legendre quadrature.

We solve the resulting finite set of ordinary differential equations using an explicit Runge-Kutta-Fehlberg scheme with constant step size $\Delta t = \log(k_\text{IR}/\Lambda)/N_\text{steps}$, formulated in RG time $t \equiv \log(k/\Lambda)$.

Arbitrarily assuming $\Lambda=\SI{50}{\MeV}$ to set a physical scale
typical values used in our numerical calculations correspond to $ k_\text{IR} = 10^{-8}\,\si{\MeV} $, $\omega_\text{max}=\SI{500}{\MeV}$, $N_\text{steps} = 2048$, $N_p=32$ and $N_{p,GL}=4$. We adjust the two parameters $\omega_\text{min}$ and $N_g$ such that all relevant structures are captured in the spectral functions.
This typically results in $N_g = 256$, $512$, and $\omega_\text{min}$ ranging from $10^{-15}\,\si{\MeV}$ (for $d=3$ Models~A,~C) over $10^{-18}\,\si{\MeV}$ (for $d=2$ Models~A,~C) to $10^{-22}\,\si{\MeV}$ (for Model B, which we have shown here only for $d=3$).

\bibliographystyle{elsarticle-num} 
\bibliography{cas-refs}

\begin{thebibliography}{10}
\expandafter\ifx\csname url\endcsname\relax
  \def\url#1{\texttt{#1}}\fi
\expandafter\ifx\csname urlprefix\endcsname\relax\def\urlprefix{URL }\fi
\expandafter\ifx\csname href\endcsname\relax
  \def\href#1#2{#2} \def\path#1{#1}\fi

\bibitem{Weldon:1990iw}
H.~A. Weldon, {Reformulation of finite temperature dilepton production}, Phys.
  Rev. D 42 (1990) 2384--2387.
\newblock \href {https://doi.org/10.1103/PhysRevD.42.2384}
  {\path{doi:10.1103/PhysRevD.42.2384}}.

\bibitem{HADES:2019auv}
J.~Adamczewski-Musch, et~al., {Probing dense baryon-rich matter with virtual
  photons}, Nature Phys. 15~(10) (2019) 1040--1045.
\newblock \href {https://doi.org/10.1038/s41567-019-0583-8}
  {\path{doi:10.1038/s41567-019-0583-8}}.

\bibitem{Tripolt:2022hhw}
R.-A. Tripolt, F.~Geurts, {Electromagnetic Probes: Theory and Experiment} (10
  2022).
\newblock \href {http://arxiv.org/abs/2210.01622} {\path{arXiv:2210.01622}}.

\bibitem{Halasz:1998qr}
A.~M. Halasz, A.~D. Jackson, R.~E. Shrock, M.~A. Stephanov, J.~J.~M.
  Verbaarschot, {On the phase diagram of QCD}, Phys. Rev. D 58 (1998) 096007.
\newblock \href {http://arxiv.org/abs/hep-ph/9804290}
  {\path{arXiv:hep-ph/9804290}}, \href
  {https://doi.org/10.1103/PhysRevD.58.096007}
  {\path{doi:10.1103/PhysRevD.58.096007}}.

\bibitem{Berges:1998rc}
J.~Berges, K.~Rajagopal, {Color superconductivity and chiral symmetry
  restoration at nonzero baryon density and temperature}, Nucl. Phys. B 538
  (1999) 215--232.
\newblock \href {http://arxiv.org/abs/hep-ph/9804233}
  {\path{arXiv:hep-ph/9804233}}, \href
  {https://doi.org/10.1016/S0550-3213(98)00620-8}
  {\path{doi:10.1016/S0550-3213(98)00620-8}}.

\bibitem{Mukherjee:2016kyu}
S.~Mukherjee, R.~Venugopalan, Y.~Yin, {Universal off-equilibrium scaling of
  critical cumulants in the QCD phase diagram}, Phys. Rev. Lett. 117~(22)
  (2016) 222301.
\newblock \href {http://arxiv.org/abs/1605.09341} {\path{arXiv:1605.09341}},
  \href {https://doi.org/10.1103/PhysRevLett.117.222301}
  {\path{doi:10.1103/PhysRevLett.117.222301}}.

\bibitem{Mukherjee:2017kxv}
S.~Mukherjee, R.~Venugopalan, Y.~Yin, {Universality regained: Kibble-Zurek
  dynamics, off-equilibrium scaling and the search for the QCD critical point},
  Nucl. Phys. A 967 (2017) 820--823.
\newblock \href {http://arxiv.org/abs/1704.05427} {\path{arXiv:1704.05427}},
  \href {https://doi.org/10.1016/j.nuclphysa.2017.06.049}
  {\path{doi:10.1016/j.nuclphysa.2017.06.049}}.

\bibitem{RevModPhys.49.435}
P.~C. Hohenberg, B.~I. Halperin,
  \href{https://link.aps.org/doi/10.1103/RevModPhys.49.435}{Theory of dynamic
  critical phenomena}, Rev. Mod. Phys. 49 (1977) 435--479.
\newblock \href {https://doi.org/10.1103/RevModPhys.49.435}
  {\path{doi:10.1103/RevModPhys.49.435}}.
\newline\urlprefix\url{https://link.aps.org/doi/10.1103/RevModPhys.49.435}

\bibitem{Son:2004iv}
D.~T. Son, M.~A. Stephanov, {Dynamic universality class of the QCD critical
  point}, Phys. Rev. D 70 (2004) 056001.
\newblock \href {http://arxiv.org/abs/hep-ph/0401052}
  {\path{arXiv:hep-ph/0401052}}, \href
  {https://doi.org/10.1103/PhysRevD.70.056001}
  {\path{doi:10.1103/PhysRevD.70.056001}}.

\bibitem{Berdnikov:1999ph}
B.~Berdnikov, K.~Rajagopal, {Slowing out-of-equilibrium near the QCD critical
  point}, Phys. Rev. D 61 (2000) 105017.
\newblock \href {http://arxiv.org/abs/hep-ph/9912274}
  {\path{arXiv:hep-ph/9912274}}, \href
  {https://doi.org/10.1103/PhysRevD.61.105017}
  {\path{doi:10.1103/PhysRevD.61.105017}}.

\bibitem{Berges:2009jz}
J.~Berges, S.~Schlichting, D.~Sexty, {Dynamic critical phenomena from spectral
  functions on the lattice}, Nucl. Phys. B 832 (2010) 228--240.
\newblock \href {http://arxiv.org/abs/0912.3135} {\path{arXiv:0912.3135}},
  \href {https://doi.org/10.1016/j.nuclphysb.2010.02.007}
  {\path{doi:10.1016/j.nuclphysb.2010.02.007}}.

\bibitem{Schlichting:2019tbr}
S.~Schlichting, D.~Smith, L.~von Smekal, {Spectral functions and critical
  dynamics of the $O(4)$ model from classical-statistical lattice simulations},
  Nucl. Phys. B 950 (2020) 114868.
\newblock \href {http://arxiv.org/abs/1908.00912} {\path{arXiv:1908.00912}},
  \href {https://doi.org/10.1016/j.nuclphysb.2019.114868}
  {\path{doi:10.1016/j.nuclphysb.2019.114868}}.

\bibitem{Schweitzer:2020noq}
D.~Schweitzer, S.~Schlichting, L.~von Smekal, {Spectral functions and dynamic
  critical behavior of relativistic $Z_2$ theories}, Nucl. Phys. B 960 (2020)
  115165.
\newblock \href {http://arxiv.org/abs/2007.03374} {\path{arXiv:2007.03374}},
  \href {https://doi.org/10.1016/j.nuclphysb.2020.115165}
  {\path{doi:10.1016/j.nuclphysb.2020.115165}}.

\bibitem{Schweitzer:2021iqk}
D.~Schweitzer, S.~Schlichting, L.~von Smekal, {Critical dynamics of
  relativistic diffusion}, Nucl. Phys. B 984 (2022) 115944.
\newblock \href {http://arxiv.org/abs/2110.01696} {\path{arXiv:2110.01696}},
  \href {https://doi.org/10.1016/j.nuclphysb.2022.115944}
  {\path{doi:10.1016/j.nuclphysb.2022.115944}}.

\bibitem{Roth:2021nrd}
J.~V. Roth, D.~Schweitzer, L.~J. Sieke, L.~von Smekal,
  \href{https://link.aps.org/doi/10.1103/PhysRevD.105.116017}{Real-time methods
  for spectral functions}, Phys. Rev. D 105 (2022) 116017.
\newblock \href {https://doi.org/10.1103/PhysRevD.105.116017}
  {\path{doi:10.1103/PhysRevD.105.116017}}.
\newline\urlprefix\url{https://link.aps.org/doi/10.1103/PhysRevD.105.116017}

\bibitem{Roder:2005vt}
D.~Roder, J.~Ruppert, D.~H. Rischke, {Selfconsistent calculations of spectral
  densities in the O(N) model: Improving the Hartree-Fock approximation by
  including nonzero decay widths}, Nucl. Phys. A 775 (2006) 127--151.
\newblock \href {http://arxiv.org/abs/hep-ph/0503042}
  {\path{arXiv:hep-ph/0503042}}, \href
  {https://doi.org/10.1016/j.nuclphysa.2006.05.007}
  {\path{doi:10.1016/j.nuclphysa.2006.05.007}}.

\bibitem{Mueller:2010ah}
J.~A. Mueller, C.~S. Fischer, D.~Nickel, {Quark spectral properties above Tc
  from Dyson-Schwinger equations}, Eur. Phys. J. C 70 (2010) 1037--1049.
\newblock \href {http://arxiv.org/abs/1009.3762} {\path{arXiv:1009.3762}},
  \href {https://doi.org/10.1140/epjc/s10052-010-1499-8}
  {\path{doi:10.1140/epjc/s10052-010-1499-8}}.

\bibitem{Fischer:2020xnb}
C.~S. Fischer, M.~Q. Huber, {Landau gauge Yang-Mills propagators in the complex
  momentum plane}, Phys. Rev. D 102~(9) (2020) 094005.
\newblock \href {http://arxiv.org/abs/2007.11505} {\path{arXiv:2007.11505}},
  \href {https://doi.org/10.1103/PhysRevD.102.094005}
  {\path{doi:10.1103/PhysRevD.102.094005}}.

\bibitem{Horak:2020eng}
J.~Horak, J.~M. Pawlowski, N.~Wink, {Spectral functions in the $\phi^4$-theory
  from the spectral DSE}, Phys. Rev. D 102 (2020) 125016.
\newblock \href {http://arxiv.org/abs/2006.09778} {\path{arXiv:2006.09778}},
  \href {https://doi.org/10.1103/PhysRevD.102.125016}
  {\path{doi:10.1103/PhysRevD.102.125016}}.

\bibitem{Floerchinger:2011sc}
S.~Floerchinger, {Analytic Continuation of Functional Renormalization Group
  Equations}, JHEP 05 (2012) 021.
\newblock \href {http://arxiv.org/abs/1112.4374} {\path{arXiv:1112.4374}},
  \href {https://doi.org/10.1007/JHEP05(2012)021}
  {\path{doi:10.1007/JHEP05(2012)021}}.

\bibitem{Kamikado:2013sia}
K.~Kamikado, N.~Strodthoff, L.~von Smekal, J.~Wambach, {Real-time correlation
  functions in the $O(N)$ model from the functional renormalization group},
  Eur. Phys. J. C 74~(3) (2014) 2806.
\newblock \href {http://arxiv.org/abs/1302.6199} {\path{arXiv:1302.6199}},
  \href {https://doi.org/10.1140/epjc/s10052-014-2806-6}
  {\path{doi:10.1140/epjc/s10052-014-2806-6}}.

\bibitem{Tripolt:2013jra}
R.-A. Tripolt, N.~Strodthoff, L.~von Smekal, J.~Wambach, {Spectral Functions
  for the Quark-Meson Model Phase Diagram from the Functional Renormalization
  Group}, Phys. Rev. D 89~(3) (2014) 034010.
\newblock \href {http://arxiv.org/abs/1311.0630} {\path{arXiv:1311.0630}},
  \href {https://doi.org/10.1103/PhysRevD.89.034010}
  {\path{doi:10.1103/PhysRevD.89.034010}}.

\bibitem{Tripolt:2014wra}
R.-A. Tripolt, L.~von Smekal, J.~Wambach, {Flow equations for spectral
  functions at finite external momenta}, Phys. Rev. D 90~(7) (2014) 074031.
\newblock \href {http://arxiv.org/abs/1408.3512} {\path{arXiv:1408.3512}},
  \href {https://doi.org/10.1103/PhysRevD.90.074031}
  {\path{doi:10.1103/PhysRevD.90.074031}}.

\bibitem{Wambach:2014vta}
J.~Wambach, R.-A. Tripolt, N.~Strodthoff, L.~von Smekal, {Spectral Functions
  from the Functional Renormalization Group}, Nucl. Phys. A 928 (2014)
  156--167.
\newblock \href {http://arxiv.org/abs/1404.7312} {\path{arXiv:1404.7312}},
  \href {https://doi.org/10.1016/j.nuclphysa.2014.04.027}
  {\path{doi:10.1016/j.nuclphysa.2014.04.027}}.

\bibitem{Strodthoff:2016pxx}
N.~Strodthoff, {Self-consistent spectral functions in the O(N) model from the
  functional renormalization group}, Phys. Rev. D 95~(7) (2017) 076002.
\newblock \href {http://arxiv.org/abs/1611.05036} {\path{arXiv:1611.05036}},
  \href {https://doi.org/10.1103/PhysRevD.95.076002}
  {\path{doi:10.1103/PhysRevD.95.076002}}.

\bibitem{Pawlowski:2017gxj}
J.~M. Pawlowski, N.~Strodthoff, N.~Wink, {Finite temperature spectral functions
  in the O(N)-model}, Phys. Rev. D 98~(7) (2018) 074008.
\newblock \href {http://arxiv.org/abs/1711.07444} {\path{arXiv:1711.07444}},
  \href {https://doi.org/10.1103/PhysRevD.98.074008}
  {\path{doi:10.1103/PhysRevD.98.074008}}.

\bibitem{Huelsmann:2020xcy}
S.~Huelsmann, S.~Schlichting, P.~Scior, {Spectral functions from the real-time
  functional renormalization group}, Phys. Rev. D 102~(9) (2020) 096004.
\newblock \href {http://arxiv.org/abs/2009.04194} {\path{arXiv:2009.04194}},
  \href {https://doi.org/10.1103/PhysRevD.102.096004}
  {\path{doi:10.1103/PhysRevD.102.096004}}.

\bibitem{Tripolt:2020irx}
R.-A. Tripolt, D.~H. Rischke, L.~von Smekal, J.~Wambach, {Fermionic excitations
  at finite temperature and density}, Phys. Rev. D 101~(9) (2020) 094010.
\newblock \href {http://arxiv.org/abs/2003.11871} {\path{arXiv:2003.11871}},
  \href {https://doi.org/10.1103/PhysRevD.101.094010}
  {\path{doi:10.1103/PhysRevD.101.094010}}.

\bibitem{Tripolt:2021jtp}
R.-A. Tripolt, C.~Jung, L.~von Smekal, J.~Wambach, {Vector and axial-vector
  mesons in nuclear matter}, Phys. Rev. D 104~(5) (2021) 054005.
\newblock \href {http://arxiv.org/abs/2105.00861} {\path{arXiv:2105.00861}},
  \href {https://doi.org/10.1103/PhysRevD.104.054005}
  {\path{doi:10.1103/PhysRevD.104.054005}}.

\bibitem{Jung:2021ipc}
C.~Jung, J.-H. Otto, R.-A. Tripolt, L.~von Smekal, {Self-consistent O(4) model
  spectral functions from analytically continued functional renormalization
  group flows}, Phys. Rev. D 104~(9) (2021) 094011.
\newblock \href {http://arxiv.org/abs/2107.10748} {\path{arXiv:2107.10748}},
  \href {https://doi.org/10.1103/PhysRevD.104.094011}
  {\path{doi:10.1103/PhysRevD.104.094011}}.

\bibitem{Braun:2022mgx}
J.~Braun, et~al., {Renormalised spectral flows} (6 2022).
\newblock \href {http://arxiv.org/abs/2206.10232} {\path{arXiv:2206.10232}}.

\bibitem{Horak:2022aza}
J.~Horak, J.~M. Pawlowski, N.~Wink, {On the quark spectral function in QCD} (10
  2022).
\newblock \href {http://arxiv.org/abs/2210.07597} {\path{arXiv:2210.07597}}.

\bibitem{Canet_2007}
L.~Canet, H.~Chat{\'e}, \href{http://dx.doi.org/10.1088/1751-8113/40/9/002}{A
  non-perturbative approach to critical dynamics}, Journal of Physics A:
  Mathematical and Theoretical 40~(9) (2007) 1937–1949.
\newblock \href {https://doi.org/10.1088/1751-8113/40/9/002}
  {\path{doi:10.1088/1751-8113/40/9/002}}.
\newline\urlprefix\url{http://dx.doi.org/10.1088/1751-8113/40/9/002}

\bibitem{Gasenzer:2007za}
T.~Gasenzer, J.~M. Pawlowski, {Towards far-from-equilibrium quantum field
  dynamics: A functional renormalisation-group approach}, Phys. Lett. B 670
  (2008) 135--140.
\newblock \href {http://arxiv.org/abs/0710.4627} {\path{arXiv:0710.4627}},
  \href {https://doi.org/10.1016/j.physletb.2008.10.049}
  {\path{doi:10.1016/j.physletb.2008.10.049}}.

\bibitem{Gasenzer:2010rq}
T.~Gasenzer, S.~Kessler, J.~M. Pawlowski, {Far-from-equilibrium quantum
  many-body dynamics}, Eur. Phys. J. C 70 (2010) 423--443.
\newblock \href {http://arxiv.org/abs/1003.4163} {\path{arXiv:1003.4163}},
  \href {https://doi.org/10.1140/epjc/s10052-010-1430-3}
  {\path{doi:10.1140/epjc/s10052-010-1430-3}}.

\bibitem{Berges:2012ty}
J.~Berges, D.~Mesterhazy, {Introduction to the nonequilibrium functional
  renormalization group}, Nucl. Phys. B Proc. Suppl. 228 (2012) 37--60.
\newblock \href {http://arxiv.org/abs/1204.1489} {\path{arXiv:1204.1489}},
  \href {https://doi.org/10.1016/j.nuclphysbps.2012.06.003}
  {\path{doi:10.1016/j.nuclphysbps.2012.06.003}}.

\bibitem{PhysRevLett.110.195301}
L.~M. Sieberer, S.~D. Huber, E.~Altman, S.~Diehl,
  \href{https://link.aps.org/doi/10.1103/PhysRevLett.110.195301}{Dynamical
  critical phenomena in driven-dissipative systems}, Phys. Rev. Lett. 110
  (2013) 195301.
\newblock \href {https://doi.org/10.1103/PhysRevLett.110.195301}
  {\path{doi:10.1103/PhysRevLett.110.195301}}.
\newline\urlprefix\url{https://link.aps.org/doi/10.1103/PhysRevLett.110.195301}

\bibitem{Mesterhazy:2013naa}
D.~Mesterh\'azy, J.~H. Stockemer, L.~F. Palhares, J.~Berges, {Dynamic
  universality class of Model C from the functional renormalization group},
  Phys. Rev. B 88 (2013) 174301.
\newblock \href {http://arxiv.org/abs/1307.1700} {\path{arXiv:1307.1700}},
  \href {https://doi.org/10.1103/PhysRevB.88.174301}
  {\path{doi:10.1103/PhysRevB.88.174301}}.

\bibitem{Mesterhazy:2015uja}
D.~Mesterh\'azy, J.~H. Stockemer, Y.~Tanizaki, {From quantum to classical
  dynamics: The relativistic $O(N)$ model in the framework of the real-time
  functional renormalization group}, Phys. Rev. D 92~(7) (2015) 076001.
\newblock \href {http://arxiv.org/abs/1504.07268} {\path{arXiv:1504.07268}},
  \href {https://doi.org/10.1103/PhysRevD.92.076001}
  {\path{doi:10.1103/PhysRevD.92.076001}}.

\bibitem{Duclut:2016jct}
C.~Duclut, B.~Delamotte, {Frequency regulators for the nonperturbative
  renormalization group: A general study and the model A as a benchmark}, Phys.
  Rev. E 95~(1) (2017) 012107.
\newblock \href {http://arxiv.org/abs/1611.07301} {\path{arXiv:1611.07301}},
  \href {https://doi.org/10.1103/PhysRevE.95.012107}
  {\path{doi:10.1103/PhysRevE.95.012107}}.

\bibitem{Corell:2019jxh}
L.~Corell, A.~K. Cyrol, M.~Heller, J.~M. Pawlowski, {Flowing with the temporal
  renormalization group}, Phys. Rev. D 104~(2) (2021) 025005.
\newblock \href {http://arxiv.org/abs/1910.09369} {\path{arXiv:1910.09369}},
  \href {https://doi.org/10.1103/PhysRevD.104.025005}
  {\path{doi:10.1103/PhysRevD.104.025005}}.

\bibitem{Rose:2015bma}
F.~Rose, F.~L\'eonard, N.~Dupuis, {Higgs amplitude mode in the vicinity of a
  $(2+1)$-dimensional quantum critical point: a nonperturbative
  renormalization-group approach}, Phys. Rev. B 91 (2015) 224501.
\newblock \href {http://arxiv.org/abs/1503.08688} {\path{arXiv:1503.08688}},
  \href {https://doi.org/10.1103/PhysRevB.91.224501}
  {\path{doi:10.1103/PhysRevB.91.224501}}.

\bibitem{PhysRevB.89.180501}
A.~Ran\ifmmode~\mbox{\c{c}}\else \c{c}\fi{}on, N.~Dupuis,
  \href{https://link.aps.org/doi/10.1103/PhysRevB.89.180501}{Higgs amplitude
  mode in the vicinity of a $(2+1)$-dimensional quantum critical point}, Phys.
  Rev. B 89 (2014) 180501.
\newblock \href {https://doi.org/10.1103/PhysRevB.89.180501}
  {\path{doi:10.1103/PhysRevB.89.180501}}.
\newline\urlprefix\url{https://link.aps.org/doi/10.1103/PhysRevB.89.180501}

\bibitem{Tan:2021zid}
Y.-y. Tan, Y.-r. Chen, W.-j. Fu, {Real-time dynamics of the $O(4)$ scalar
  theory within the fRG approach}, SciPost Phys. 12 (2022) 026.
\newblock \href {http://arxiv.org/abs/2107.06482} {\path{arXiv:2107.06482}},
  \href {https://doi.org/10.21468/SciPostPhys.12.1.026}
  {\path{doi:10.21468/SciPostPhys.12.1.026}}.

\bibitem{Fu:2022gou}
W.-j. Fu, {QCD at finite temperature and density within the fRG approach: An
  overview} (5 2022).
\newblock \href {http://arxiv.org/abs/2205.00468} {\path{arXiv:2205.00468}}.

\bibitem{Wetterich:1992yh}
C.~Wetterich, {Exact evolution equation for the effective potential}, Phys.
  Lett. B 301 (1993) 90--94.
\newblock \href {http://arxiv.org/abs/1710.05815} {\path{arXiv:1710.05815}},
  \href {https://doi.org/10.1016/0370-2693(93)90726-X}
  {\path{doi:10.1016/0370-2693(93)90726-X}}.

\bibitem{Berges:2000ew}
J.~Berges, N.~Tetradis, C.~Wetterich, {Nonperturbative renormalization flow in
  quantum field theory and statistical physics}, Phys. Rept. 363 (2002)
  223--386.
\newblock \href {http://arxiv.org/abs/hep-ph/0005122}
  {\path{arXiv:hep-ph/0005122}}, \href
  {https://doi.org/10.1016/S0370-1573(01)00098-9}
  {\path{doi:10.1016/S0370-1573(01)00098-9}}.

\bibitem{Pawlowski:2005xe}
J.~M. Pawlowski, {Aspects of the functional renormalisation group}, Annals
  Phys. 322 (2007) 2831--2915.
\newblock \href {http://arxiv.org/abs/hep-th/0512261}
  {\path{arXiv:hep-th/0512261}}, \href
  {https://doi.org/10.1016/j.aop.2007.01.007}
  {\path{doi:10.1016/j.aop.2007.01.007}}.

\bibitem{Wilson:1973jj}
K.~G. Wilson, J.~B. Kogut, {The Renormalization group and the epsilon
  expansion}, Phys. Rept. 12 (1974) 75--199.
\newblock \href {https://doi.org/10.1016/0370-1573(74)90023-4}
  {\path{doi:10.1016/0370-1573(74)90023-4}}.

\bibitem{Pawlowski:2015mia}
J.~M. Pawlowski, N.~Strodthoff, {Real time correlation functions and the
  functional renormalization group}, Phys. Rev. D 92~(9) (2015) 094009.
\newblock \href {http://arxiv.org/abs/1508.01160} {\path{arXiv:1508.01160}},
  \href {https://doi.org/10.1103/PhysRevD.92.094009}
  {\path{doi:10.1103/PhysRevD.92.094009}}.

\bibitem{kamenev_2011}
A.~Kamenev, Field Theory of Non-Equilibrium Systems, Cambridge University
  Press, 2011.
\newblock \href {https://doi.org/10.1017/CBO9781139003667}
  {\path{doi:10.1017/CBO9781139003667}}.

\bibitem{doi:10.1143/JPSJ.12.570}
R.~Kubo, \href{https://doi.org/10.1143/JPSJ.12.570}{Statistical-mechanical
  theory of irreversible processes. i. general theory and simple applications
  to magnetic and conduction problems}, Journal of the Physical Society of
  Japan 12~(6) (1957) 570--586.
\newblock \href {http://arxiv.org/abs/https://doi.org/10.1143/JPSJ.12.570}
  {\path{arXiv:https://doi.org/10.1143/JPSJ.12.570}}, \href
  {https://doi.org/10.1143/JPSJ.12.570} {\path{doi:10.1143/JPSJ.12.570}}.
\newline\urlprefix\url{https://doi.org/10.1143/JPSJ.12.570}

\bibitem{PhysRev.115.1342}
P.~C. Martin, J.~Schwinger,
  \href{https://link.aps.org/doi/10.1103/PhysRev.115.1342}{Theory of
  many-particle systems. i}, Phys. Rev. 115 (1959) 1342--1373.
\newblock \href {https://doi.org/10.1103/PhysRev.115.1342}
  {\path{doi:10.1103/PhysRev.115.1342}}.
\newline\urlprefix\url{https://link.aps.org/doi/10.1103/PhysRev.115.1342}

\bibitem{Wang:1998wg}
E.~Wang, U.~W. Heinz, {A Generalized fluctuation dissipation theorem for
  nonlinear response functions}, Phys. Rev. D 66 (2002) 025008.
\newblock \href {http://arxiv.org/abs/hep-th/9809016}
  {\path{arXiv:hep-th/9809016}}, \href
  {https://doi.org/10.1103/PhysRevD.66.025008}
  {\path{doi:10.1103/PhysRevD.66.025008}}.

\bibitem{Sieberer:2015hba}
L.~M. Sieberer, A.~Chiocchetta, A.~Gambassi, U.~C. T\"auber, S.~Diehl,
  {Thermodynamic Equilibrium as a Symmetry of the Schwinger-Keldysh Action},
  Phys. Rev. B 92~(13) (2015) 134307.
\newblock \href {http://arxiv.org/abs/1505.00912} {\path{arXiv:1505.00912}},
  \href {https://doi.org/10.1103/PhysRevB.92.134307}
  {\path{doi:10.1103/PhysRevB.92.134307}}.

\bibitem{Litim:2001up}
D.~F. Litim, {Optimized renormalization group flows}, Phys. Rev. D 64 (2001)
  105007.
\newblock \href {http://arxiv.org/abs/hep-th/0103195}
  {\path{arXiv:hep-th/0103195}}, \href
  {https://doi.org/10.1103/PhysRevD.64.105007}
  {\path{doi:10.1103/PhysRevD.64.105007}}.

\bibitem{Gies:2006wv}
H.~Gies, {Introduction to the functional RG and applications to gauge
  theories}, Lect. Notes Phys. 852 (2012) 287--348.
\newblock \href {http://arxiv.org/abs/hep-ph/0611146}
  {\path{arXiv:hep-ph/0611146}}, \href
  {https://doi.org/10.1007/978-3-642-27320-9_6}
  {\path{doi:10.1007/978-3-642-27320-9_6}}.

\bibitem{CALDEIRA1983587}
A.~Caldeira, A.~Leggett,
  \href{http://www.sciencedirect.com/science/article/pii/0378437183900134}{Path
  integral approach to quantum brownian motion}, Physica A: Statistical
  Mechanics and its Applications 121~(3) (1983) 587 -- 616.
\newblock \href {https://doi.org/https://doi.org/10.1016/0378-4371(83)90013-4}
  {\path{doi:https://doi.org/10.1016/0378-4371(83)90013-4}}.
\newline\urlprefix\url{http://www.sciencedirect.com/science/article/pii/0378437183900134}

\bibitem{Canet:2011wf}
L.~Canet, H.~Chate, B.~Delamotte, {General framework of the non-perturbative
  renormalization group for non-equilibrium steady states}, J. Phys. A 44
  (2011) 495001.
\newblock \href {http://arxiv.org/abs/1106.4129} {\path{arXiv:1106.4129}},
  \href {https://doi.org/10.1088/1751-8113/44/49/495001}
  {\path{doi:10.1088/1751-8113/44/49/495001}}.

\bibitem{Marguet:2021gab}
B.~Marguet, E.~Agoritsas, L.~Canet, V.~Lecomte, {Supersymmetries in
  nonequilibrium Langevin dynamics}, Phys. Rev. E 104~(4) (2021) 044120.
\newblock \href {http://arxiv.org/abs/2101.08766} {\path{arXiv:2101.08766}},
  \href {https://doi.org/10.1103/PhysRevE.104.044120}
  {\path{doi:10.1103/PhysRevE.104.044120}}.

\bibitem{Crossley:2015evo}
M.~Crossley, P.~Glorioso, H.~Liu, {Effective field theory of dissipative
  fluids}, JHEP 09 (2017) 095.
\newblock \href {http://arxiv.org/abs/1511.03646} {\path{arXiv:1511.03646}},
  \href {https://doi.org/10.1007/JHEP09(2017)095}
  {\path{doi:10.1007/JHEP09(2017)095}}.

\bibitem{Glorioso:2017fpd}
P.~Glorioso, M.~Crossley, H.~Liu, {Effective field theory of dissipative fluids
  (II): classical limit, dynamical KMS symmetry and entropy current}, JHEP 09
  (2017) 096.
\newblock \href {http://arxiv.org/abs/1701.07817} {\path{arXiv:1701.07817}},
  \href {https://doi.org/10.1007/JHEP09(2017)096}
  {\path{doi:10.1007/JHEP09(2017)096}}.

\bibitem{Gao:2018bxz}
P.~Gao, P.~Glorioso, H.~Liu, {Ghostbusters: Unitarity and Causality of
  Non-equilibrium Effective Field Theories}, JHEP 03 (2020) 040.
\newblock \href {http://arxiv.org/abs/1803.10778} {\path{arXiv:1803.10778}},
  \href {https://doi.org/10.1007/JHEP03(2020)040}
  {\path{doi:10.1007/JHEP03(2020)040}}.

\bibitem{Florio:2021jlx}
A.~Florio, E.~Grossi, A.~Soloviev, D.~Teaney, {Dynamics of the $O(4)$ critical
  point in QCD} (11 2021).
\newblock \href {http://arxiv.org/abs/2111.03640} {\path{arXiv:2111.03640}}.

\bibitem{Rennecke:2015lur}
F.~Rennecke, {The Chiral Phase Transition of QCD.}, Ph.D. thesis, U. Heidelberg
  (main) (2015).
\newblock \href {https://doi.org/10.11588/heidok.00019205}
  {\path{doi:10.11588/heidok.00019205}}.

\bibitem{PhysRevA.8.423}
P.~C. Martin, E.~D. Siggia, H.~A. Rose,
  \href{https://link.aps.org/doi/10.1103/PhysRevA.8.423}{Statistical dynamics
  of classical systems}, Phys. Rev. A 8 (1973) 423--437.
\newblock \href {https://doi.org/10.1103/PhysRevA.8.423}
  {\path{doi:10.1103/PhysRevA.8.423}}.
\newline\urlprefix\url{https://link.aps.org/doi/10.1103/PhysRevA.8.423}

\bibitem{Hertz_2016}
J.~A. Hertz, Y.~Roudi, P.~Sollich,
  \href{https://doi.org/10.1088/1751-8121/50/3/033001}{Path integral methods
  for the dynamics of stochastic and disordered systems}, Journal of Physics A:
  Mathematical and Theoretical 50~(3) (2016) 033001.
\newblock \href {https://doi.org/10.1088/1751-8121/50/3/033001}
  {\path{doi:10.1088/1751-8121/50/3/033001}}.
\newline\urlprefix\url{https://doi.org/10.1088/1751-8121/50/3/033001}

\bibitem{Sinner:2007ws}
A.~Sinner, N.~Hasselmann, P.~Kopietz, {Functional renormalization group in the
  broken symmetry phase: Momentum dependence and two-parameter scaling of the
  self-energy}, J. Phys. Condens. Matter 20 (2008) 075208.
\newblock \href {http://arxiv.org/abs/0707.4110} {\path{arXiv:0707.4110}},
  \href {https://doi.org/10.1088/0953-8984/20/7/075208}
  {\path{doi:10.1088/0953-8984/20/7/075208}}.

\bibitem{Son:1999pa}
D.~T. Son, {Hydrodynamics of nuclear matter in the chiral limit}, Phys. Rev.
  Lett. 84 (2000) 3771--3774.
\newblock \href {http://arxiv.org/abs/hep-ph/9912267}
  {\path{arXiv:hep-ph/9912267}}, \href
  {https://doi.org/10.1103/PhysRevLett.84.3771}
  {\path{doi:10.1103/PhysRevLett.84.3771}}.

\bibitem{Son:2002ci}
D.~T. Son, M.~A. Stephanov, {Real time pion propagation in finite temperature
  QCD}, Phys. Rev. D 66 (2002) 076011.
\newblock \href {http://arxiv.org/abs/hep-ph/0204226}
  {\path{arXiv:hep-ph/0204226}}, \href
  {https://doi.org/10.1103/PhysRevD.66.076011}
  {\path{doi:10.1103/PhysRevD.66.076011}}.

\bibitem{Fujii:2004za}
H.~Fujii, M.~Ohtani, {Soft modes at the critical end point in the chiral
  effective models}, Prog. Theor. Phys. Suppl. 153 (2004) 157--164.
\newblock \href {http://arxiv.org/abs/hep-ph/0401028}
  {\path{arXiv:hep-ph/0401028}}, \href {https://doi.org/10.1143/PTPS.153.157}
  {\path{doi:10.1143/PTPS.153.157}}.

\bibitem{Nakano:2011re}
E.~Nakano, V.~Skokov, B.~Friman, {Transport coefficients of O(N) scalar field
  theories close to the critical point}, Phys. Rev. D 85 (2012) 096007.
\newblock \href {http://arxiv.org/abs/1109.6822} {\path{arXiv:1109.6822}},
  \href {https://doi.org/10.1103/PhysRevD.85.096007}
  {\path{doi:10.1103/PhysRevD.85.096007}}.

\bibitem{Jeon:2015dfa}
S.~Jeon, U.~Heinz, {Introduction to Hydrodynamics}, Int. J. Mod. Phys. E
  24~(10) (2015) 1530010.
\newblock \href {http://arxiv.org/abs/1503.03931} {\path{arXiv:1503.03931}},
  \href {https://doi.org/10.1142/S0218301315300106}
  {\path{doi:10.1142/S0218301315300106}}.

\bibitem{Scavenius:2000qd}
O.~Scavenius, A.~Mocsy, I.~N. Mishustin, D.~H. Rischke, {Chiral phase
  transition within effective models with constituent quarks}, Phys. Rev. C 64
  (2001) 045202.
\newblock \href {http://arxiv.org/abs/nucl-th/0007030}
  {\path{arXiv:nucl-th/0007030}}, \href
  {https://doi.org/10.1103/PhysRevC.64.045202}
  {\path{doi:10.1103/PhysRevC.64.045202}}.

\bibitem{Fujii:2003bz}
H.~Fujii, {Scalar density fluctuation at critical end point in NJL model},
  Phys. Rev. D 67 (2003) 094018.
\newblock \href {http://arxiv.org/abs/hep-ph/0302167}
  {\path{arXiv:hep-ph/0302167}}, \href
  {https://doi.org/10.1103/PhysRevD.67.094018}
  {\path{doi:10.1103/PhysRevD.67.094018}}.

\bibitem{Guida:1998bx}
R.~Guida, J.~Zinn-Justin, {Critical exponents of the N vector model}, J. Phys.
  A 31 (1998) 8103--8121.
\newblock \href {http://arxiv.org/abs/cond-mat/9803240}
  {\path{arXiv:cond-mat/9803240}}, \href
  {https://doi.org/10.1088/0305-4470/31/40/006}
  {\path{doi:10.1088/0305-4470/31/40/006}}.

\bibitem{tauber}
U.~C. Täuber, Critical Dynamics: A Field Theory Approach to Equilibrium and
  Non-Equilibrium Scaling Behavior, Cambridge University Press, 2014.
\newblock \href {https://doi.org/10.1017/CBO9781139046213}
  {\path{doi:10.1017/CBO9781139046213}}.

\bibitem{Canet:2002gs}
L.~Canet, B.~Delamotte, D.~Mouhanna, J.~Vidal, {Optimization of the derivative
  expansion in the nonperturbative renormalization group}, Phys. Rev. D 67
  (2003) 065004.
\newblock \href {http://arxiv.org/abs/hep-th/0211055}
  {\path{arXiv:hep-th/0211055}}, \href
  {https://doi.org/10.1103/PhysRevD.67.065004}
  {\path{doi:10.1103/PhysRevD.67.065004}}.

\bibitem{Canet:2003qd}
L.~Canet, B.~Delamotte, D.~Mouhanna, J.~Vidal, {Nonperturbative renormalization
  group approach to the Ising model: A Derivative expansion at order
  partial**4}, Phys. Rev. B 68 (2003) 064421.
\newblock \href {http://arxiv.org/abs/hep-th/0302227}
  {\path{arXiv:hep-th/0302227}}, \href
  {https://doi.org/10.1103/PhysRevB.68.064421}
  {\path{doi:10.1103/PhysRevB.68.064421}}.

\bibitem{Kos:2016ysd}
F.~Kos, D.~Poland, D.~Simmons-Duffin, A.~Vichi, {Precision Islands in the Ising
  and $O(N)$ Models}, JHEP 08 (2016) 036.
\newblock \href {http://arxiv.org/abs/1603.04436} {\path{arXiv:1603.04436}},
  \href {https://doi.org/10.1007/JHEP08(2016)036}
  {\path{doi:10.1007/JHEP08(2016)036}}.

\bibitem{Komargodski:2016auf}
Z.~Komargodski, D.~Simmons-Duffin, {The Random-Bond Ising Model in 2.01 and 3
  Dimensions}, J. Phys. A 50~(15) (2017) 154001.
\newblock \href {http://arxiv.org/abs/1603.04444} {\path{arXiv:1603.04444}},
  \href {https://doi.org/10.1088/1751-8121/aa6087}
  {\path{doi:10.1088/1751-8121/aa6087}}.

\bibitem{onsager}
L.~Onsager, \href{https://link.aps.org/doi/10.1103/PhysRev.65.117}{Crystal
  statistics. i. a two-dimensional model with an order-disorder transition},
  Phys. Rev. 65 (1944) 117--149.
\newblock \href {https://doi.org/10.1103/PhysRev.65.117}
  {\path{doi:10.1103/PhysRev.65.117}}.
\newline\urlprefix\url{https://link.aps.org/doi/10.1103/PhysRev.65.117}

\bibitem{pft-2d}
V.~Prudnikov, A.~Ivanov, A.~Fedorenko, Critical dynamics of spin systems in the
  four-loop approximation, Journal of Experimental and Theoretical Physics
  Letters 66 (1997) 835--840.
\newblock \href {https://doi.org/10.1134/1.567606}
  {\path{doi:10.1134/1.567606}}.

\bibitem{PhysRevB.71.144406}
M.~J. Dunlavy, D.~Venus,
  \href{https://link.aps.org/doi/10.1103/PhysRevB.71.144406}{Critical slowing
  down in the two-dimensional ising model measured using ferromagnetic
  ultrathin films}, Phys. Rev. B 71 (2005) 144406.
\newblock \href {https://doi.org/10.1103/PhysRevB.71.144406}
  {\path{doi:10.1103/PhysRevB.71.144406}}.
\newline\urlprefix\url{https://link.aps.org/doi/10.1103/PhysRevB.71.144406}

\bibitem{zhong_critical_2018}
W.~Zhong, G.~T. Barkema, D.~Panja, R.~C. Ball,
  \href{https://link.aps.org/doi/10.1103/PhysRevE.98.062128}{Critical dynamical
  exponent of the two-dimensional scalar $ \phi^4 $ model with local moves},
  Physical Review E 98~(6) (2018) 062128.
\newblock \href {https://doi.org/10.1103/PhysRevE.98.062128}
  {\path{doi:10.1103/PhysRevE.98.062128}}.
\newline\urlprefix\url{https://link.aps.org/doi/10.1103/PhysRevE.98.062128}

\bibitem{PhysRevB.62.1089}
M.~P. Nightingale, H.~W.~J. Bl\"ote,
  \href{https://link.aps.org/doi/10.1103/PhysRevB.62.1089}{Monte carlo
  computation of correlation times of independent relaxation modes at
  criticality}, Phys. Rev. B 62 (2000) 1089--1101.
\newblock \href {https://doi.org/10.1103/PhysRevB.62.1089}
  {\path{doi:10.1103/PhysRevB.62.1089}}.
\newline\urlprefix\url{https://link.aps.org/doi/10.1103/PhysRevB.62.1089}

\bibitem{Adzhemyan:2021hvo}
L.~T. Adzhemyan, D.~A. Evdokimov, M.~Hnati\v{c}, E.~V. Ivanova, M.~V.
  Kompaniets, A.~Kudlis, D.~V. Zakharov, {The dynamic critical exponent z for
  2d and 3d Ising models from five-loop \ensuremath{\varepsilon} expansion},
  Phys. Lett. A 425 (2022) 127870.
\newblock \href {http://arxiv.org/abs/2111.04719} {\path{arXiv:2111.04719}},
  \href {https://doi.org/10.1016/j.physleta.2021.127870}
  {\path{doi:10.1016/j.physleta.2021.127870}}.

\bibitem{Hasenbusch_2020}
M.~Hasenbusch,
  \href{https://link.aps.org/doi/10.1103/PhysRevE.101.022126}{Dynamic critical
  exponent $z$ of the three-dimensional ising universality class: Monte carlo
  simulations of the improved blume-capel model}, Phys. Rev. E 101 (2020)
  022126.
\newblock \href {https://doi.org/10.1103/PhysRevE.101.022126}
  {\path{doi:10.1103/PhysRevE.101.022126}}.
\newline\urlprefix\url{https://link.aps.org/doi/10.1103/PhysRevE.101.022126}

\bibitem{DePolsi:2020pjk}
G.~De~Polsi, I.~Balog, M.~Tissier, N.~Wschebor, {Precision calculation of
  critical exponents in the $O(N)$ universality classes with the
  nonperturbative renormalization group}, Phys. Rev. E 101~(4) (2020) 042113.
\newblock \href {http://arxiv.org/abs/2001.07525} {\path{arXiv:2001.07525}},
  \href {https://doi.org/10.1103/PhysRevE.101.042113}
  {\path{doi:10.1103/PhysRevE.101.042113}}.

\bibitem{Grossi:2019urj}
E.~Grossi, N.~Wink, {Resolving phase transitions with Discontinuous Galerkin
  methods} (3 2019).
\newblock \href {http://arxiv.org/abs/1903.09503} {\path{arXiv:1903.09503}}.

\bibitem{Koenigstein:2021syz}
A.~Koenigstein, M.~J. Steil, N.~Wink, E.~Grossi, J.~Braun, M.~Buballa, D.~H.
  Rischke, {Numerical fluid dynamics for FRG flow equations: Zero-dimensional
  QFTs as numerical test cases - Part I: The $O(N)$ model} (8 2021).
\newblock \href {http://arxiv.org/abs/2108.02504} {\path{arXiv:2108.02504}}.

\bibitem{Koenigstein:2021rxj}
A.~Koenigstein, M.~J. Steil, N.~Wink, E.~Grossi, J.~Braun, {Numerical fluid
  dynamics for FRG flow equations: Zero-dimensional QFTs as numerical test
  cases -- Part II: Entropy production and irreversibility of RG flows} (8
  2021).
\newblock \href {http://arxiv.org/abs/2108.10085} {\path{arXiv:2108.10085}}.

\bibitem{Steil:2021cbu}
M.~J. Steil, A.~Koenigstein, {Numerical fluid dynamics for FRG flow equations:
  Zero-dimensional QFTs as numerical test cases - Part III: Shock and
  rarefaction waves in RG flows reveal limitations of the $N \rightarrow
  \infty$ limit in $O(N)$-type models} (8 2021).
\newblock \href {http://arxiv.org/abs/2108.04037} {\path{arXiv:2108.04037}}.

\bibitem{Ihssen:2023qaq}
F.~Ihssen, F.~R. Sattler, N.~Wink, {Numerical RG-time integration of the
  effective potential: Analysis and Benchmark} (2 2023).
\newblock \href {http://arxiv.org/abs/2302.04736} {\path{arXiv:2302.04736}}.

\end{thebibliography}

\end{document}